# How quickly can sodium-ion learn? Assessing scenarios for techno-economic competitiveness against lithium-ion batteries

Adrian Yao[1,3], Sally M. Benson[2], and William C. Chueh[*,1,2,3]

[1]Department of Materials Science & Engineering, Stanford University, Stanford, CA 94305, USA.
[2]Department of Energy Science & Engineering, Stanford University, Stanford, CA 94305, USA.
[3]Applied Energy Division, SLAC National Accelerator Laboratory, Menlo Park, CA 94025, USA.

**ABSTRACT**

Sodium-ion batteries have garnered significant attention as a potentially low-cost alternative to lithium-ion batteries, which have experienced supply shortages and pricing volatility of key minerals. Here we assess their techno-economic competitiveness against incumbent lithium-ion batteries using a modeling framework incorporating componential learning curves constrained by minerals prices and engineering design floors. We compare projected sodium-ion and lithium-ion price trends across over 6,000 scenarios while varying Na-ion technology development roadmaps, supply chain scenarios, market penetration, and learning rates. Assuming substantial progress can be made along technology roadmaps via targeted R&D, we identify several sodium-ion pathways that might reach cost-competitiveness with low-cost lithium-ion variants in the early 2030s. Additionally, we show timelines are highly sensitive to movements in critical minerals supply chains—namely that of lithium, graphite, and nickel. Modeled outcomes suggest increasing sodium-ion energy densities to decrease materials intensity to be among the most impactful ways to improve competitiveness.

The energy transition requires massive deployment of batteries for electric vehicles (EVs) and stationary energy storage systems (ESS). Lithium-ion (Li-ion) batteries have been responsible for nearly all new deployments of storage in recent years[1–3], largely enabled by the tremendous cost declines over the past three decades of commercialization characterized by an aggressive learning rate where prices have fallen by more than 97% since they were first commercialized in 1991[4,5]. This trend is reflected in the average Li-ion cell prices shown in Supplementary Discussion 1 aggregated from the industry data contributors to this work[1–3,6], which also resolves the diverging price trajectories of NMC-type (nickel-manganese-cobalt oxide) and LFP-type (lithium iron phosphate) cathode chemistries. However, rapid Li-ion demand growth has recently placed a significant burden on the minerals supply chain—namely that of lithium, nickel, graphite, and cobalt—resulting in a first-ever increase in the average Li-ion cell price index in 2022[3] which subsequently fell again as minerals prices plummeted[7–9]. This has prompted some to escalate concerns regarding the possible overdependence on Li-ion and risks of production bottlenecks, supply chain shocks, and geopolitical constraints.

Sodium-ion (Na-ion) batteries present a potentially viable near-term substitute for Li-ion for two primary reasons: (1) increased abundance and availability of sodium suggests lower prices and (2) drop-in compatibility with Li-ion manufacturing infrastructure suggests rapid scaling timelines. Therefore, in response to severe post-COVID lithium price spikes, manufacturers recently announced over 240 GWh of Na-ion cell manufacturing pipeline through 2030[10], promising lower prices than Li-ion. However, exactly if, when, and by how much Na-ion batteries will be price advantageous is still largely a matter of speculation.

---

[*] Corresponding Author. E-mail: wchueh@stanford.edu

Learning curves have been widely used in forecasting technological progress and price since patterns of improvement were first postulated by Wright for manufacturing airplanes[11]. The typical learning curve (Wright's Law) predicts price reductions as a function of cumulative production ("learning-by-doing"), and has been statistically shown to produce accurate forecasts across wide-ranging sectors[12], including energy technologies[13]. Given the industrial importance of Li-ion batteries, several studies have sought to characterize their price trends using the conventional form of Wright's Law[4,14–16] (Equation (1) in Methods) whereas others have ascribed additional descriptors, such as learning-by-researching or economies-of-scale, to understand their historic trends[4,5,17]. However, conventional learning curves unrealistically assume infinite cost reductions approaching zero, which neglects limits set by underlying minerals. Therefore, yet other approaches have sought to additionally constrain the learning curve with a minerals price floor to produce more realistic projections[18,19]. It is worth noting that not all price floors or additional parameters can be assumed to be effective. Price floors based upon expert opinion, for example, have historically been shown to be unrealistic[20], and extra parameters have been shown to reduce model performance due to overfitting[12].

In this work, we propose an approach that combines componential, floor-constrained learning curves for individual material components with technical development roadmaps that also exhibit learning behavior to better forecast future price trends—balancing model simplicity and physical accuracy. We then use it to forecast battery price trends for Li-ion and Na-ion in various technological, market, and supply chain conditions to identify strategies that improve the techno-economic competitiveness of Na-ion.

**Combining floor-constrained learning curves with physically-accurate technology roadmaps**

We construct historical and forecasted aggregate price curves by incorporating knowledge of both (a) material component prices (in \$-kg$^{-1}$ or \$-m$^{-2}$) which evolve via learning-by-doing as a function of cumulative experience and (b) materials intensities (kg-kWh$^{-1}$) which evolve due to continued cell engineering and materials advancements as a function of time. We institute practical floors for both—using minerals price floors for the former and engineering limits for the latter. This approach thus combines a modified-Wright's Law and a modified-Moore's Law, represented by Equations (3) and (4) in Methods, respectively.

For each material component in a battery cell, we construct individual learning curves and institute a price floor based upon their known elemental composition of benchmarkable key minerals which dynamically evolve with time. Therefore, each of our material component costs comprise two elements: (1) a fluctuating materials price floor representing the weighted sum of underlying minerals from a time series of historical and forecasted prices, and (2) decreasing costs of non-mineral inputs represented by a learning curve that scales with historical and forecasted market share. Non-mineral inputs include costs associated with capital equipment, operating costs, production scrap from yield, chemical conversion inefficiencies, and others. The downward trajectory of the learning curve component is consistent with expectations of technological improvements due to processing innovation and optimization as well as implicit assumptions of economies-of-scale. In contrast, market-driven volatility is expected in the materials price floor component. This approach is analogous to prior literature decomposing coal-fired electricity generation costs into a "noisy" price floor of fuel commodities and improving costs of plant construction due to technological learning[19]. In this case, however, we do not disaggregate the materials component learning curves further into the technical factors identified above, and we highlight this area for future work.

Once the floor-constrained learning curves are obtained for each material component, they are subsequently summed according to their mass- or areal-intensities derived from physically-accurate cell modeling. To capture the steady decreases in materials intensities over time (which have contributed greatly to Li-ion price declines to date[4,5]), we model cell designs in a bottoms-up fashion that are representative of past commercial products, present state-of-art, and future designs considering anticipated improvements for materials performance and cell engineering. Based upon differences in materials intensities between cell designs which are each ascribed to a point in time, we construct roadmaps by fitting curves with asymptotic limits defined by theoretical or practical engineering limits. This approach is similar to prior literature disaggregating solar photovoltaic device-level price reductions based on observable technical factors such as silicon consumption and wafer sizes[21]. However, in this work, we add resolution by fitting materials intensity curves for each component based on bottoms-up modelling, and we do not decompose the costs of manufacturing further into other technical factors. We justify this based on knowing battery cell costs are dominated by their bill-of-materials, and we seek to specifically probe how materials and cell design improvements can alter price trajectories. Our modeled manufacturing overhead cost trends therefore include cost savings from improved processing yield and economies-of-scale. To obtain bill-of-materials and manufacturing costs, we leverage the Battery Performance and Cost (BatPaC) process-based model[22] developed by Argonne National Laboratory to recreate cell designs (see Methods). As BatPaC specifically models state-of-art manufacturing, we only use the manufacturing overhead cost component of present-day cell designs and subsequently fit learning curves against historical data on manufacturing costs to estimate future trends.

Our models rely upon a wide-ranging collection of sources obtained from industry data contributors to ensure accuracy and relevance. They include Benchmark Mineral Intelligence, Wood Mackenzie, BloombergNEF, Avicenne Energy, S&P Global, and other commercial reports[23,24]. Additionally, we validate and confirm accuracy and relevance by surveying a panel of industry experts to obtain inputs on realistic price figures, state-of-art cell designs, and production costs[25].

**Establishing Li-ion baselines**

We divide the effort to construct Li-ion baseline curves into two parts as discussed above: (a) establishing floor-constrained materials component learning curves and (b) establishing physically-accurate technology roadmaps. We define 2023 as present-day. Prior years are designated historical and static which inform parameter fitting. Forward years are projected, driven by the inputs of minerals pricing and market growth forecasts.

*Li-ion materials component learning curves*

An example procedure for establishing the floor-constrained materials component learning curves is illustrated in Figure 1 for NMC cathode material. Historical price assessments were tabulated from industry and literature sources to establish a time series starting at or before 2010[2,23,5,26,27]. We also collected data on historical production quantities to enable calculation of cumulative "experience". Both are shown in Figure 1a. Historical pricing for key minerals were obtained from industry data contributors[1,2,28,29]. Historical prices were also collected from the United States Geological Survey (USGS) mineral commodity summaries dating back to 1991[30], and all price figures were inflation-adjusted to USD2023. Aggregated minerals prices are shown in Supplementary Discussion 2, but a subset of the benchmarkable minerals comprising NMC— lithium, nickel, manganese, and cobalt—are shown in Figure 1b. Subsequently, the weighted sum of each mineral component is added according to their stoichiometric ratios to produce a material component price floor, shown in Figure 1c (see Supplementary Discussion 3 for a comprehensive list of material components and their compositions). For NMC-type materials that have historically experienced an evolution in stoichiometry, we take a volume-weighted average of each "nodal" composition to reflect a representative blend at every time step (see Supplementary Discussion 4). Finally, with knowledge of component price floors, actual historical price assessments, and cumulative production experience, this sufficiently constrains the modified-Wright's Law to leave the learning rate as the only fitting parameter, per Equation (3) in Methods. The learning curve then represents the non-mineral costs associated with the material component, and is fit in a way so that the total (price floor + learning curve) matches historical price assessments, shown in Figure 1d. See Supplementary Discussion 5.1 for additional comments on sources of uncertainty given data scarcity.

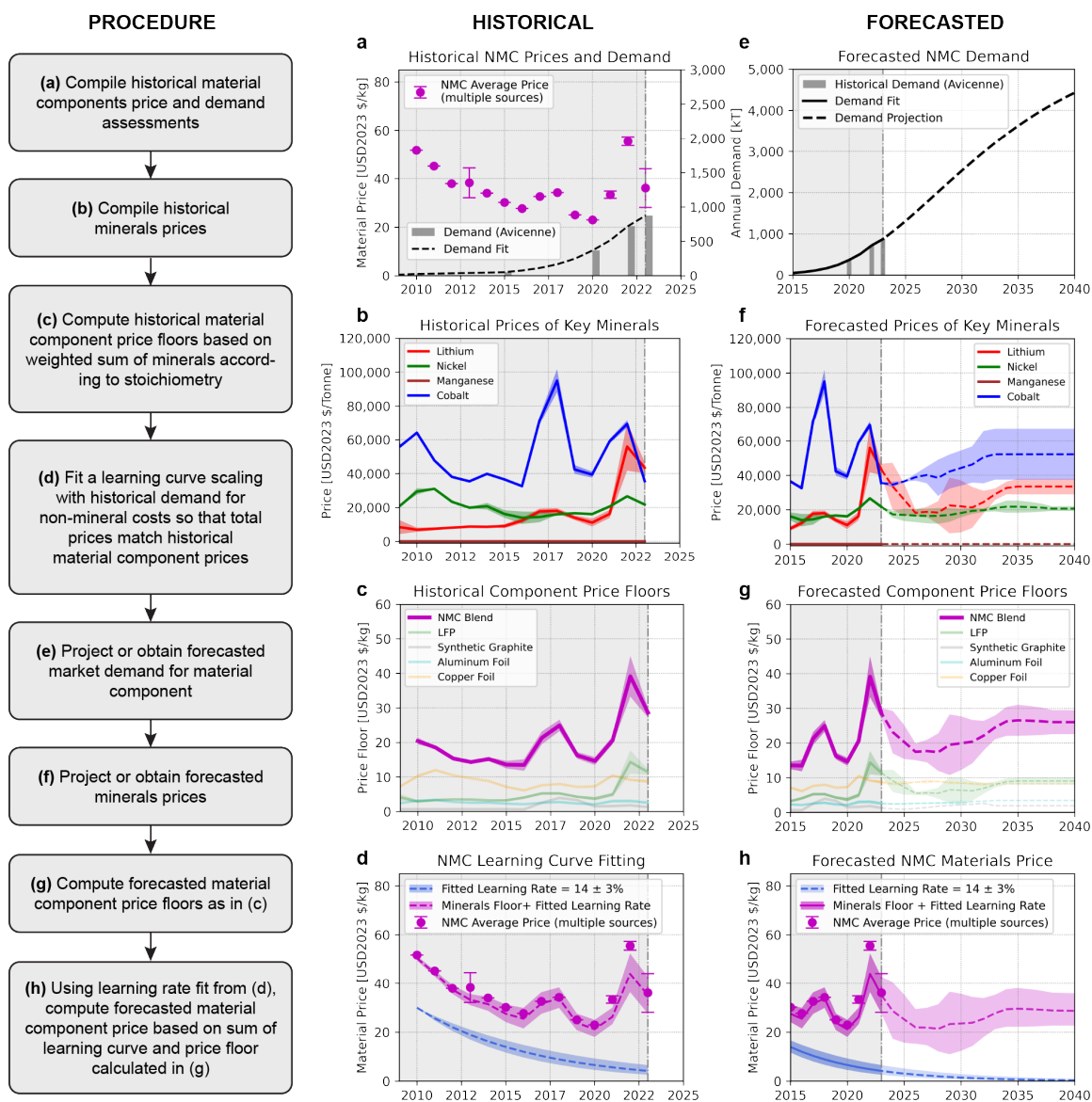

*Figure 1: **Establishing floor-constrained materials component learning curves.** A step-by-step process (flowchart in left column) of fitting material component learning rates based upon historical data (center column) and then projecting future materials prices based on forecasted demand and minerals prices (right column).* **a.** *Historical material component price assessments for NMC cathode active material mapped against historical material demand obtained from industry sources. Error bars show 95% confidence interval for years where multiple data sources were available.* **b.** *Historical prices of key benchmarkable minerals in NMC—namely lithium, nickel, manganese, and cobalt—obtained from industry sources. Shaded regions show 95% confidence interval.* **c.** *Historical material component price floors representing a weighted sum of underlying minerals according to their stoichiometric ratios. See Supplementary Discussion 4 for details on capturing the evolution in NMC stoichiometry—represented here as a "NMC blend". Price floors of other material components also shown.* **d.** *Fitting a learning curve for the non-mineral cost components of NMC. Combining the learning curve with the price floor shown in* **(c)***, a total material component price can be obtained which agrees with historical price assessments.* **e.** *Forecasted demand for NMC material based upon market projections obtained from industry partners.* **f.** *Forecasted minerals prices for the same key elements in* **(b)** *obtained from industry partners.* **g.** *Calculated material component price floors based on forecasted minerals prices.* **h.** *Forecasted material component prices obtained by combining the price floor shown in* **(g)** *with a continuation of the same learning curve obtained in* **(d)** *while keeping the learning rate parameter fixed. We recognize here our assumptions that learning rates remain fixed over a forecasted timespan longer than the fitted historical data. Long-term minerals price forecasts are also based upon expert opinion of industry partners. Therefore, these forecasted prices represent a simulated future based upon best knowledge today. See Supplementary Discussion 2.1 for additional high/mid/low minerals price simulations.*

Once a characteristic learning rate is established, forecasting future prices can be performed based upon forecasted market growth and minerals pricing, assuming learning rates do not change. It is worth noting that endogeneity is expected between the two, but we leave them independent. Additionally, we recognize the forecasted timespan is longer than that of the historical data used to fit the constant learning rate, but this provides a first-order approximation sufficient for future simulation purposes. Forecasted Li-ion market growth from several market intelligence sources were approximated with Gompertz functions to obtain continuous curves (see Methods and Supplementary Discussion 6), where the "LIB Base Case" scenario assumes no appreciable Na-ion deployment and LFP market penetration reaching 50% market share saturation by 2050 (see Supplementary Discussion 6.1). This can then be translated to NMC demand in Figure 1e by mass. Forecasted minerals prices are again provided by data contributors and aggregated in Figure 1f. As predicting future price movements of key minerals is inherently challenging, we also separately model fixed-price high/mid/low forecasted scenarios for lithium and nickel to test sensitivity (see Supplementary Discussion 2.1). Material component price floors can again be calculated (Figure 1g). To ensure continuity between historical and forecasted materials price time series, both were constrained to converge on present-day (2023) pricing, which were obtained from various industry sources and aggregated for robustness[2,23,25–27] (a summary of Li-ion component price assessments is presented in Supplementary Discussion 7). The final result is a material component price forecast that is responsive to dynamic minerals price movements as well as market penetration (Figure 1h).

We repeat the above procedure for a full set of material components used in Li-ion batteries, including various cathode material chemistries. Whereas NMC was used as an illustrative example above, hereafter we focus on LFP in the Main Text as it is widely considered the more relevant comparison with Na-ion with respect to performance and cost. Using the same methodology, the price forecast for LFP material is shown in Figure 2a. The remaining material price curves are presented in Supplementary Discussion 5, and their fitted learning rates are summarized in Figure 2b in comparison with average historical Li-ion learning rates aggregated from prior literature[3,4,14,15,17].

*Li-ion technology roadmaps*

To establish Li-ion technology roadmaps, physically-accurate models of NMC and LFP cell designs representing past and present were developed. Present cell designs were modeled based upon recent teardowns of state-of-art commercial cells published in literature[31–33] (see Supplementary Discussion 8). Past cell designs were also modeled based upon previously published commercial cell teardowns in literature and industry reports documenting engineering progress at the time of their publication[34–36]. This enabled fitting of materials intensity trends that correspond with development roadmaps (see Supplementary Discussion 9). Note here that we do not model next-generation Li-ion roadmaps (lithium manganese iron phosphate cathodes, silicon anodes, etc.) which could significantly reduce materials intensity. Thus, our analysis presents a conservative case of anticipated price reductions for Li-ion.

Finally, combining material component price forecasts and materials intensity roadmaps per Equation (12), complete Li-ion price curves were constructed as shown in Figure 2c for LFP (NMC component curve stack-up shown in Supplementary Discussion 10). The final LFP price curve is shown in Figure 2d with predicted prices reaching \$51 kWh$^{-1}$ by 2030 assuming lithium prices at the time are just under \$22,000 tonne$^{-1}$ (and as low as \$45 kWh$^{-1}$ if lithium prices are ~\$10,500 tonne$^{-1}$ per the low price scenario in Supplementary Discussion 2.1).

**A closer look at learning**

Examining the fitted componential learning rates, learning rates appear correlated with processing complexity, which may be anecdotally associated with the number of required control parameters during production (a clearer definition and correlation of complexity to learning rate is a subject of future work). Higher complexity processes (e.g. cathode synthesis, cell assembly) tend to have higher rates of learning whereas lower complexity processes (e.g. metal foil and anode production) tend to have lower rates (see

Supplementary Discussion 5.2 for further discussion). However, in comparison to previous literature on Li-ion learning rates, we observe that our componential learning rates are all lower than the aggregate average rate experienced by historical cell prices (Figure 2b) of 21.1 ± 3.7%[3,4,14,15,17], which suggests there must be another key contributor responsible for the aggressive price reductions in Li-ion. This can be explained by the technical advancements in the design of Li-ion cells over the past several decades, now requiring substantially lower materials intensity on a per-kWh basis than before[4,5]. Specifically, the shift from cells optimized for power to those optimized for energy enabled higher active material utilization[5,16,36]. The steady improvements to material specific capacities have also contributed to this phenomenon.

Therefore, the aggressive Li-ion price trends appears to be a story of modest price reductions in materials from learning-by-doing compounded by significant engineering advancements and learnings. This deconvolution of the traditional learning model into key constituent contributors is an important nuance to forecast realistic price trends of incumbent and emerging technologies like Li-ion and Na-ion, respectively. To illustrate, Figure 2d additionally compares price forecasts of LFP cells using our method—accounting for minerals price and engineering design floors—against the typical unconstrained single-factor learning curve form of Wright's Law (Equation (1) in Methods) utilizing the historical average Li-ion rate. This curve nearly intersects the calculated LFP minerals price floor in 2050, which may be unrealistic and overly optimistic.

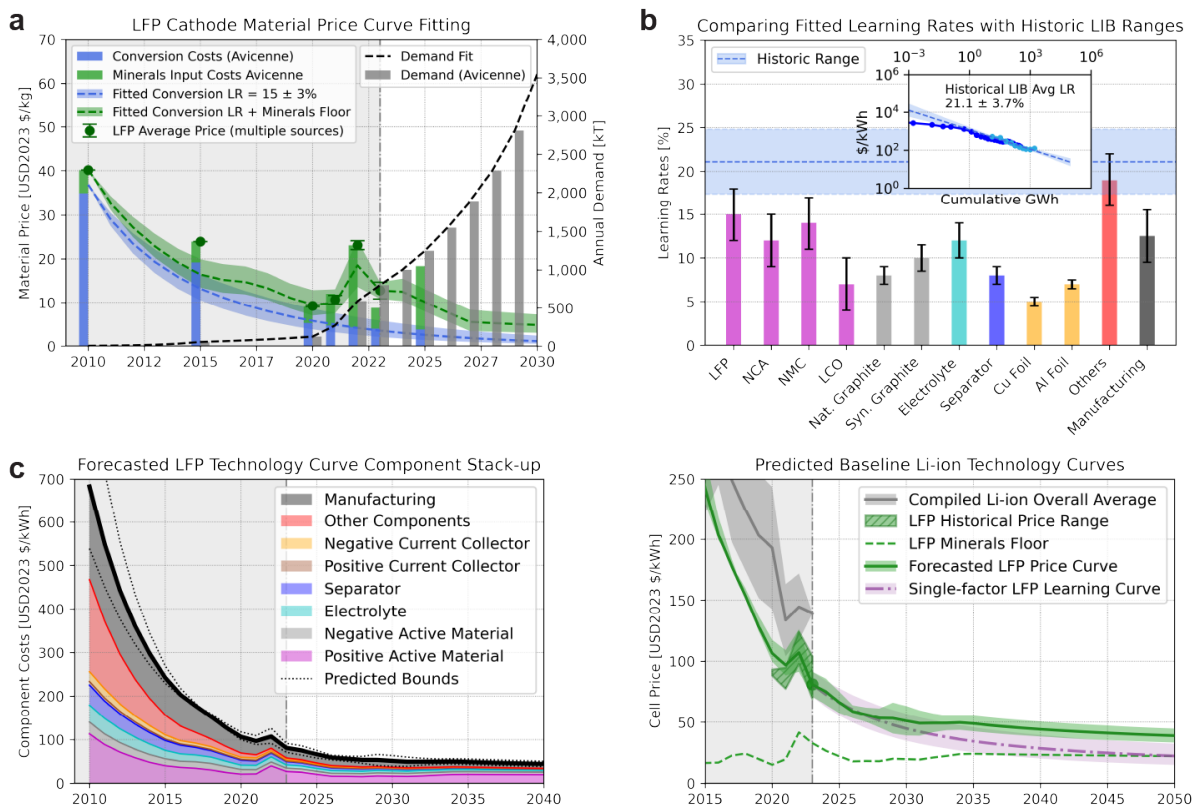

*Figure 2: Componential floor-constrained construction of technology price curves.* **a.** *Example material component price curve generation for LFP material. A learning curve (in blue) is fit against historical conversion cost data to capture experience as cumulative production increases (gray bars and dotted black line) while a minerals price floor (in green) rides upon the learning curve. Reasonable agreement between the generated price curve and average prices collected from industry sources is observed.* **b.** *Fitted learning rates for key Li-ion material components (NCA=nickel-cobalt-aluminum oxide, LCO=lithium cobalt oxide, Nat. Graphite=natural graphite, Syn. Graphite=synthetic graphite) compared against the historical average learning rate (LR) for Li-ion compiled from literature and industry reports—shown in inset. See Supplementary Discussion 5 for details.* **c.** *Total price curve for LFP-type cells constructed by summing up all constituent material components based upon cell design models that inform materials intensity per energy content stored. See Supplementary Discussion 9 for equivalent plots for NMC.* **d.** *Forecasted LFP Li-ion price curve for the baseline scenario, showing good agreement with historical averages. A typical single-factor learning curve using the historical average learning rate (established in* **b***) is also shown for comparison with potentially overly optimistic outcomes. Confidence intervals capture underlying uncertainty in minerals prices, starting materials prices, and learning rates.*

**Forecasting Na-ion cell development roadmaps**

Like Li-ion, Na-ion is an umbrella term encompassing several material chemistries generally categorized by cathode material classes. They include transition metal layered oxides, polyanions, and Prussian Blue analogues (PBAs)[37]. Recent commercialization efforts indicate significant momentum behind (i) layered oxides with a general formula of $Na_x(M)O_2$ (x~1, M=Ni, Fe, Mn, Cu) with $NaNi_xMn_y(M)_{1-x-y}O_2$ (NaNM) being the most common and (ii) vanadium-free mixed-polyanions with the formula $Na_4Fe_3(PO_4)_2(P_2O_7)$ (NFPP)[25]. In almost all instances, disordered hard carbons (HC) are used as Na-ion anode materials instead of the crystalline graphite used in Li-ion batteries due to their low sodium storage capacities[38]. Alternative anodes to hard carbons include emerging technology directions such as (i) metallic tin (Sn) that alloys with sodium[39] or (ii) "anode-free" configurations that electrodeposit and strip sodium metal directly on current collectors[40,41]. Here, in reference to NaNM, we specifically model $NaNi_{0.33}Mn_x(M)_{0.67-x}O_2$ where M includes Mg and Ti dopants and that are currently being commercialized[42–44]. As the non-nickel transition metals contribute negligible minerals costs and redox activity[44], our NaNM archetype also functionally represents compositions such as $NaNi_{0.33}Fe_{0.33}Mn_{0.33}O_2$ (NFM111). Given the recent interest in replacing nickel with earth-abundant elements like iron and manganese due high costs, we also model a material such as $Na_{0.67}Fe_{0.5}Mn_{0.5}O_2$ (NaFM). However, as these materials still suffer from unresolved performance challenges, we model technical roadmaps that gradually reduce nickel to 0% by ~2040 (see Supplementary Discussion 11)—analogous to the gradual reduction of cobalt content in NMC cathodes over the years. This is modeled separately from roadmaps that maintain a static 33% nickel stoichiometry.

The three key market segments modeled were stationary ESS, two/three-wheelers and micromobility (TTW/MM), and EVs. Specifically, we consider three hypothetical Na-ion market growth scenarios with increasing degrees of penetration, ranging from the most conservative: 100% Na-ion penetration into ESS by 2040, to the most aggressive: 100% penetration into ESS by 2035, 100% of TTW/MM by 2040, and 25% EVs by 2040. Note that the most conservative market penetration scenario matches more closely to forecasts provided by industry data contributors.

Given the importance of capturing engineering advancements on techno-economics, we model a multitude of Na-ion cell development roadmaps for NaNM|HC, NaFM|HC, NFPP|HC, NaNM|Sn, NaFM|Sn, NFPP|Sn, NaNM|Anode-free, and NaFM|Anode-free cell designs (see Supplementary Discussion 8 for details on 18 different Na-ion cell designs). These roadmaps assume gradual improvements from state-of-art 2024 designs toward future technical advancements by 2030 that increase energy densities and thereby reduce prices. See Methods and Supplementary Discussion 9 for details on fitting trends with a modified Moore's Law. A summary of the modeled technical roadmaps is tabulated in Table 1.

*Table 1: Summary of modeled Na-ion technical development roadmaps including gravimetric and volumetric energy densities (GED and VED) of modeled large-format pouch cells.*

| Roadmap Name | Starting Cell Design (2024) | | Improved Cell Design (2030) | GED / VED [Wh/kg] / [Wh/L] | |
|---|---|---|---|---|---|
| | | | | 2024 | 2030 |
| NaNM\|HC 0 | NaNM \| HC Baseline | ⇒ | NaNM \| HC Baseline (No improvements) | 134 / 272 | |
| NaNM\|HC 1 | NaNM \| HC Baseline | ⇒ | Increased electrode loadings | 134 / 272 | 142 / 295 |
| NaNM\|HC 2 | NaNM \| HC Baseline | ⇒ | Anode capacity increase to 400 mAh-g$^{-1}$ | 134 / 272 | 157 / 325 |
| NaNM\|HC 3 | NaNM \| HC Baseline | ⇒ | Cathode 20% capacity increase | 134 / 272 | 160 / 314 |
| NaNM|HC 4 | NaNM | HC Baseline | ⇒ | Anode & cathode capacity increase | 134 / 272 | 174 / 347 |
| NaNM|HC 5 | NaNM | HC 4.2V Baseline | ⇒ | Anode & cathode capacity increase @ 4.2V | 134 / 272 | 196 / 376 |
| NaNM|HC 6 | NaNM | HC Baseline | ⇒ | Anode capacity increase to 400 mAh-g$^{-1}$ @ 4.2V | 134 / 272 | 181 / 359 |
| NaNM|HC 7 | NaNM | HC Baseline | ⇒ | Anode & cathode capacity increase @ 4.2V | 155 / 303 | 196 / 376 |
| NaFM|HC 1 | NaFM | HC Baseline | ⇒ | Increased electrode loadings | 134 / 272 | 142 / 295 |
| NaFM|HC 2 | NaFM | HC Baseline | ⇒ | Anode capacity increase to 400 mAh-g$^{-1}$ | 134 / 272 | 157 / 325 |
| NaFM|HC 3 | NaFM | HC Baseline | ⇒ | Cathode 20% capacity increase | 134 / 272 | 160 / 314 |
| NaFM|HC 4 | NaFM | HC Baseline | ⇒ | Anode & cathode capacity increase | 134 / 272 | 174 / 347 |
| NaFM|HC 5 | NaFM | HC 4.2V Baseline | ⇒ | Anode & cathode capacity increase @ 4.2V | 134 / 272 | 196 / 376 |
| NaFM|HC 6 | NaFM | HC Baseline | ⇒ | Anode capacity increase to 400 mAh-g$^{-1}$ @ 4.2V | 134 / 272 | 181 / 359 |
| NaNM|Sn 0 | NaNM | Sn Baseline | ⇒ | NaNM | Sn Baseline (No improvements) | 185 / 476 | |
| NaNM|Sn 1 | NaNM | Sn Baseline | ⇒ | Cathode 20% capacity increase | 185 / 476 | 216 / 551 |
| NaFM|Sn 0 | NaFM | Sn Baseline | ⇒ | NaFM | Sn Baseline (No improvements) | 185 / 476 | |
| NaFM|Sn 1 | NaFM | Sn Baseline | ⇒ | Cathode 20% capacity increase | 185 / 476 | 216 / 551 |
| NFPP|HC 0 | NFPP | HC Baseline | ⇒ | NFPP | HC Baseline (No improvements) | 111 / 210 | |
| NFPP|HC 1 | NFPP | HC Baseline | ⇒ | Increased electrode loadings | 111 / 210 | 119 / 226 |
| NFPP|HC 2 | NFPP | HC Baseline | ⇒ | Anode capacity increase to 400 mAh-g$^{-1}$ | 111 / 210 | 126 / 241 |
| NFPP|Sn 0 | NFPP | Sn Baseline | ⇒ | NFPP | Sn Baseline (No improvements) | 137 / 305 | |
| NFPP|Sn 1 | NFPP | Sn Baseline | ⇒ | Increased electrode loadings | 137 / 305 | 147 / 330 |
| NaNM|AF 0 | NaNM | Anode-free Base. | ⇒ | NaNM | Anode-free Baseline (No improvements) | 207 / 547 | |
| NaNM|AF 1 | NaNM | Anode-free Base. | ⇒ | Increased electrode loadings | 207 / 547 | 239 / 635 |
| NaNM|AF 2 | NaNM | Anode-free Base. | ⇒ | Cathode 20% capacity increase | 207 / 547 | 283 / 734 |
| NaFM|AF 0 | NaFM | Anode-free Base. | ⇒ | NaFM | Anode-free Baseline (No improvements) | 207 / 547 | |
| NaFM|AF 1 | NaFM | Anode-free Base. | ⇒ | Cathode 20% capacity increase | 207 / 547 | 283 / 734 |

Evaluating Roadmap "NaNM|HC 7" as an example: a NaNM|HC Baseline cell design (see Supplementary Discussion 8.7) evolves to experience an increase in hard carbon specific capacity from 330 mAh-g$^{-1}$ to 400 mAh-g$^{-1}$, an increase in cell operating voltage from 4.0 to 4.2V, and a 20% increase to NaNM specific capacity (to ~175 mAh-g$^{-1}$ measured from 2.0 to 4.25V) (see Supplementary Discussion 8.13)—all by 2030. The resulting materials intensity trends are shown in Figure 3a. Pricing for key Na-ion materials components are obtained by surveying industry expert sources to ensure commercial relevance[25] (see Methods). These values are summarized in Figure 3b. Applying various learning rates and market growth assumptions, we forecast Na-ion price curves via our componential construction methodology. Figure 3c illustrates a scenario with the most conservative market penetration (100% Na-ion penetration into ESS by 2040, see Supplementary Discussion 6) and with novel cumulative experience applied to the anode material only (see Supplementary Discussion 12).

It is worth highlighting our key assumptions. Here, we assume: (1) drop-in compatibility with existing Li-ion manufacturing infrastructure, which implies immediate high-yield production and no new manufacturing learnings (see Supplementary Discussion 12), (2) decoupled market penetration and technology pricing, whereby endogeneity captured within a typical demand curve is ignored for the sake of modeling and understanding intervention opportunities, and (3) transferable learning rates from Li-ion to Na-ion, where cathodes, anodes, and other components learn at their same respective rates between the two alkali chemistries. This last assumption may be supported by the earlier observation that materials production with similar degrees of process complexity have similar rates of learning (see Supplementary Discussion 5.2), but more aggressive learning rates are also evaluated in our scenarios analysis below. We note that the first two assumptions enable a best-case scenario to model Na-ion's techno-economic competitiveness as a starting point. Additionally, we do not include next-generation Li-ion roadmaps (lithium manganese iron phosphate cathodes, silicon anodes, etc.) to again evaluate a best-case scenario for Na-ion. We finally emphasize that here we are only assessing the technologies at the cell-level with the price of stored energy ($-kWh$^{-1}$) as the primary figure of merit. Therefore, this work does not capture systems-level considerations (e.g. potential savings on pack integration due to enhanced safety or reduced thermal management) or alternative performance-enabled economic considerations (e.g. potentially lower levelized price of storage due to increased cycle life). We address these additional complexities in future work.

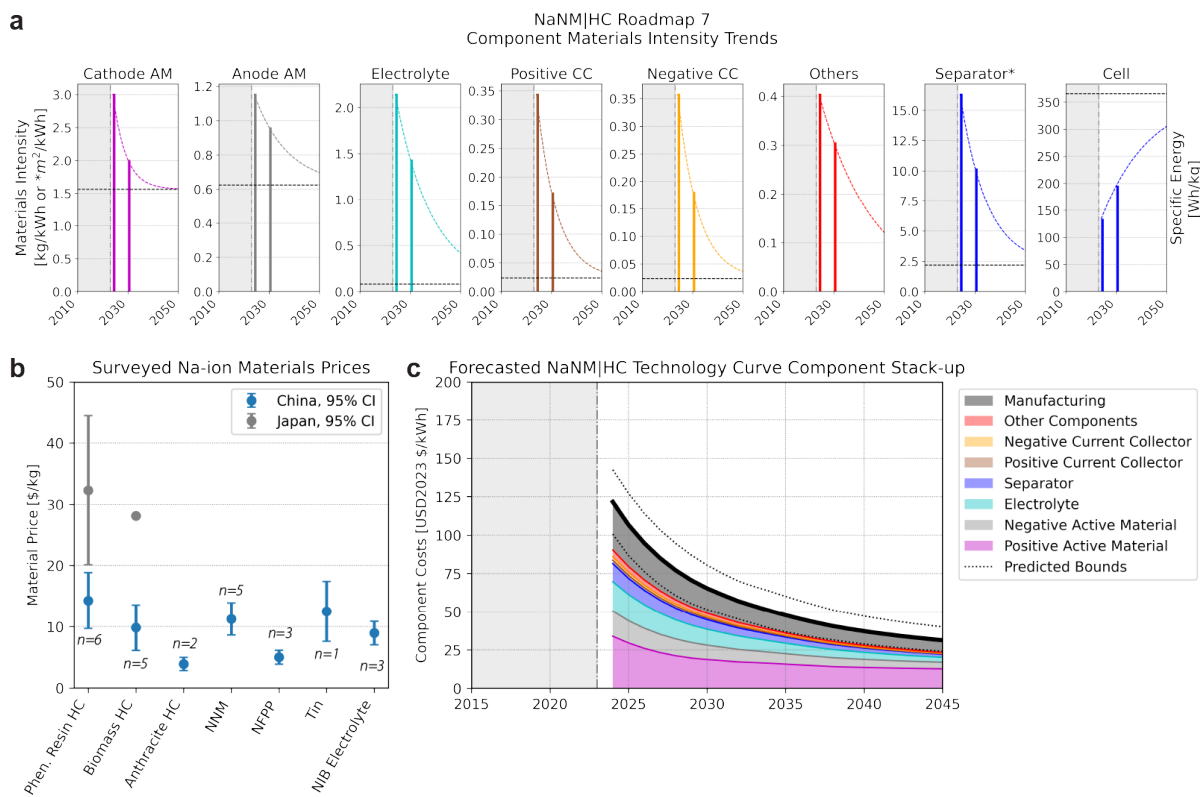

*Figure 3: Exemplary Na-ion price curve construction based on cell modeling, road mapping, and realistic materials price assessments.* **a.** *Fitted materials intensity trends that approach asymptotic engineering floors (horizontal dashed lines), shown here for Roadmap "NaNM|HC 7". AM=Active Material, CC=Current Collector.* **b.** *Surveyed price assessments for present-day Na-ion materials. The number of surveyed sources per material component (n) are also shown, and the 95% confidence interval indicates the high- and low-range estimates of prices. Note that each source may provide more than one value if also providing a range (see Methods).* **c.** *Componential construction of Na-ion price curve for given technical roadmap. Predicted bounds illustrate confidence intervals on the total curve from underlying uncertainty in minerals prices, starting materials prices, and learning rates.*

**Modeling scenarios to find conditions for competitiveness**

Using the above approach, we evaluate 6,048 combinations of scenarios, all assuming GWh-scale production for Na-ion commences in 2024. The scenario variables evaluated are detailed in Supplementary Discussion 13. For each scenario, the Li-ion and Na-ion price curves are calculated and compared, producing plots exemplified by those shown in Figure 4a. This example illustrates a "NaFM|HC 2" roadmap where an initially 33% nickel stoichiometry cathode evolves to ~0% by 2040 while maintaining the same specific capacity, and a hard carbon anode evolves to increase its specific capacity to 400 mAh-g$^{-1}$ by 2030. Due to the reduction in nickel content, the minerals price floor decreases substantially between 2024 and 2035.

Given that the forecasted curves carry uncertainty from confidence intervals of minerals prices (Supplementary Discussion 2), starting materials prices (see Figure 3b), and learning rates (see Figure 2b), we employ methods established in prior forecasting literature[45] to quantify expected timelines for price parity and advantage by calculating the probability at each time step that one technology is lower priced than the other. These probability plots are shown in the right-hand side of Figure 4a and Figure 4b comparing LFP against the modeled example Na-ion scenarios. We further define a "Price Parity" condition as the point in which Na-ion has ≥ 20% probability of being lower priced than LFP, and a "Price Advantage" condition when that probability exceeds 80%. The yellow hatched regions in Figure 4a and Figure 4b therefore mark the periods in which a Na-ion technology and LFP are competitive in pricing and may be considered substitutes, performance notwithstanding.

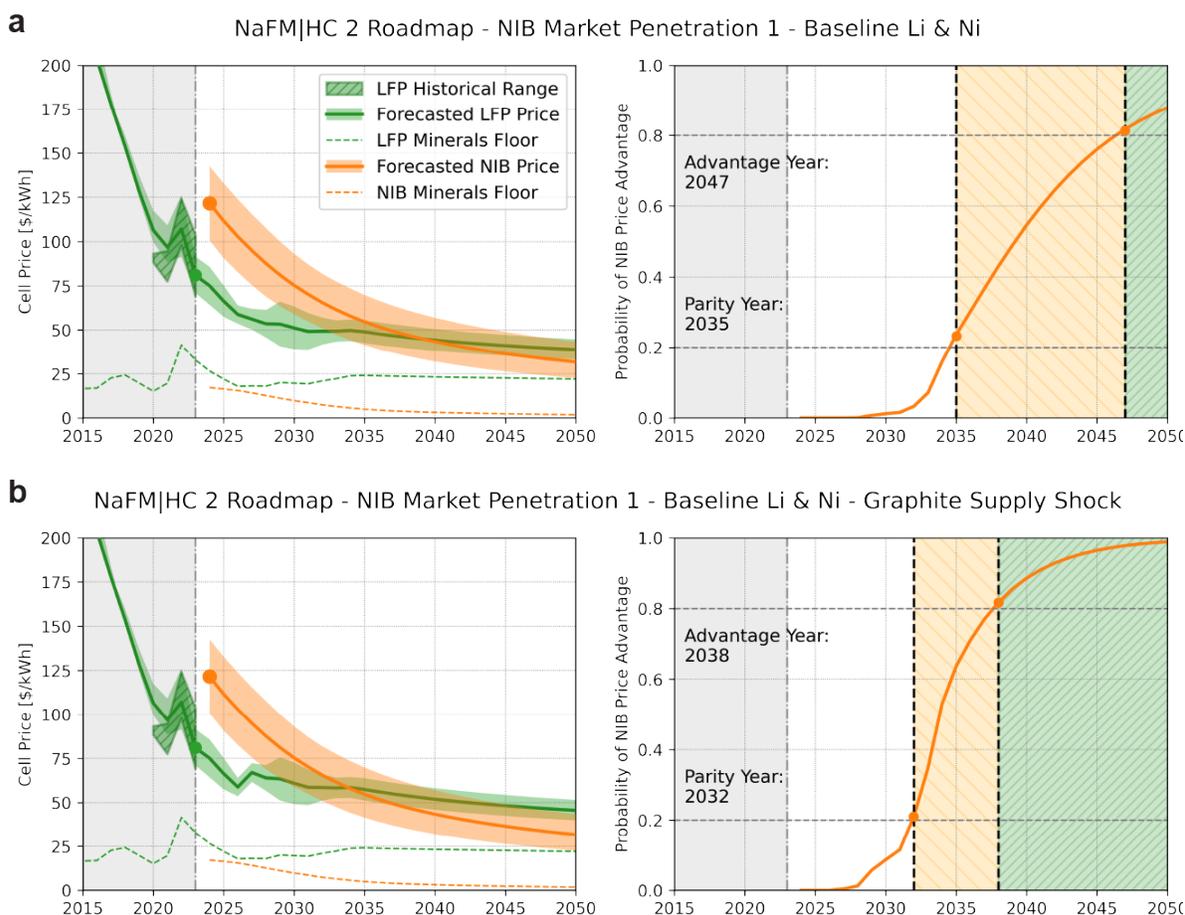

*Figure 4: Comparative price curve assessments to evaluate techno-economic competitiveness considering hypothetical Li-ion supply chain disruptions.* Example of Roadmap "NaFM|HC 2" with a conservative market penetration (NIB Market Penetration 1), baseline lithium and nickel prices, and **a**. no graphite supply shock or **b.** a hypothetical graphite supply shock in 2027. Comparative price curves (left) and probability profiles (right) allow evaluation of Price Parity and Price Advantage timelines. The example shown in **b** has a 4-year timeframe (yellow hatched) where Na-ion and LFP are closely competitive, whereas the example in **a** has a 12-year competitive timeframe.

Figure 4b models an identical set of assumptions as Figure 4a with the exception of a hypothetical graphite supply shock in 2027 that causes a permanent offset in prices but continues to learn at the same rate (see Supplementary Discussion 14 for details). This enables analysis of techno-economic competitiveness and risk considering potential supply chain disruptions. In the absence of a graphite supply shock condition as illustrated in Figure 4a, the Price Parity timeline is extended from 2032 to 2035, and the Price Advantage timeline is delayed substantially from 2038 to 2047. This large disparity is a result of the similar slopes of the price curves, which are nearly coinciding during this period.

**Discussion / Conclusions**

It is worth emphasizing that whereas our approach enables calculation of timelines to Price Advantage, the primary value proposition is understanding the impacts of various market scenarios on the viability of competing technology roadmaps instead of predicting specific years in which we anticipate crossover. The uncertainty associated with market forces—minerals pricing, demand growth, geopolitical conflicts—simply make it too challenging to predict exact timelines with accuracy and is not the key objective. In contrast, our approach enables quantitative evaluation of the relative robustness of certain technology roadmap propositions to withstand potentially disruptive market forces, thereby increasing confidence or challenging assumptions in the risk-reward profile motivating pursuit of technology development.

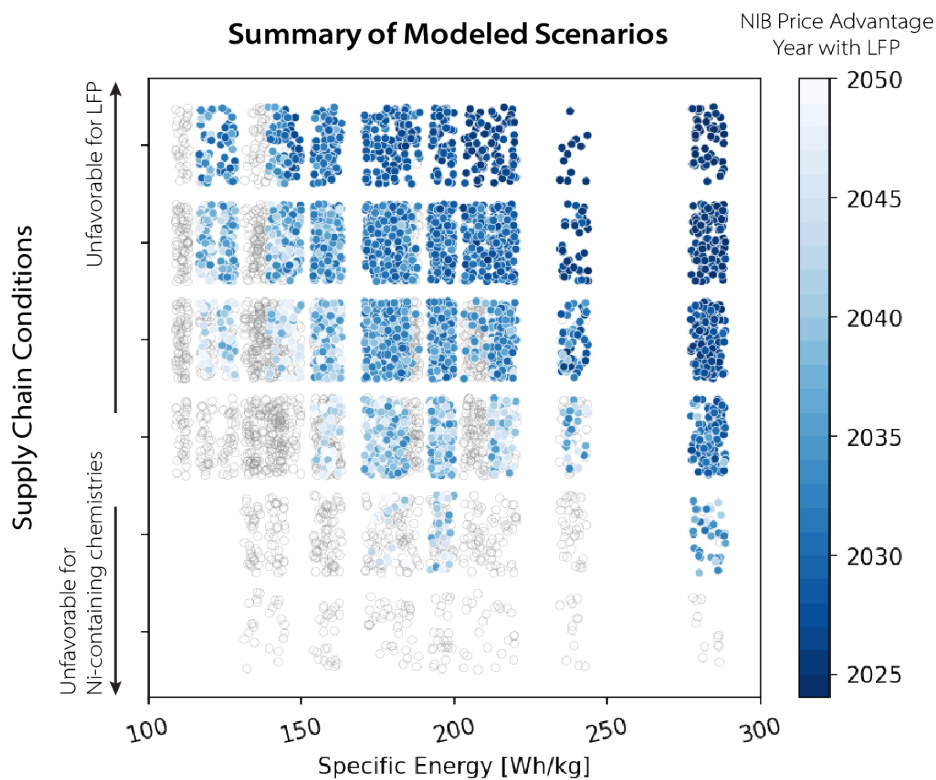

*Figure 5: Na-ion techno-economic competitiveness highly dependent on supply chain conditions and technology development roadmaps toward higher energy densities. Heatmap showing timelines to Price Advantage. Supply chain conditions ranked categorically here via summation of integer increments/decrements of lithium, graphite, and nickel supply chain conditions according to Na-ion favorability—with increased nickel commodity pricing being unfavorable to only nickel-containing Na-ion layered oxide cathodes. For example, a high lithium price scenario (+1) coinciding with a graphite supply shock (+1) will produce a supply chain score of +2. The x-axis is ranked by the specific energy density targets achieved by 2030 in the various roadmaps.*

Thus, a key outcome from our scenarios modeling is to reveal key trends in the conditions enabling maximum Na-ion techno-economic competitiveness. Figure 5 shows a clustered plot of all 6,048 scenarios categorized by gravimetric energy density and supply chain conditions, where a strong dependence on both factors reasonably describes the modeled timelines to Na-ion Price Advantage over LFP. When supply chain conditions are unfavorable for Li-ion (specifically LFP)—such as high lithium prices, graphite supply shocks, or both—Na-ion competitiveness is accelerated across the board. In contrast, when supply chain conditions are unfavorable for Na-ion (specifically nickel-containing chemistries), competitiveness is significantly hindered. As indicated by the trends from left to right, pursuing technology development roadmaps that maximize energy densities is important to accelerate timelines to competitiveness. Note that a large subset (2,522) of scenarios do not produce conditions for Na-ion Price Advantage before 2050, represented by the open circles in Figure 5. Importantly, however, this does not mean Na-ion is not competitive. A similar clustered plot as Figure 5 in Supplementary Discussion 13.1 shows the timelines to Na-ion Price Parity with LFP, illustrating that over 40% of all modeled scenarios reach a Price Parity condition on or before 2030, and the average "parity period" spans 5.6 ± 3.6 years. With Na-ion being a competitive, viable substitute to Li-ion having similar price curves (assuming performance parity is achieved), any disruptions to the Li-ion supply chain will likely present Na-ion as an immediately price advantageous alternative.

For example, if lithium prices increase between now and 2027 and remain high (~$50,000 tonne$^{-1}$ lithium carbonate equivalent [LCE], see Supplementary Discussion 2.1), over 55% of all Na-ion technical roadmaps lead to a Price Advantage condition before 2035. In contrast, if lithium minerals prices remain low (around $10,000 tonne$^{-1}$ LCE, see Supplementary Discussion Section 2.1), there will be virtually no Na-ion development scenarios that will result in a Price Advantage condition without a coinciding supply chain disruption in graphite (or potentially other Li-ion specific materials). For reference, the spot price of lithium at the time of this writing (2024) averages $10,000 – 15,000 tonne$^{-1}$—the result from oversupply of battery metals due to a slump in electric vehicle sales in the latter half of 2023[7,46]. Whether lithium prices can continue to stay low is thus a question with profound implications, especially on the competitiveness of Na-ion batteries.

The fastest and most certain way for Na-ion to be price advantageous is to reduce materials intensity by increasing materials and cell-level energy densities. This is supported quantitatively in the parameter sensitivity analysis shown in Figure 6, where some of the biggest drivers of forecasted Na-ion cell prices in 2030 and 2040 are accessible upper voltage cutoffs, cathode and anode specific capacities, and electrode thicknesses. Increasing accessible upper voltage cutoffs (a capability specific to layered oxide cathodes with solid-solution intercalation behavior) simultaneously increases available specific capacity and the nominal voltage, which unsurprisingly compounds to provide the largest feature importance. However, this is not always an available option due to material limitations, gas release, or challenges with power electronics upon systems integration (see Supplementary Discussion 8.7)[25]. Therefore, increasing specific capacity alone is an important strategy. In nickel-containing cathodes, this can ideally be done in conjunction with minimizing nickel content due to its relative feature importance. It is worth noting here that the uncertainties associated with fitted learning rates and starting prices of material components (which are further elaborated in Supplementary Discussion 5.1) are unlikely to affect the conclusions significantly due to the low variability in present-day price assessments compounded with the lower sensitivity of both learning rates and starting costs to the forecasted outcomes as illustrated in Figure 6.

Increasing the specific capacity of hard carbon anodes is yet another critical design direction identified. With low tap and calender densities, hard carbons impose practical thickness limitations that relegate a balanced cathode to low areal capacity loadings (mAh-cm$^{-2}$). Knock-on effects include requiring more electrolyte volume to saturate increased porosity within electrodes. Additionally, the aggressively slopey voltage profiles of hard carbons limits the deliverable cell energy, which negatively impacts cost performance[47]. For these reasons, hard carbons may enable the first generation of Na-ion batteries to demonstrate commercial viability but may require considerable specific capacity improvements (>400 mAh-g$^{-1}$) or outright replacement to enable long-term techno-economic competitiveness with Li-ion.

Replacing hard carbons with alloying-anodes, such as tin, is one approach to significantly increase anode specific capacity. In our models however, despite a NaNM|Sn baseline cell starting off ~30% lower price than a NaNM|HC cell, its observed price curve appears to fall at a slower rate. This is explained by the minerals price floor and engineering design floor of metallic tin both leaving little room for further improvements on cost and material-level performance. Therefore, alloy-based anodes appear to be effective tools to lower costs in the short-term but may require cathode improvements to stay competitive in the long-term.

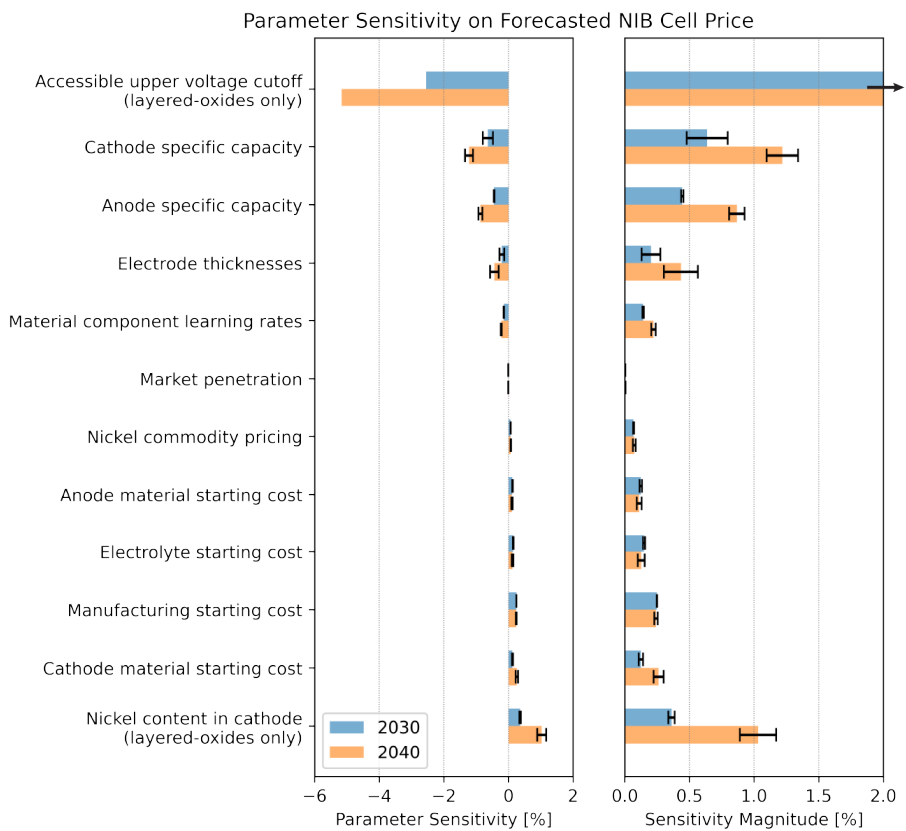

*Figure 6: Increasing energy densities and reducing critical minerals content are biggest levers to affect Na-ion cell prices. Parameter sensitivity analysis on the relative importance of features affecting Na-ion cell prices in 2030 (blue) and 2040 (orange). Error bars show standard deviation of parameter sensitivity between multiple one-at-a-time (OAT) perturbations, and the right-hand plot shows the absolute values of parameter sensitivities to enable comparison. Certain features (e.g. reducing nickel content in cathodes and increasing accessible voltage cutoff) are specific to layered oxide cathodes only. A pattern emerges indicating design directions that increase energy densities (accessible voltage, cathode and anode specific capacities, and electrode thicknesses) are key drivers of price. Additionally, the reduction of nickel-content in cathodes has a significant impact on the minerals price floor—more so than the absolute price of nickel commodities. These provide guidance on road mapping directions to maximize the techno-economic competitiveness of Na-ion.*

An alternative design direction is to forgo an anode material altogether and opt for an anode-free cell configuration. This of course would require higher-risk innovation but may not require substantial improvements to current-generation Na-ion cathode technology. Roadmap "NaNM|AF 1", for example, does not assume any improvements to cathode specific capacity and only assumes an increase in areal capacity loadings of state-of-art nickel-based layered oxides while maintaining a modest upper cutoff voltage of 4.0V. However, advanced separators, electrolytes, and/or current collectors required to enable anode-free designs must do so without reversing the techno-economic argument. Here, we highlight the importance of using such a techno-economic toolkit to *drive* research decision making as opposed to applying it *post facto*.

Finally, despite NFPP cathodes being less than half the cost of NNM and exhibiting a low minerals floor, the confluence of three technical parameters—low specific capacity, low tap density, and low voltage—present challenges to outright advantage on a $-kWh$^{-1}$ basis. Its volumetric capacity (mAh-cm$^{-3}$) is even lower than that of hard carbons, which results in low areal capacity loadings. Necessary routes to compete therefore likely require dry electrode processing to enable ultra-thick electrodes and incorporating alloy-based anodes to increase energy density. However, we again note that cell-level prices do not adequately capture other critical performance features, and NFPP-based Na-ion cells may prove more competitive at the systems-level if safety, thermal robustness, and cycle life metrics can be demonstrated to exceed those of LFP.

Both Figure 5 and Figure 6 illustrate that technical development roadmaps are not only beneficial but imperative to Na-ion's techno-economic competitiveness. Relying solely on learning-by-doing from scaling the production of material components is insufficient to compete with the rapidly evolving Li-ion trends. Therefore, adjusting between the most conservative and aggressive market penetration scenarios yielded minimal feature importance on future Na-ion prices. Similarly, adjusting material component learning rates to higher-than-typical ("Aggressive") values of $20 \pm 3\%$ for the Na-ion cathode, anode, and electrolyte also had modest feature importance. These observations suggest that engineering advancements are a greater lever to affect prices than the limited room for materials price reductions via learning-by-doing. This increases the likelihood of competitiveness, as price reductions do not rely on a chicken-and-egg problem with capturing market share.

In closing, Na-ion deserves significant research, development, and commercialization attention. We caution against assumptions and promises of immediate or near-term (pre-2030) price advantage against Li-ion, specifically LFP, but believe it has a critical role to play in the energy transition as a viably scalable Li-ion substitute under circumstances of supply chain volatility. Therefore, we believe investing in Na-ion development roadmaps to maintain a competitive track with Li-ion would be prudent. In the same vein, investments in Li-ion supply chain security should not be foregone given the demonstrated sensitivity to supply shocks.

The ultimate objective of developing low-cost batteries is to enable rapid deployment of energy storage in vehicular and stationary applications to meet the needs of the energy transition. These goals must be met even in the face of rising geopolitical tensions and supply chain volatilities. It is therefore not only important to accurately forecast price trends for incumbent technologies, but also critical to understand competitive or substitutional opportunities for emerging technologies. Our proposed modeling framework herein enables such an approach. Modeled outcomes help guide research endeavors and inform strategic investments commensurate with the probability of techno-economic competitiveness and commercial success.

**Methods.**

**Componential floor-constrained learning curves.** The typical Wright's learning curve follows the form:

$$Y = Ax^{-b} \quad (1)$$

where the price of a technology *Y* at a given cumulative experience *x* is described by the price of the first unit *A* and a rate of price reduction *b*. Here, the Learning Rate (or Experience Rate) is defined as the percentage price reduction after every doubling of cumulative experience, given by:

$$LR = 1 - 2^{-b} \quad (2)$$

As established above, these curves have been demonstrated to accurately capture the price trends across various sectors in a technology-agnostic manner, but they can also overestimate price reductions when technologies approach their price floors dictated by physical limits. In such scenarios, the equation can be modified to incorporate a floor constraint, shown by:

$$Y = \left(A_0 - A_{floor,0}\right)\left(\frac{x_t}{x_0}\right)^{-b} + A_{floor,t} \quad (3)$$

Here $A_{floor,0}$ and $A_{floor,t}$ illustrate that the price floors can evolve dynamically with time. Cumulative experience is also normalized for a more practical implementation of the general form in (1) that enables one to use the known price and cumulative "incurred" experience of components in the present and backcast and forecast accordingly. It also enables us to use actual quantities produced (in kT or Mm², for example) in fitting historical data where such production data is available, and switch to using anticipated demand (in GWh) in projecting future price trends. Agreement on present pricing is therefore all that is required to yield continuity between historical and forecasted prices. In this paper, we implement dynamically varying minerals price floors obtained from historical and forecasted price trends to capture the price evolution of individual component costs of a battery. This enables us to capture individual learning rates, market growth rates, and price floors between material components, which are unlikely to always be the same.

**Physically-accurate technology roadmaps.** To capture the evolution of materials intensities within cell designs due to technical advancements in materials performance and cell engineering, we implement a modified Moore's Law to capture improvements as a function of time. We first model physically-accurate cell designs via bottoms-up modeling to obtain exact materials intensities for each component (in kg-kWh$^{-1}$ or m²-kWh$^{-1}$) based upon the bill-of-materials. Cell designs are then assigned to specific years that represent the requisite technological progress, and we fit a curve with an asymptotic limit defined by theoretical or practical engineering limits. The curve follows the form:

$$M = M_{min} + Ae^{-b*(t-t_0)} \quad (4)$$

where $M_{min}$ is defined by Equations (6) – (11) for each material component. Here, the cathode and anode minimum materials intensities are defined by their respective maximum theoretical specific capacities and allowable electrode thicknesses. The minimum electrolyte materials intensity is defined by the total pore volume within the electrodes only. The minimum separator materials intensity (in m²-kWh$^{-1}$) is defined by a maximum cathode thickness and therefore capacity loading. The minimum positive and negative current collector materials intensities are then calculated from the separator materials intensity but converted to a mass-basis (kg-kWh$^{-1}$) based upon the densities and minimum practical thicknesses of the foils. For the Other material components category, the asymptotic limit $M_{min}$ is omitted as there is no basis for a minimum limit. In a similar fashion, the maximum theoretical cell-level specific energy is calculated by summing up a balanced anode- or cathode-limited unit-cell based on the provided limits, where Equation (4) is modified to approach an asymptotic maximum as opposed to a minimum, with the form:

$$SE = SE_{max} - Ae^{-b*(t-t_0)} \quad (5)$$

The equations defining the calculation of minimum materials intensities is as follows:

$$M_{CAT_{theo}}\left[\frac{kg}{kWh}\right] = \frac{1}{q_{CAT_{theo}}}\left[\frac{g}{mAh}\right] * \frac{1}{E_{cell}}\left[\frac{1}{V}\right] \quad (6)$$

$$M_{AND_{theo}}\left[\frac{kg}{kWh}\right] = \frac{1}{q_{AND_{theo}}}\left[\frac{g}{mAh}\right] * \frac{1}{E_{cell}}\left[\frac{1}{V}\right] \quad (7)$$

$$M_{ELY_{theo}}\left[\frac{kg}{kWh}\right] = \rho_{ELY}\left[\frac{g}{cm^3}\right] * \left\{ \frac{\varepsilon_{CAT_{min}}[\%]}{\rho_{CAT_{theo}}\left[g/cm^3\right] * q_{CAT_{theo}}[mAh/g]} + NP\_Ratio * \frac{\varepsilon_{AND_{min}}[\%]}{\rho_{AND_{theo}}\left[g/cm^3\right] * q_{AND_{theo}}[mAh/g]} \right\} * \frac{1}{E_{cell}}\left[\frac{1}{V}\right] \quad (8)$$

$$M_{\text{SEP}_{\text{theo}}}\left[\frac{\text{m}^2}{\text{kWh}}\right] = \frac{2}{t_{\text{CAT}_{\text{max}}}[\mu\text{m}] * \rho_{\text{CAT}_{\text{theo}}}\left[\frac{\text{g}}{\text{cm}^3}\right] * q_{\text{CAT}_{\text{theo}}}[\text{mAh/g}] * (1 - \varepsilon_{\text{CAT}_{\text{min}}}[\%])} * \frac{1}{E_{\text{cell}}}\left[\frac{1}{V}\right] \qquad (9)$$

$$M_{\text{PCC}_{\text{theo}}}\left[\frac{\text{kg}}{\text{kWh}}\right] = M_{\text{SEP}_{\text{theo}}}\left[\frac{\text{m}^2}{\text{kWh}}\right] * \rho_{\text{PCC}_{\text{theo}}}\left[\frac{\text{g}}{\text{cm}^3}\right] * t_{\text{PCC}_{\text{min}}}[\mu\text{m}] \qquad (10)$$

$$M_{\text{NCC}_{\text{theo}}}\left[\frac{\text{kg}}{\text{kWh}}\right] = M_{\text{SEP}_{\text{theo}}}\left[\frac{\text{m}^2}{\text{kWh}}\right] * \rho_{\text{NCC}_{\text{theo}}}\left[\frac{\text{g}}{\text{cm}^3}\right] * t_{\text{NCC}_{\text{min}}}[\mu\text{m}] \qquad (11)$$

For each material component, the *A* and *b* parameters in Equation (4) are fitted after the $M_{min}$ parameter is calculated. Occasionally, when one parameter is assumed to not experience significant improvements, such as the specific capacity of graphite (current values ~360 mAh-g$^{-1}$ are already near the theoretical capacity), the fitted parameter *b* may end up < 0, making the exponential term > 0 with a very shallow slope. This is an artifact of fitting, and we do not expect actual material performance to decline with time. Therefore, in cases where the fitted *b* parameter ends up negative, we instead use a flat line with zero slope centered at the average of the fitted values.

**Combining the modified Wright's Law with the modified Moore's Law.** After having established (a) floor-constrained learning curves and (b) materials intensities for each component scaling as a function of cumulative experience and time, respectively, we can combine them to produce a generalized equation which captures the overall cell price. This is represented in general form as:

$$Y_{\text{cell}} = \left(\sum_n^k M_{n,t} \times \left[\left(a_{n_0} - a_{n_{\text{floor},0}}\right)\left(\frac{x_{n,t}}{x_{n,0}}\right)^{-b_n} + a_{n_{\text{floor},t}}\right]\right) + a_{\text{mfg}}\left(\frac{x_{\text{mfg},t}}{x_{\text{mfg},0}}\right)^{-b_{\text{mfg}}} \qquad (12)$$

The first term of the right-hand-side aggregates the individual material component costs of a battery cell, each defined by its own learning rate, $-b_n$, normalized cumulative experience, $\left(\frac{x_{n,t}}{x_{n,0}}\right)$, price floor, $a_{n_{\text{floor},t}}$, and a material intensity scalar, $M_{n,t}$, that captures the kg-kWh$^{-1}$ or m$^2$-kWh$^{-1}$ contribution required for the representative modeled cell design. Here, $n \in$ Positive Active Material, Negative Active Material, Electrolyte, Separator, Positive Current Collector, Negative Current Collector, and Others. While in practice, all material components will have hard physical limits of a price floor, some components have negligible minerals costs, including polyolefins and conductive carbons. Therefore, for those materials, $a_{n_{\text{floor},t}}$ and $a_{n_{\text{floor},0}}$ are zeroed. The second term of the right-hand-side captures the costs associated with manufacturing, including equipment depreciation, labor, scrap, SG&A, other overheads, warranty, and profit. Notably, this second term is an unconstrained learning curve as there is no direct physical basis to institute a hard price floor. One may consider implementing manufacturing CapEx as a potential price floor, but as we do not yet have a clear methodology to firmly establish a minimum, we leave it unconstrained in this paper. Note that the learning parameter associated with manufacturing costs inherently captures other technical factors such as processing yield improvements and economies-of-scale, which we do not further resolve in our current work. The materials intensity scalar is used to capture the improvements to cell design and active materials specific capacities, both of which are key contributors to historic cell price declines. This componental approach captures the nuance that different components of a battery will experience different materials improvements, learn at different rates, experience different market growth scenarios, and be constrained by respectively different floors.

The above learning curves can be correlated with time by defining individual component market growth rates. Here, we use Gompertz sigmoidal functions to describe the annual demand of individual battery components, as they provide better fits to market projections than standard logistic functions[18]. The annual production capacity at year *t* can be defined as:

$$q_{n,t} = q_{n,\text{base}} \exp\left[\ln\left(\frac{q_{n,\text{sat}}}{q_{n,\text{base}}}\right)(1 - \exp[-r_n t])\right] \qquad (13)$$

where $q_{n,\text{base}}$, $q_{n,\text{sat}}$, and $r_n$ represent the starting annual production capacity, the annual production capacity upon market saturation, and growth rate of component *n*, respectively. Notably, not all battery components experience the same market growth conditions. For example, whereas the demand for graphite materials in tonnes per annum scales closely with the cumulative demand of total Li-ion batteries due to it being the predominant anode chemistry, LFP and NMC materials will each scale at a lower rate due to their fractional market share. With this, the cumulative capacity can be obtained from

$$x_{n,t} = \sum_{t=0}^{T} q_{n,t} \qquad (14)$$

**Minerals pricing.** Each mineral dataset was converted into aggregate averages with a sample size of $n \geq 3$ if including proprietary industry sources, or $n = 1$ if data was only available from USGS. Due to the proprietary nature of each of the industry-supplied forecasts, the minerals pricing datasets are averaged at each time step with 95% confidence intervals to prevent traceability to any one data source using the formula:

$$\mu \pm 1.96 \frac{\sigma}{\sqrt{n}} \qquad (15)$$

While USGS datasets do not provide minerals price forecasts, industry intelligence firms do, and we use their 2023 forecasts as our baseline scenario for the key minerals (lithium, nickel, cobalt, etc.) looking forward. As expected, there is good agreement on historical minerals pricing but notably larger variances between forecasts within the next decade. Due to inherently large uncertainties associated with forecasting prices of volatile minerals, we also perform analysis of self-generated fixed price scenarios (e.g. high/mid/low) on the key commodities to evaluate the sensitivity of outcomes (see Supplementary Discussion 2.1).

**Historical material component pricing.** All prices obtained from industry and literature were inflation-adjusted to USD2023. For any year with multiple data points, averages and confidence intervals were calculated using the same methodology above. For tabulating annual production quantities (in kT or Mm$^2$), gaps in data were interpolated using the requisite CAGRs established by the bounding years for which data was available, where CAGR is defined by:

$$\text{CAGR} = \left(\frac{P_2}{P_1}\right)^{1/t} - 1 \qquad (16)$$

**Cell modeling.** To accurately model cells, representative half-cell voltage vs. specific capacity (mAh-g$^{-1}$) curves of positive and anode material candidates (e.g, LFP, NMC622, NMC811, graphite) were extracted from literature and were mathematically scaled to meet target electrode coating mass loadings, active mass fractions, areal capacity loadings, and calender densities (and therefore porosities). Importantly, first cycle (de)lithiation or (de)sodiation curves were utilized in order capture differences in first cycle efficiencies between positive and negative electrode pairings, and full cell voltage curves were obtained by subtracting negative from positive. This approach is critical to obtaining accurate predictions of realizable energy densities[47]. Cell modeling details are discussed at length in Supplementary Discussion 8, along with validation against experimental curves.

Once the electrode balancing procedure is completed to obtain accurate unit cell designs, the electrode parameters were inputted to the Battery Performance and Cost (BatPaC, v5.1) spreadsheet model produced by Argonne National Lab[22] to obtain exact mass- and areal-quantities per stored energy content (kg-kWh$^{-1}$ or m$^2$-kWh$^{-1}$). Additionally, we leverage the detailed manufacturing cost calculations within BatPaC to obtain the present-day manufacturing costs on a per-kWh basis. We take this approach for the following reasons: (1) Despite being highly detailed in modeling manufacturing-related costs, BatPaC has shortcomings in capturing the true voltage, capacity, and energy characteristics of cells due to inherent limitations of a spreadsheet approach. Especially considering occasional mismatched first cycle efficiencies between cathode and anode pairings and uniquely slopey or stepped voltage curves of emerging Na-ion materials, the true energy (Wh)—which is the area underneath the voltage curve between capacity and voltage windows—can often be miscalculated. This may result in errors in calculating true $-kWh$^{-1}$ [47]. (2) We note that BatPaC only models large format (>60 Ah) pouch cells, whereas some of the cells modeled (e.g. Tesla 4680 cylindrical cells) do not share the same format. However, we replicate all cell designs using BatPaC to enable systematic comparisons across generations and chemistries, and we also note that materials intensity, energy densities, and manufacturing costs at GWh scale will not deviate substantially between formats. This general approach allows us to systematically evaluate new cell designs and obtain cell manufacturing costs associated with each design while being more nuanced in electrode balancing. See Supplementary Discussion 15 for details on BatPaC Modifications made in our modeling efforts.

**Current Na-ion Material Price Assessment Survey.** To obtain present-day pricing of materials used in the nascent industry of Na-ion batteries, we take the approach of surveying industry experts and reports with insights on actual current-day pricing[24,25]. We survey from *n=11* sources on the key material components used in Na-ion batteries. Not all sources were able to provide estimates for every material component, but all material components, with the exception of tin anode powder, had at least *n=3* sources. If a given source provided a range of pricing corresponding to a low and high estimate, both values were used to appropriately weight the samples instead of the mid-range value. Specifically for biomass- and phenolic resin-based hard carbon materials which can currently be sourced from producers in China and Japan at significantly disparate costs, we also delineate which region the prices are quoting. In our modeling, we only use prices from China given the substantially higher concentration of Na-ion commercialization activity in that region. The results of our pricing survey are shown in Figure 3b.

## Acknowledgements

This work was supported by three offices within the U.S. Department of Energy through the STEER Program, a partnership between the SLAC-Stanford Battery Center and the Stanford Precourt Institute for Energy. Specifically, the three offices are: the Office of Electricity, the Vehicle Technologies Office under the Office of Energy Efficiency & Renewable Energy, and the Office of Technology Transitions. Additionally, this work was supported in part by the National Science Foundation Graduate Research Fellowship Program (award number DGE-2146755).

The authors thank the panel of industry consultants/experts cited throughout the work, including those that provided written feedback/critique of the manuscript draft (Supplementary Discussion 16). They include: Jerry Barker, Colin Wessells (Natron Energy), Brandon Kelly (Peak Energy), Spencer Gore (Bedrock Materials), Andrew Wang, Darren Tan (Unigrid), Linda Jing (Tesla), Tyler Evans (Mana Battery), Nicholas Singstock (Mana Battery), Heather A.S. Platt (Mana Battery), Katherine He (TDK Ventures), Fabrice Renard (Avicenne Energy), Christophe Pillot (Avicenne Energy), Wayne Yen (LongTime Technologies), William Hwang (LongTime Technologies), Kara Rodby (Volta Energy Technologies), Simon Lux (Universität Münster), Jannis Wesselkämper (Universität Münster), Philipp Voß (Universität Münster), Charlie Parker (Ratel Consulting), Mrigi Munjal (MIT), the Boston Consulting Group, and Brian Spatocco (Lucid Motors). The written feedback/critique from those that consented to our publishing is included in the Supplementary Information.

The authors further thank those attendees who attended the Na-ion Commercialization Forum hosted by STEER on April 15[th], 2024 on Stanford University campus centered around the pre-print of this manuscript. All discussions were held under Chatham House Rule with productive discussions that led to important revisions for this work. A summary of the key learnings from the Forum are included in Supplementary Discussion 17.

The authors finally thank Karan Bhuwalka, Hari Ramachandran, Julia Frohmann, and Alexis Geslin for their thoughtful comments and feedback on the manuscripts. The authors especially thank Karan Bhuwalka and Hari Ramachandran for collecting formal responses and summarizing learnings from the STEER Commercialization Forum.

## Author Contributions

**Adrian Yao:** conceptualization, methodology, modeling, data gathering, data contributor relations, data curation, analysis, visualization, writing, project administration, funding acquisition. **Sally Benson:** conceptualization, resources, review, editing, supervision, funding acquisition. **William Chueh:** conceptualization, resources, review, editing, supervision, funding acquisition.

## Declarations of Interests

A.Y. is among the founders of EnPower, a manufacturer of Li-ion batteries. W.C.C. is among the founders of Mitra Chem, a developer of LFP cathode materials.

## Supplementary Information

Supplementary Information to this work is available and attached separately.

# How quickly can sodium-ion learn? Assessing scenarios for techno-economic competitiveness against lithium-ion batteries

Adrian Yao[1,3], Sally M. Benson[2], and William C. Chueh[*,1,2,3]

[1]Department of Materials Science & Engineering, Stanford University, Stanford, CA 94305, USA.
[2]Department of Energy Science & Engineering, Stanford University, Stanford, CA 94305, USA.
[3]Applied Energy Division, SLAC National Accelerator Laboratory, Menlo Park, CA 94025, USA.

## SUPPLEMENTARY INFORMATION

### Section 1: Historical Li-ion price trends

Aggregate average Li-ion cell prices have been well-documented in literature and market intelligence reports[1–4]. In recent years, two dominant families of Li-ion cathode material chemistries have emerged to capture most of the Li-ion market share—NMC-type (nickel-manganese-cobalt oxide) and LFP-type (lithium iron phosphate) cathodes. Due to their diverging price trajectories resulting from differing minerals makeup, market intelligence firms have begun distinguishing chemistry-specific cell prices of NMC and LFP circa 2019. This is reflected in the historic NMC and LFP cell prices shown in Figure 1.1 aggregated from the industry data contributors to this work[1,2,5,6]. As NMC and LFP will approach different price floors, we find it important to resolve the price differences in our modeling.

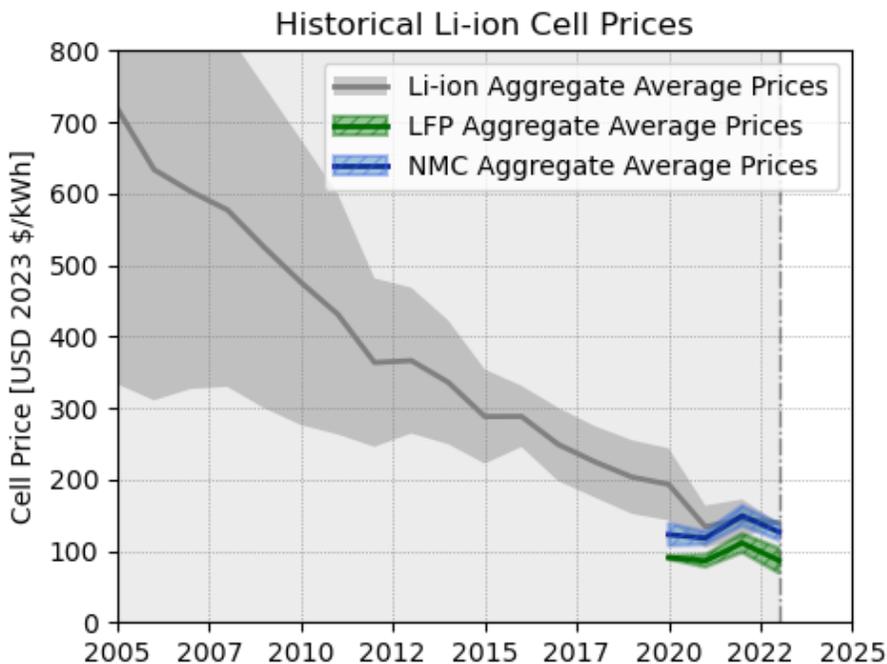

*Figure 1.1: Aggregated historical average Li-ion cell prices with 95% confidence intervals. NMC- and LFP-specific cell prices also shown for recent years.*

---

[*] Corresponding Author. E-mail: wchueh@stanford.edu

**Section 2: Establishing minerals price scenarios**

Minerals and commodities prices are known to be volatile, especially for specialty materials which have historically included key ingredients for Li-ion batteries such as lithium and cobalt. Additionally, many other minerals and metals have also seen significant recent price fluctuations due to a confluence of the energy transition, concentration of supply chains, and supply shocks such as those induced by COVID-19. Therefore, it is important to institute a technology price floor dictated by the dynamically evolving minerals prices.

In this work, we collect historical and forecasted prices for minerals from industry data contributors, such as Benchmark Mineral Intelligence, Wood Mackenzie, S&P Global, and additionally include historical prices from the United States Geological Survey (USGS) dating back to 1991. Aggregated price trends of key minerals are shown in Figure 2.1, showing averages and confidence intervals to protect the proprietary nature of price forecasts from individual data sources.

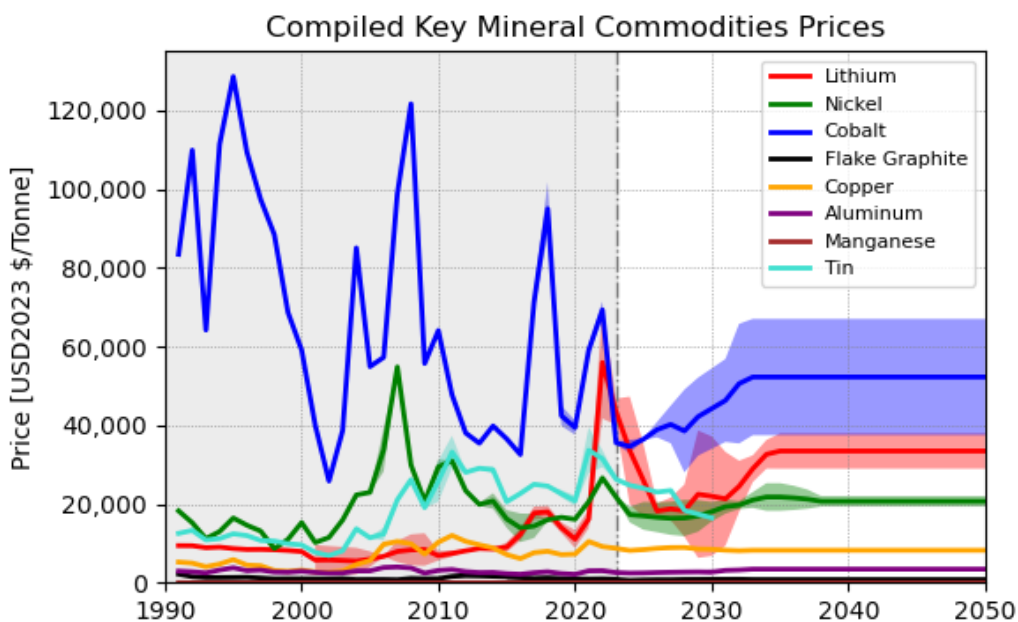

*Figure 2.1: Compiled commodities prices, historical and forecasted, for key minerals. Averages are shown with 95% confidence intervals.*

A summary of the tabulated minerals and commodities along with their data sources are detailed in Table 2.1. All prices were inflation-adjusted to USD2023. For all minerals/commodities price data sets that are incomplete, the first price figure in the time-series is extended backwards to 1991, and the last price figure in the time-series is extended forwards to 2050.

*Table 2.1: Summary of tabulated minerals/commodities and data sources*

| Mineral / Commodity | Source(s) |
|---|---|
| Lithium Carbonate Equivalents (LCE), Lithium Carbonate, Lithium Hydroxide | Benchmark Mineral Intelligence<br>Wood Mackenzie<br>S&P Global<br>Avicenne Energy<br>United States Geological Survey |
| Nickel, Nickel Sulfate | Benchmark Mineral Intelligence |

|  | Wood Mackenzie |
|  | S&P Global |
|  | United States Geological Survey |
| Cobalt, Cobalt Sulfate | Benchmark Mineral Intelligence |
|  | Wood Mackenzie |
|  | S&P Global |
|  | United States Geological Survey |
| Flake Graphite, Spherical Graphite, Needle Coke | Benchmark Mineral Intelligence |
|  | Wood Mackenzie |
|  | United States Geological Survey |
| Manganese, Manganese Sulfate | Wood Mackenzie |
|  | United States Geological Survey |
| Copper | Wood Mackenzie |
|  | S&P Global |
|  | United States Geological Survey |
| Aluminum | Wood Mackenzie |
|  | S&P Global |
|  | United States Geological Survey |
| Phosphate Rock | United States Geological Survey |
| Fluorspar | United States Geological Survey |
| Soda Ash | United States Geological Survey |
| Tin | S&P Global |
|  | United States Geological Survey |
| Iron Ore | Wood Mackenzie |

**2.1: Fixed High/Mid/Low minerals price forecasts**

Due to the inherent uncertainty associated with forecasting the prices of key minerals, we also evaluate scenarios where prices are fixed at a high/mid/low value. Specifically, we institute this for lithium and nickel, as they are the biggest contributors to the outcome of LFP-type Li-ion batteries and nickel-containing layered oxide cathode Na-ion batteries. We note that while the forecasted cobalt prices have significant variance between multiple sources and has historically been a metal subject to scrutiny, it only affects the pricing of NMC-type Li-ion batteries, which have also managed to significantly reduce cobalt intensity through the years. As the primary comparison of Na-ion batteries is against LFP, we ignore scenarios that may affect cobalt and NMC.

To calculate High/Mid/Low scenarios, we calculate the $90^{th}$, $50^{th}$, and $10^{th}$ percentiles of the total range of historic pricing for lithium and nickel. We fix the prices starting in 2027 and implement a linear ramp from 2023 actual prices and include a confidence interval representing ±10% of the total historic range for each scenario. We then compare our High/Mid/Low scenarios with the Base Cases obtained from aggregating forecasts from data contributors and drop the scenarios that most resemble the Base Cases.

For lithium prices, we drop the Mid case and keep the High/Low scenarios as the Base Case settles around $33,440 ± 4,497 $tonne^{-1}$. The High scenario is fixed from 2027 onwards at $50,880 ± 5,043 $tonne^{-1}$, and the Low scenario at $10,534 ± 5,043 $tonne^{-1}$. See Figure 2.2.

For nickel prices, we drop the Low case and keep the High/Mid scenarios as the Base Case settles around $20,659 ± 1,355 $tonne^{-1}$. The High scenario is fixed from 2027 onwards at $50,190 ± 4,613 $tonne^{-1}$, and the Mid scenario at $31,737 ± 4,613 $tonne^{-1}$. See Figure 2.3.

For reference, at the time of this writing (1Q 2024), the spot price of lithium averages $10,000 – 15,000 $tonne^{-1}$, and the spot price of nickel averages near historic lows at less than $20,000 $tonne^{-1}$, both largely the result from oversupply of battery metals due to a slump in electric vehicle sales in the latter half of 2023[7–9].

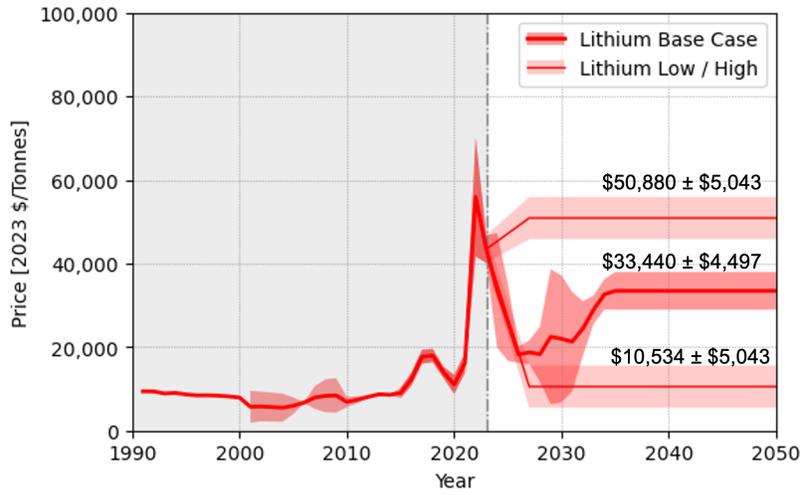

*Figure 2.2: Lithium price scenarios.*

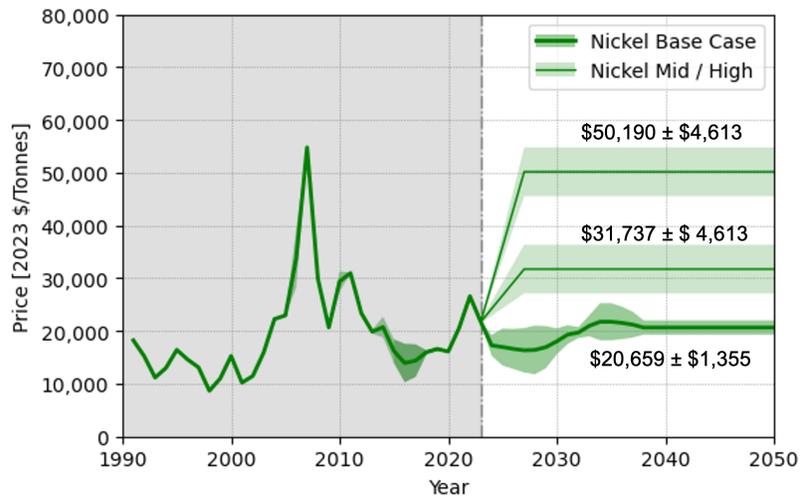

*Figure 2.3: Nickel price scenarios.*

**Section 3: Establishing material component price floors**

Each material component used in a Li-ion or Na-ion battery is synthesized from certain underlying minerals and commodities. The material component price floor can thus be calculated based on the useful molar mass fraction of the mineral/commodity contributing to the molar mass fraction of the material composition.

Taking NMC532 as an example, the key underlying minerals/commodities include lithium hydroxide (which is captured within the Lithium Carbonate Equivalent, LCE, commodity), nickel, manganese, and cobalt. Oxygen is ignored here due to abundance. The lithium content within lithium carbonate is 18.79%, which represents its useful molar fraction. In contrast, nickel, manganese, and cobalt are largely traded based on the metal itself, and therefore have useful molar fractions of 100% for the sake of pricing. Within NMC532, the stoichiometric molar mass fractions of lithium, nickel, manganese, and cobalt are 7.19%, 30.39%, 17.07%, and 12.21%, respectively, with oxygen comprising the remainder. Thus, by multiplying the underlying commodity prices of each mineral by its useful molar fraction and its mass fraction within a given material component, the mineral's contribution to the material component's price floor can be obtained. The total price floor of a material component is then obtained by summing all contributions from its key minerals. The useful molar fraction of minerals/commodities along with their mass fraction within material components are tabulated below in Table 3.1.

*Table 3.1: Summary of mineral and elemental compositions of key material components.*

| Material | Mineral / Commodity | Useful Molar Fraction [%] | Percent Mass of Material [%] |
|---|---|---|---|
| NMC111, $LiNi_{0.33}Mn_{0.33}Co_{0.33}O_2$ | Lithium Carbonate Equivalent | 18.79 | 7.20 |
| | Nickel | 100.00 | 20.27 |
| | Manganese | 100.00 | 18.98 |
| | Cobalt | 100.00 | 20.36 |
| NMC532, $LiNi_{0.50}Mn_{0.30}Co_{0.20}O_2$ | Lithium Carbonate Equivalent | 18.79 | 7.19 |
| | Nickel | 100.00 | 30.39 |
| | Manganese | 100.00 | 17.07 |
| | Cobalt | 100.00 | 12.21 |
| NMC622, $LiNi_{0.60}Mn_{0.20}Co_{0.20}O_2$ | Lithium Carbonate Equivalent | 18.79 | 7.16 |
| | Nickel | 100.00 | 36.33 |
| | Manganese | 100.00 | 11.34 |
| | Cobalt | 100.00 | 12.16 |
| NMC811, $LiNi_{0.80}Mn_{0.10}Co_{0.10}O_2$ | Lithium Carbonate Equivalent | 18.79 | 7.14 |
| | Nickel | 100.00 | 48.27 |
| | Manganese | 100.00 | 5.65 |
| | Cobalt | 100.00 | 6.06 |
| NCA, $LiNi_{0.82}Co_{0.14}Al_{0.04}O_2$ | Lithium Carbonate Equivalent | 18.79 | 7.20 |
| | Nickel | 100.00 | 49.93 |
| | Cobalt | 100.00 | 8.56 |
| | Aluminum | 100.00 | 1.12 |
| LFP, $LiFePO_4$ | Lithium Carbonate Equivalent | 18.79 | 4.40 |
| | Iron Ore | 55.00 | 35.40 |
| | Phosphate Rock | 5.00 | 60.20 |
| LCO, $LiCoO_2$ | Lithium Carbonate Equivalent | 18.79 | 7.09 |
| | Cobalt | 100.00 | 60.21 |
| NaNM (Faradion O3-type Gen2), $NaNi_{0.30}Mn_{0.30}Mg_{0.17}Ti_{0.17}O_2$ | Soda Ash | 43.38 | 21.92 |
| | Nickel | 100.00 | 18.64 |
| | Manganese | 100.00 | 17.44 |
| | Magnesium Hydroxide | 41.67 | 3.87 |
| | Titanium Dioxide | 59.93 | 7.62 |

| | Soda Ash | 43.38 | 15.05 |
|---|---|---|---|
| NaNM (Faradion P2-type Gen2) | Nickel | 100.00 | 17.28 |
| | Manganese | 100.00 | 32.34 |
| | Magnesium Hydroxide | 41.67 | 0.79 |
| | Titanium Dioxide | 59.93 | 3.15 |
| NaNM (Faradion 0.67-0.33 O3-P2 Composite Blend Gen2) | Soda Ash | 43.38 | 19.65 |
| | Nickel | 100.00 | 18.19 |
| | Manganese | 100.00 | 22.36 |
| | Magnesium Hydroxide | 41.67 | 6.15 |
| | Titanium Dioxide | 59.93 | 2.85 |
| NaFM, $Na_{0.67}Fe_{0.5}Mn_{0.5}O_2$ | Soda Ash | 43.38 | 14.92 |
| | Iron Ore | 55.00 | 27.18 |
| | Manganese | 100.00 | 26.74 |
| NFPP, $Na_4Fe_3(PO_4)_2(P_2O_7)$ | Soda Ash | 43.38 | 15.11 |
| | Iron Ore | 55.00 | 25.11 |
| | Phosphate Rock | 5.00 | 31.21 |
| Synthetic Graphite | Needle Coke | 100.00 | 100.00 |
| Natural Graphite | Flake Graphite | 100.00 | 100.00 |
| 1M $LiPF_6$ in Carbonate Solvent | Lithium Carbonate Equivalent | 18.79 | 0.55 |
| | Fluorspar | 49.00 | 9.01 |
| 1M $NaPF_6$ in Carbonate Solvent | Soda Ash | 43.38 | 1.64 |
| | Fluorspar | 49.00 | 8.14 |
| Copper Foil | Copper | 100.00 | 100.00 |
| Aluminum Foil | Aluminum | 100.00 | 100.00 |
| Tin Powder | Tin | 100.00 | 100.00 |

A representative set of material component price floors are shown in Figure 3.1, dynamically varying within the time frame from 2010 to 2050. Due to global supply chain shocks resulting from the COVID-19 pandemic exacerbated by record electric vehicle demand, the prices of many key material components saw dramatic spikes in the year 2022. The biggest contributor by far in these price spikes is the cost of lithium, which rose 1,000% between 2021 and 2022 exceeding \$80,000 tonne$^{-1}$ at its peak in certain markets[10]. A noticeable, albeit smaller lithium price spike is also observed circa 2018. An "NMC Blend" material is shown as a representative NMC-type cathode in Figure 3.1 (see Section 4: for discussion on a volume-weighted NMC average), and it shows a substantially higher price spike relative to LFP in 2022. This is because NMC also suffered from the coinciding price spikes within the nickel market upon the Russian invasion of Ukraine in 2022, as Russia accounts for 17% of the world's production of high-purity nickel[11] to which LFP is immune. This nickel spike is also manifested in the price floor of Na-ion nickel-containing layered oxide cathodes, NaNM, (comprising primarily sodium, nickel and manganese and abbreviated NNM), though to a much lesser degree due to a lower overall nickel content.

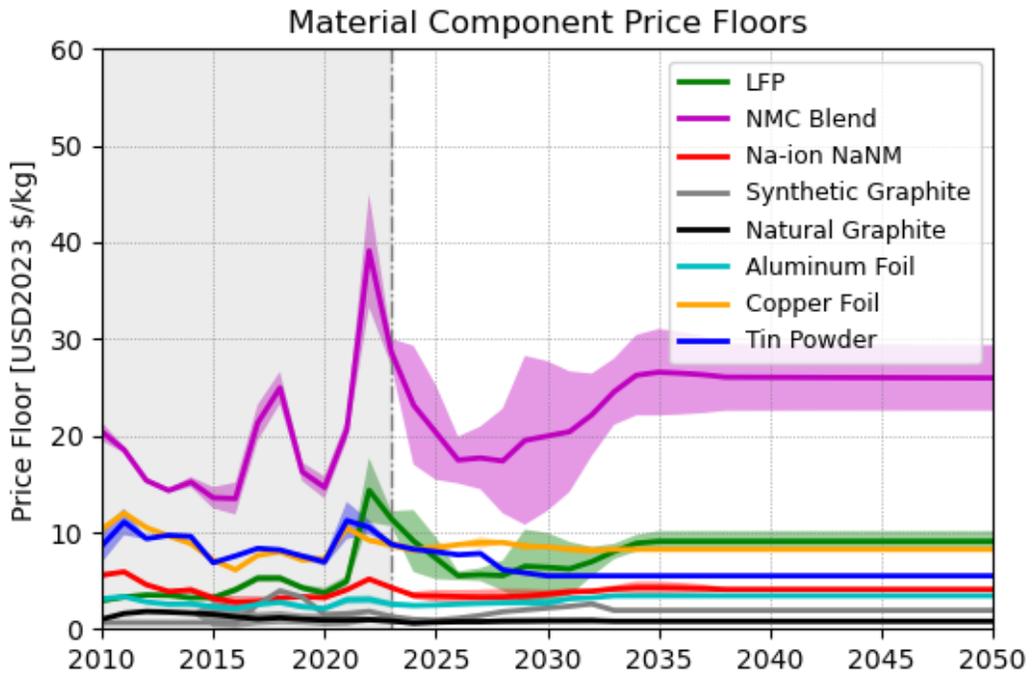

*Figure 3.1: Material component price floors for key components.*

## Section 4: Capturing evolution of NMC

NMC-type cathodes are a family of transition metal layered oxides that primarily comprise lithium, nickel, manganese, and cobalt. Throughout the years, the stoichiometry of NMC has evolved, following the trend of increasing nickel content to maximize specific capacity (mAh-g$^{-1}$) and reducing cobalt content to minimize reliance on an expensive, volatile mineral with ethical sourcing concerns. While the exact stoichiometry varies between cathode material component manufacturers, the evolution of NMC can be mapped to technology "nodes" with rough estimates on composition. Starting with equal-parts nickel, manganese, and cobalt in $LiNi_{0.33}Mn_{0.33}Co_{0.33}O_2$ (NMC333—but more traditionally referred to as NMC111), the stoichiometry continued to evolve through $LiNi_{0.50}Mn_{0.30}Co_{0.20}O_2$ (NMC532), $LiNi_{0.60}Mn_{0.20}Co_{0.20}O_2$ (NMC622), and $LiNi_{0.80}Mn_{0.10}Co_{0.10}O_2$ (NMC811). Intermediate stoichiometries also exist between nodes (e.g. NMC433 and NMC721) and cathode chemistries with $Ni_x$, $x > 0.90$ continues to be developed and commercialized, but the above are generally considered primary.

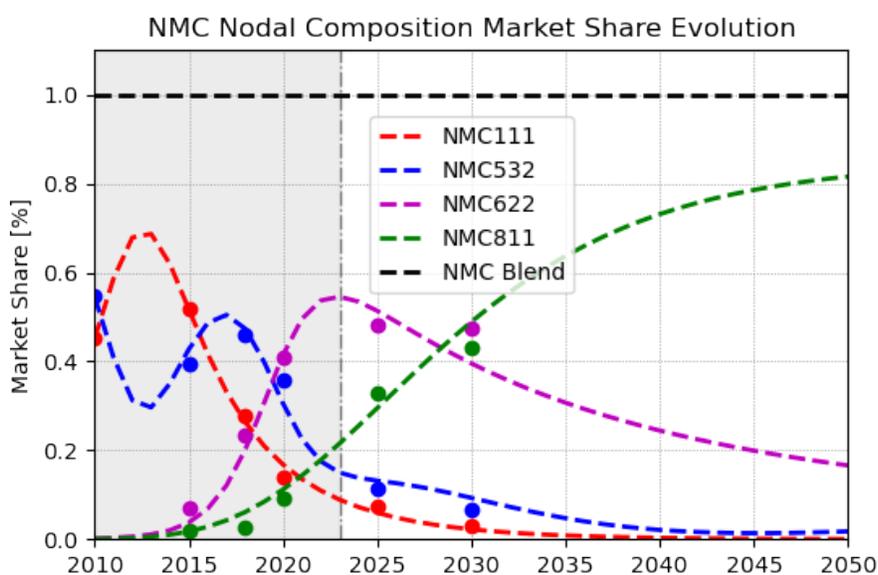

*Figure 4.1: Fitted NMC nodal composition market penetration curves.*

To accurately describe the prices of the NMC family using one metric, we calculate a volume-weighted average at each year from 2010 onwards based upon the individual minerals price floors (using methods described in Section 3:) and historical price assessments of each nodal composition. Here, volume refers to total market share obtained from industry-specific reports with relative penetration of each NMC nodal composition[1]. We fit continuous curves to match the data as closely as possible so that we can compute a volume-weighted average at each time step (see Figure 4.1). Here, NMC811 is fitted with a Gompertz sigmoidal function to capture its gradual market share growth, whereas NMC111 and NMC622 are fitted with a piecewise function that has a logistic growth phase and an exponential decay phase smoothed via a sigmoid weighting function. This allows us to capture the sunrise and sunset phases characteristic of any technology as they gain and lose market share relative to incumbents and new entrants. Despite knowing higher-Ni materials are currently being developed, we do not sunset NMC811 here as we lack credible projections, and the ultimate impact on technoeconomics is likely to be trivial. Finally, the market share of NMC532 is taken to be the difference between 100% and the sum of the other three nodal compositions. Note that even though state-of-art (in 2023) corresponds to high-nickel NMC811 or greater, the market share of earlier nodal compositions is still substantial, bringing the volume-weighted average nickel stoichiometry closer to 0.6 in 2023 (Figure 4.2). Mapping the nickel stoichiometry to realizable specific capacities[12], we also show the increasing trends in specific capacities throughout time (Figure 4.2). The minor dip in nickel stoichiometry and hence specific capacity between 2010-2015 is believed to be an artifact of fitting the available dataset—which shows competition between NMC111 and NMC532 in that period—and not truly reflective of an anomaly in actual trends.

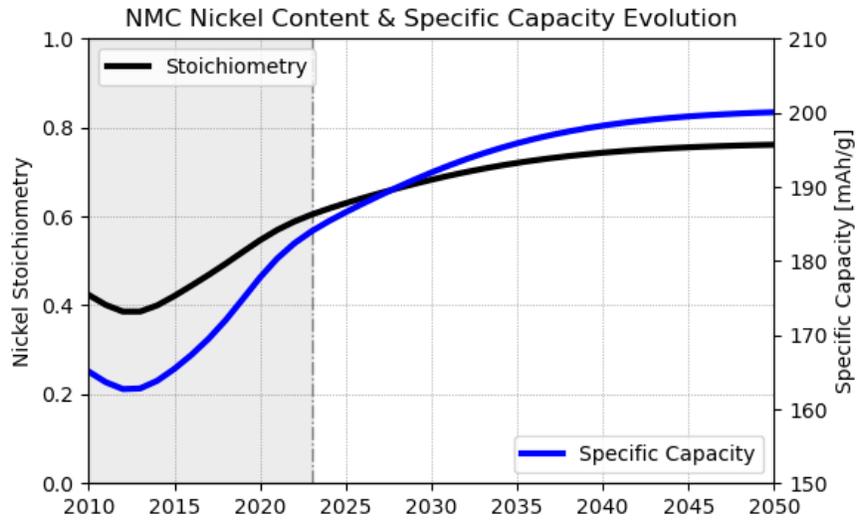

*Figure 4.2: NMC volume-weighted nickel content and specific capacity.*

**Section 5: Material component learning curves**

As described in the main text, we combine knowledge of historical demand (in kT or Mm$^2$) and pricing for each material component (sometimes with knowledge of conversion costs) to fit individual learning curves. This enables a novel look at the actual learning rates of materials versus devices and informs anticipated learnings in emerging technologies, such as Na-ion. We show all the fitted learning curves below.

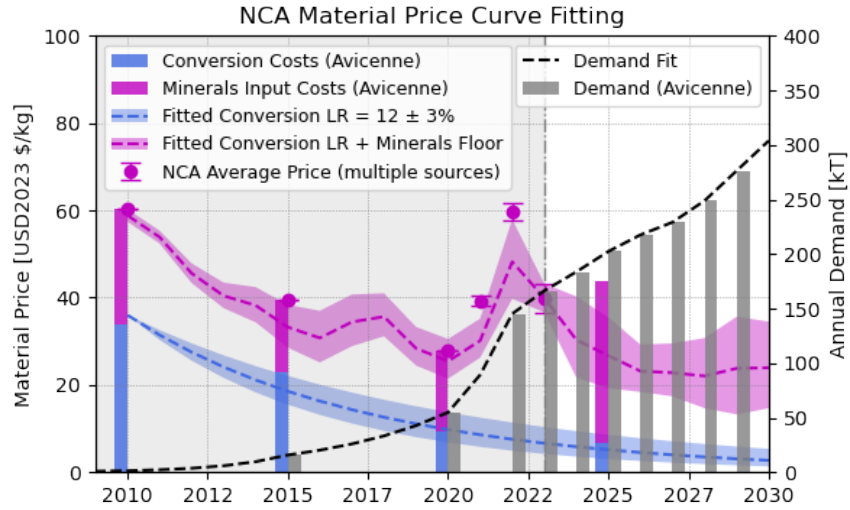

*Figure 5.1: NCA component price predictions.*

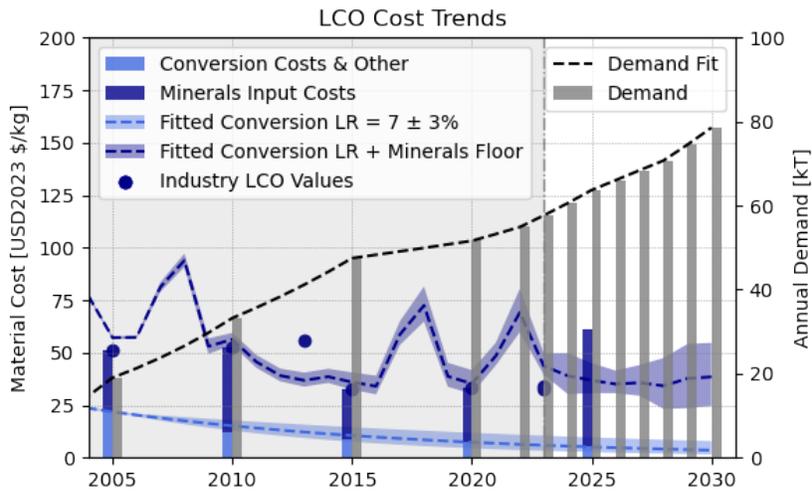

*Figure 5.2: LCO component price predictions.*

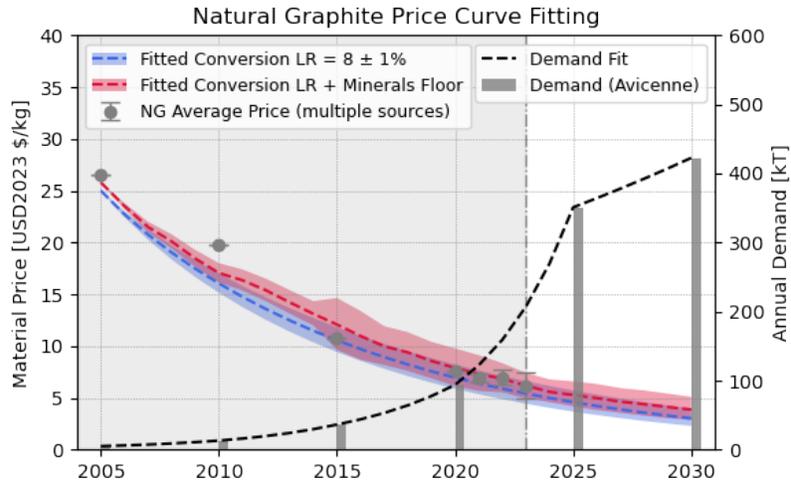

*Figure 5.3: Natural graphite component price predictions.*

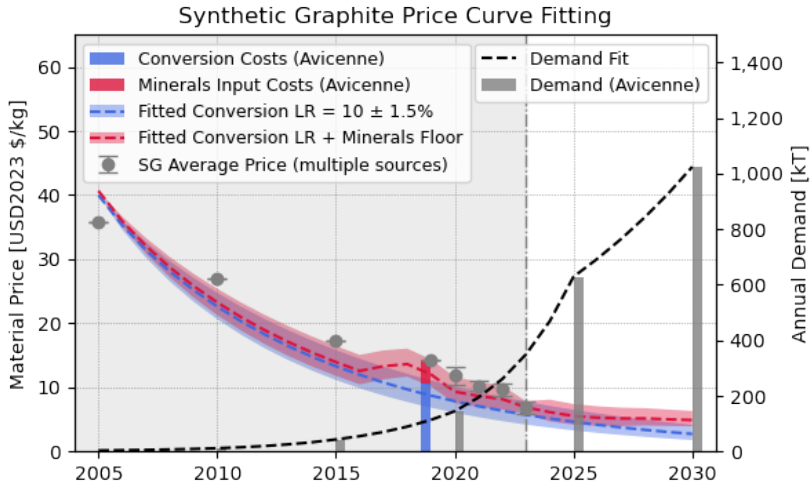

*Figure 5.4: Synthetic graphite component price predictions.*

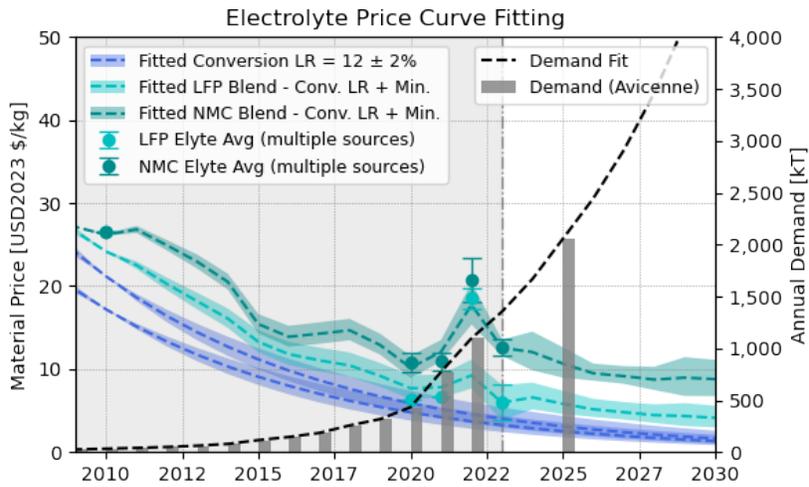

*Figure 5.5: Electrolyte component price predictions.*

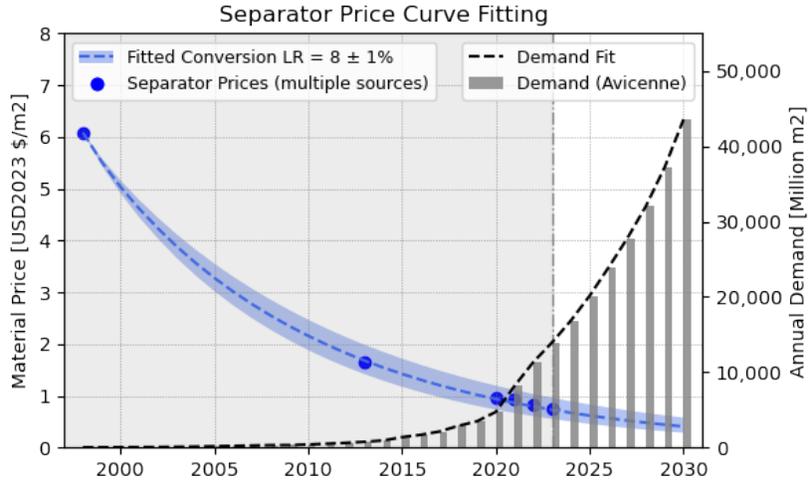
*Figure 5.6: Separator component price predictions.*

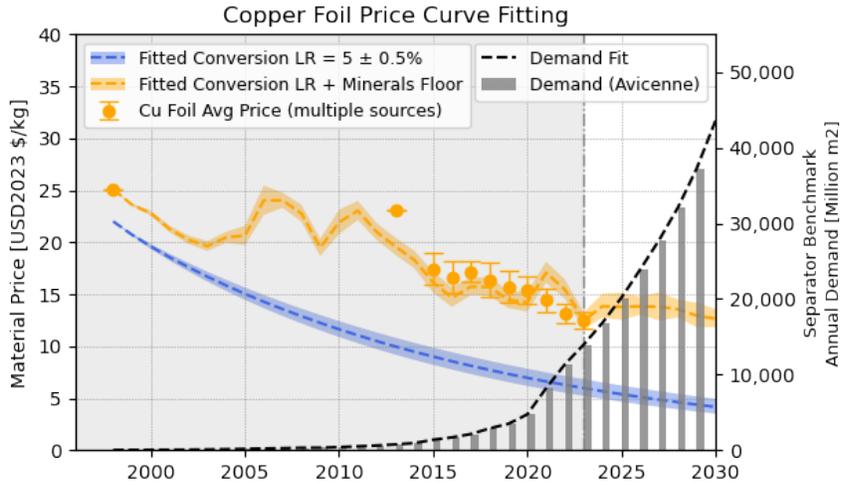
*Figure 5.7: Copper foil component price predictions.*

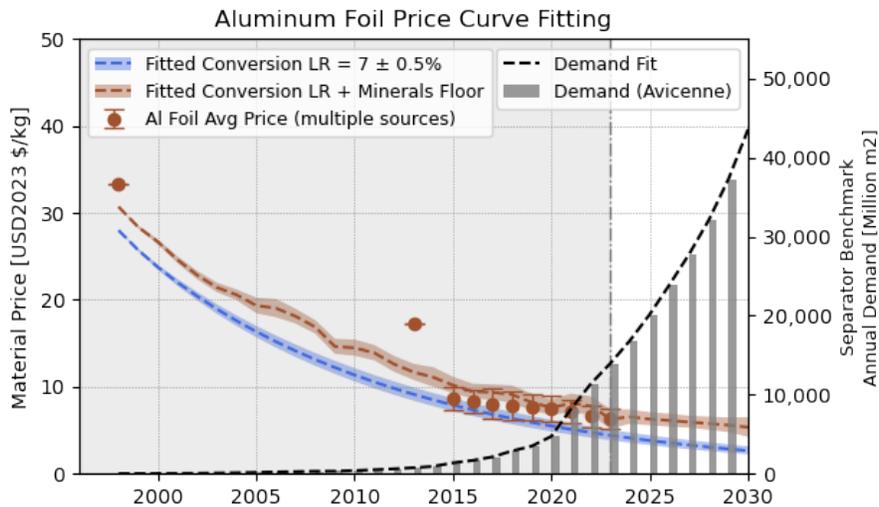
*Figure 5.8: Aluminum foil component price predictions.*

### 5.1: Sources of uncertainty

Given the scarcity in historical price assessments from which the above learning curves are fitted, it is worth discussing the potential sources of uncertainty in our modeling.

*Data uncertainty* – As evident in the data points shown in the above plots (Supplemental Figure 5.1Figure 5.8) and Main Text (Figures 1 and 2a), it is apparent that price assessments for each material is sporadic over many years. Where possible, we have presented the averages and confidence intervals of component pricing from multiple sources in a given year, but it is evident that many data points are obtained from a single source. Additionally, the historical annual demand (in kT or $Mm^2$) of material components are sourced soley from Avicenne[1,13]. This can thus be subject to biases inherent to the price assessment methodologies adopted by the data source, which may be regionally biased (e.g. China- or European-centric pricing) or simply inaccurate due to limited sampling. However, where multiple data sources are available above, the confidence intervals appear to be constrained within relatively tight spreads—suggesting the compiled dataset above to be a reasonably accurate representation of historical trends.

*Structural uncertainty (model uncertainty)* – Implicit in our floor-constrained learning curve modeling approach is the assumed Wright's Law relationship between prices and cumulative experience. Whereas prior literature has established this to be a reasonable descriptor for many different industries, including chemical and energy products[14], it is not guaranteed all material components follow this trend. From the historical price assessments presented above, it is not immediately clear that all material components follow a power law. This of course is convoluted by the volatility in minerals pricing that we seek to deconvolute. Whereas our fitted floor-constrained learning curves above may appear to reasonably fit historical price assessments, it is still worth clarifying that our modeling approach is only an estimation of the price trends of each material component.

## 5.2: Fitting learning rates for Manufacturing and Others components

To fit the learning rates of Manufacturing and Other components, historical component price breakdowns for the two categories were obtained from Avicenne Energy tracking the price evolution of NMC-type cells[1,13]. Unconstrained learning curves were then backcast and fitted with known starting cost contributions from the bill-of-materials (for the Others component) and BatPaC cost calculations (for the Manufacturing component). For the Others component, the materials intensity trends established in Section 9: were used as a multiplier to capture the reduction in the Others component due to improvements in cell design efficiency. Similarly, the evolution of increasing cell-level gravimetric energy densities over time (also established in Section 9:) is factored to ensure the Manufacturing cost component falls as a function of increasing energy densities. Both of these measures enable deconvolution of improvements due to learning-by-doing versus cell design efficiency. The fitted curves are shown in Figure 5.9 below. We assume the learning rates fitted for NMC-type Li-ion are transferrable to LFP-type Li-ion as well as Na-ion batteries.

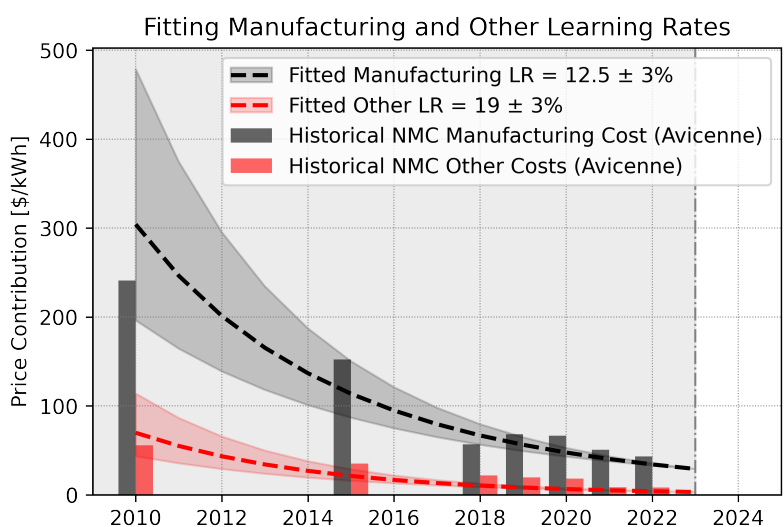

*Figure 5.9: Backcasted fitting against historical NMC cell price data to obtain manufacturing and other learning rates.*

## 5.3: Patterns in observed learning rates

Based on the fitted learning rates (Main Text, Figure 1b), there appears to be an observable pattern when evaluating the color coding. In magenta are cathode active materials, in gray are anode active materials, in orange are metallic foils, and the electrolyte and separator are within their own categories. Among the material components, it appears that cathodes tend to exhibit a higher learning rate). On the opposite end of the spectrum, metallic foils have the lowest learning rates. Qualitatively envisioning a "degree of transformation" for the two categories, we can note that cathode synthesis exceeds foil production by far. We postulate that learning rates may be a function of manufacturing process complexity and such degree of transformation. Between these two extremes, graphite anode production seems to have an intermediate learning rate. Considering that the conversion of natural mined flake graphite into battery grade spherical graphite or the conversion of needle coke into synthetic graphite might appear to have an "intermediate" degree of transformation, the hypothesis holds true qualitatively. We note that fitted cell manufacturing learning rates are also high, on the same order as cathode learning rates. Considering the complex assembly of many components to produce a battery cell—a very high degree of transformation—the high manufacturing learning rate for cell production also makes sense.

Being able to predict characteristic learning rates for new industries would have profound implications. This would enable forecasts of emerging technologies that do not have as clear-cut analogues as does Na-ion with Li-ion. We highlight this as a subject of future studies.

## Section 6: Market growth rates

Modeled market growth and penetration scenarios are shown below. Reasonable agreement was found between three independent sources[1,2,5] on market growth trajectories to 2050 for the total battery market, EV segment, and stationary ESS segment. There was noticeably more variance in predicted sizes and growth rates of emerging markets, such as two/three-wheelers & micromobility (TTW/MM). Note, however, that emerging markets are dwarfed by EV demand.

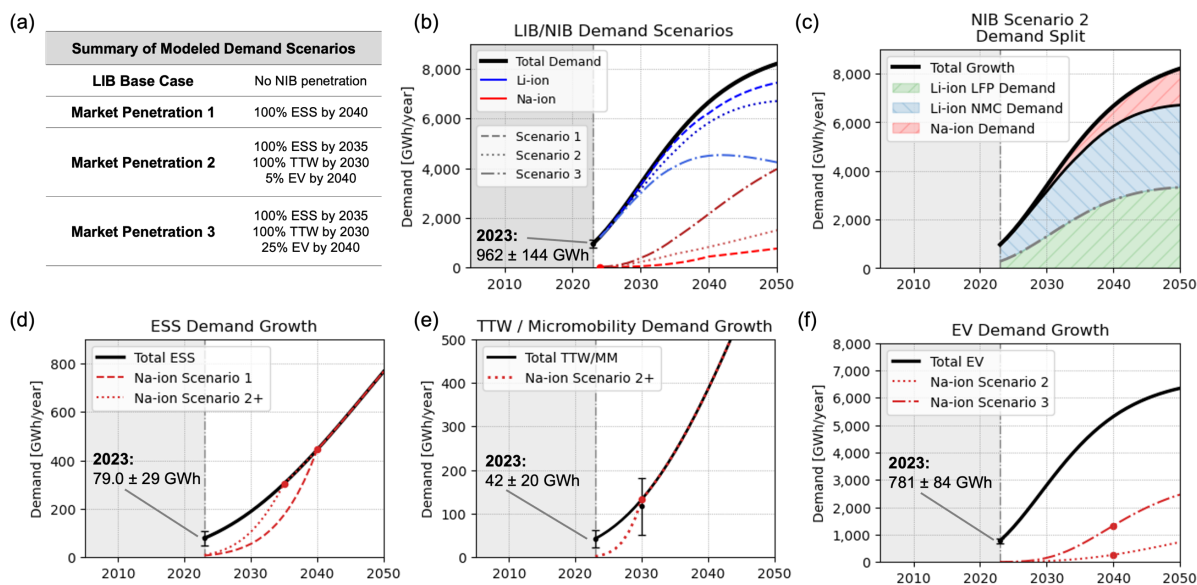

*Figure 6.1: (a) Modeled market demand scenarios. (b) Overall demand curves for Li-ion and Na-ion based upon the three modeled market penetration scenarios that each cannibalize Li-ion market share but do not change the overall market demand for batteries. (c) Demand split between NMC, LFP, and a hypothetical Na-ion chemistry based upon scenario "Market Penetration 2". (d-f) Modeled market growth curves for the ESS, TTW/MM, and EV markets showing various degrees of Na-ion penetration. Industry average and 95% CI for current market sizes are shown with generally good agreement between independent sources, with the exception of TTW/MM where a larger variance in forecasts is shown in 2030.*

### 6.1: Nickel- vs. iron-based Li-ion chemistries market penetration

NMC- and LFP-type Li-ion cells, more broadly referred to as nickel- and iron-based chemistries to capture future evolution, are the two competing sub-classes of Li-ion that capture the majority of the Li-ion market share[5,13]. How they evolve with respect to each other, however, is still unclear. With the lower cost, improved safety, and more stable cycling performance of LFP, predictions of LFP market share continue to be revised[15,16]. As the market share of LFP is directly related to its cumulative demand and hence learning curve, we want to accurately capture the market segmentation between nickel- and iron-based chemistries within Li-ion.

Here, we chart three different market penetration scenarios for iron (Fe)-based chemistries. The "Conservative" case is aligned with the aforementioned industry estimates of LFP penetration. The "Moderate" case assumes ~50% saturation by 2050, and the "Aggressive" case assumes ~65% saturation by 2050. These are summarized in Figure 6.2.

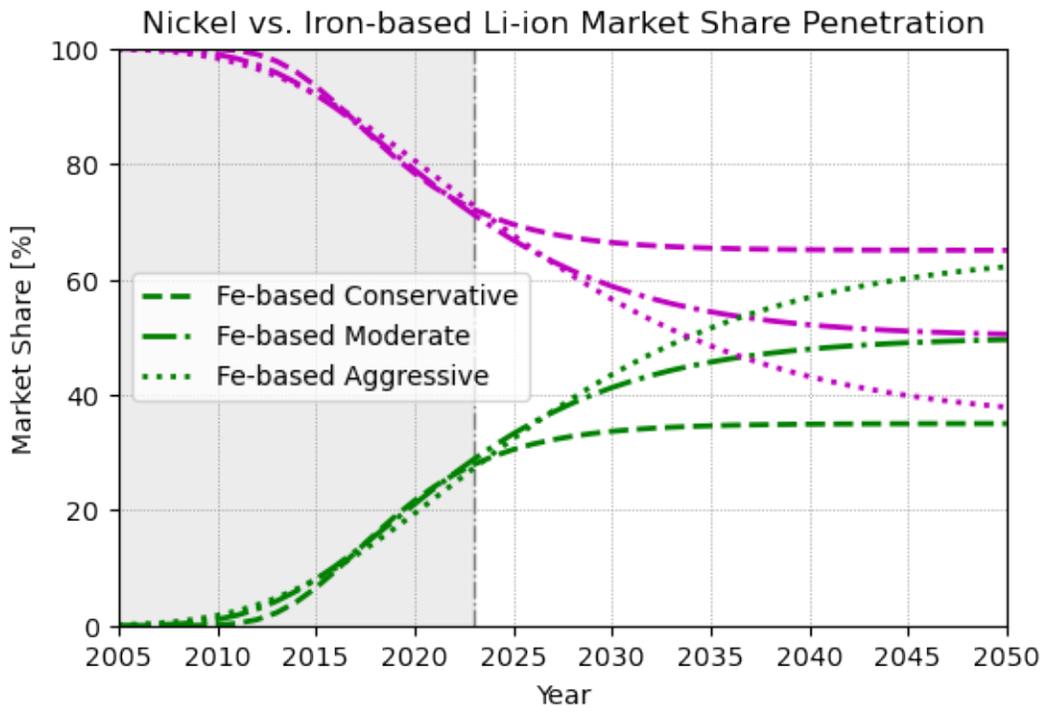

*Figure 6.2: Market penetration scenarios for Fe-based vs. Ni-based Li-ion chemistries.*

**Section 7: Li-ion Material Component Price Assessment**

A summary of present-day pricing for key Li-ion material components is shown below. Note the relatively tight spread in materials prices for most components except for NMC and NCA. The higher variance for these materials is attributed to (1) their nickel content, which is currently experiencing more market volatility, and (2) the multiple commercially available stoichiometries categorized within their respectively broad family labels.

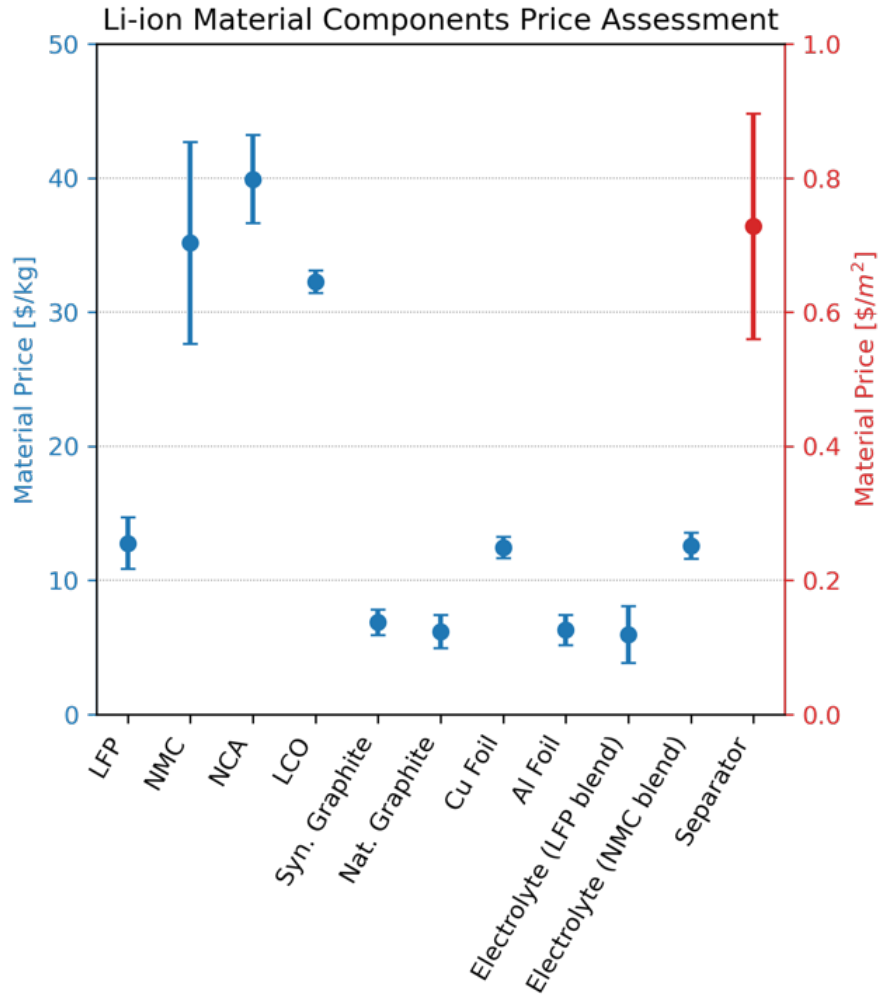

*Figure 7.1: Present-day price assessments of key Li-ion material components. LFP=Lithium iron phosphate polyanion cathodes. NMC=Nickel-manganese-cobalt layered oxide cathodes. NCA=Nickel-cobalt-aluminum layered oxide cathodes. LCO=Lithium cobalt oxide layered cathodes. Syn. Graphite=Synthetic Graphite. Nat. Graphite=Natural Graphite.*

## Section 8: Modeling cells

Accurately modeling the technoeconomic performance of battery technologies requires detailed modeling of commercial cell designs. To do this, it is most important to capture the unit cell stack design. This includes electrode design parameters for the cathode and anode, their relative balancing, and the resulting voltage profiles they generate after accounting for first cycle Coulombic efficiencies (assuming the majority of solid electrolyte interphase (SEI) side reactions have been formed by then). To do this, we mathematically scale voltage curves obtained from half-cells measured against $Li/Li^+$ for Li-ion materials and $Na/Na^+$ for Na-ion materials. We obtain the first two half-cycles (e.g. one (de)lithiation cycle or (de)sodiation cycle) measured as a function of active material specific capacity (mAh-$g^{-1}$) for a positive (cathode) and negative (anode) electrode couple obtained at sufficiently low rates to avoid excessive polarization (e.g. C/20), multiply by their respective active mass fractions corresponding to their formulation within realistic electrode composite designs, and then multiply by their areal mass loadings (mg-$cm^{-2}$) to obtain areal capacity loadings (mAh-$cm^{-2}$). Based upon known total cathode and anode electrode footprints (in $cm^2$) derived from large-format cell designs generated by the industry-validated Battery Performance and Cost (BatPaC) model developed by Argonne National Lab[17], we obtain true cathode and anode capacities within a cell. Calculation for the electrode capacity is given in Equation (1).

$$Q_{electrode}[Ah] = q_{active\ material}\left[\frac{mAh}{g}\right] * x_{fraction}[\%] * m_{electrode}\left[\frac{mg}{cm^2}\right] * A_{electrode}[cm^2] \qquad (1)$$

Anode capacities are oversized according to realistic N/P ratios based on areal capacity (in addition to being dimensionally oversized with an electrode "overhang"), and electrode calender press densities (in g-$cm^{-3}$) are also modeled based upon industry expert opinion to obtain practical electrode thicknesses and hence porosities[18]. Finally, the difference between the cathode and anode half-cell voltage curves are obtained via curve interpolation to derive a theoretical full cell voltage profile. The electrode parameters are iteratively tweaked to obtain the desired capacity between the operating voltage limits defined by the lower- and upper-voltage cutoffs. The true energy contained within a cell is calculated based upon the integrated area under the first discharge (second half cycle) full-cell voltage curve bounded by the window of operation (top/bottom defined by voltage limits, left/right defined by irreversible capacity loss and cell capacity). This approach ensures an accurate calculation of energy density, as the characteristically slopey voltage curves of some material chemistries (e.g. Na-ion layered oxide cathodes and hard carbon anodes) can have substantially lower expected energy within a balanced full-cell than their individual specific capacities suggest. See Figure 8.1 for representative considerations for full cell voltage curve derivation. Since the N/P ratio can be defined in various ways, here we clarify the N/P ratio is defined within this work as the ratio of the *irreversible* areal capacity loadings of the negative versus the positive electrode, as shown in Figure 8.1.

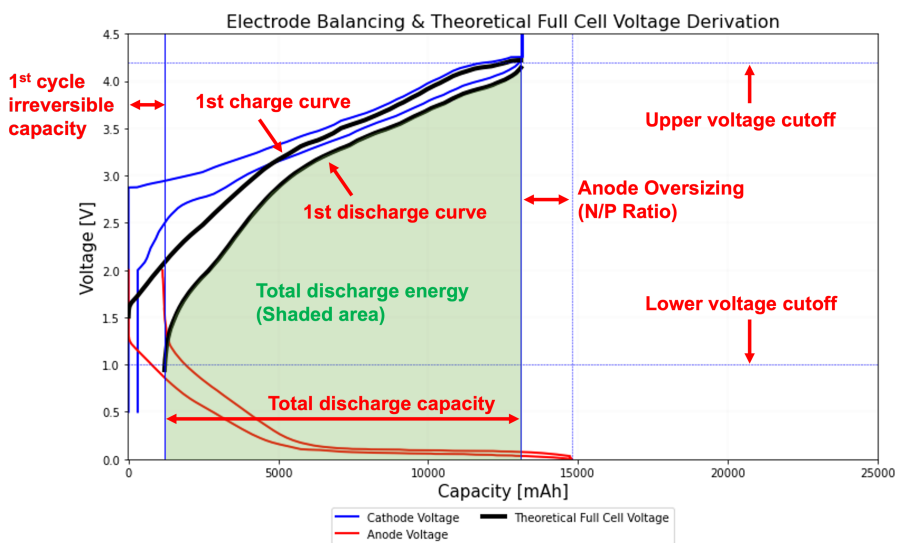

*Figure 8.1: Representative figure showing cathode (blue) and anode (red) curves scaled from extracted half-cell data, producing a theoretical full cell curve from their difference. Other cell design parameters labeled in red. Note the green shaded region representing the total available discharge energy is substantially less than what the cathode can provide.*

To calculate relevant electrolyte quantities, we calculate the total pore volume within the cathode, the anode, and the separator and add an additional overhead of 80% electrolyte fill to account for additional "dead" volume to arrive at practical mL-Ah$^{-1}$ values. While an 80% overhead amount may be considered a more generous amount specifically for pouch cells that tend to be electrolyte starved systems, a higher fill quantity is required for hard-cased cylindrical and prismatic formats[18]. Here, we choose 80% to capture an "average" electrolyte fill quantity that spans all three cell formats. Knowledge of electrolyte density enables subsequent calculation of g-Ah$^{-1}$ quantities.

## 8.1: Cell modeling validation with experimental data

To demonstrate the utility and accuracy of the above modeling approach, we validate against publicly available information on 12Ah large-format pouch cells produced by UK-based Na-ion battery developer Faradion (now Reliance-owned). Faradion's "Gen 2" layered oxide cathode is a 2/3 – 1/3 blend of an O3- and P2-type material[19]. The O3-type end member has a stoichiometry of $NaNi_{0.333}Mn_{0.333}Mg_{0.167}Ti_{0.167}O_2$ and the P2-type end member has a stoichiometry of $Na_{0.667}Ni_{0.300}Mn_{0.600}Mg_{0.033}Ti_{0.067}O_2$, delivering close to 150 mAh-g$^{-1}$ when discharged from 4.25V to 2.0V vs Na/Na$^+$ [20]. Additionally, Faradion's internally-developed hard carbon has been demonstrated to have up to 330 mAh-g$^{-1}$ capacity with 91.9% first cycle efficiency[20]. Half-cell voltage curves for the first desodiation-sodiation (charge-discharge) for the Gen 2 cathode were extracted from published data[20]. Similarly, half-cell voltage curves for the first sodiation-desodiation for Faradion's hard carbon material were also extracted from published data[20]. Electrode composite formulations for positive and negative electrodes were informed by yet other publicly available literature published by the same company, along with realizable mass loadings, calender densities, and cathode/anode mass balancing[21]. Finally, dimensional analysis was performed on images presented by Faradion in publicly available symposiums[19] on their 12Ah pouch cell platform to obtain anode and cathode electrode areal footprints (Figure 8.2).

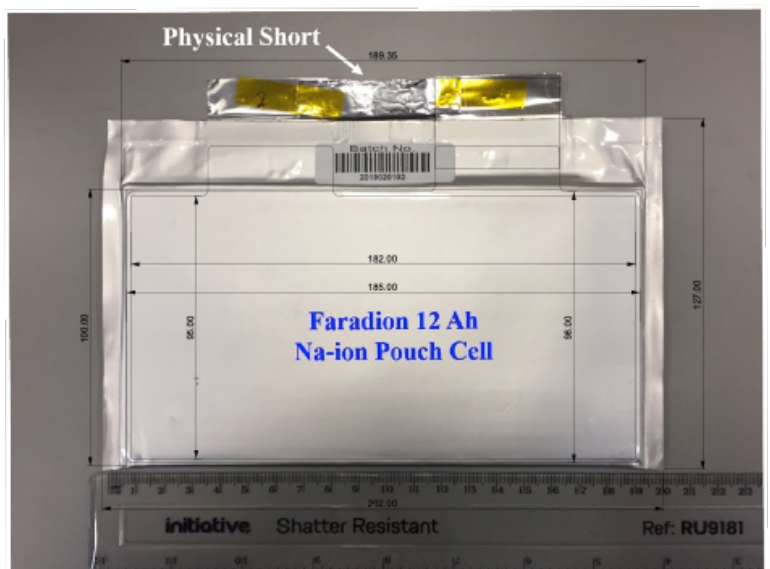

*Figure 8.2: Dimensional analysis of Faradion's 12Ah pouch cell platform with estimated electrode footprint design.*

Based on this information, we produce full cell predictions summarized in Figure 8.3 with unit cell details summarized in Table 8.1, showing good agreement with Faradion's published results. Not only do predicted cell parameters match Faradion reported values, but our predicted theoretical voltage curves (black) also match near-identically with Faradion's full-cell voltage profiles (orange). Especially given the slopey voltage curves of the layered oxide cathode and hard carbon anode, matching full-cell voltage profiles this accurately gives confidence in our ability to capture realistic cell designs and parameters. Comparing between Figure 8.3a and Figure 8.3b, we see the effect of derating the cell upper cutoff voltage from 4.2V to 4.0V. Given various degradation mechanisms that dominate at higher operating voltages, Faradion demonstrated the ability to extend cycle life from 1,000 cycles to 3,000 cycles by lowering the cutoff voltage within the same cell design (i.e. no change to cathode loadings or cathode/anode balancing)[20]. However, the same unit cell system will therefore result in much higher N/P ratios, which is penalizing from an energy density perspective, summarized in Figure 8.3c, which will in-turn negatively affect cost in $-kWh$^{-1}$. See Section 8.7: for additional commentary on voltage ranges as it relates to systems integration with power conversion electronics.

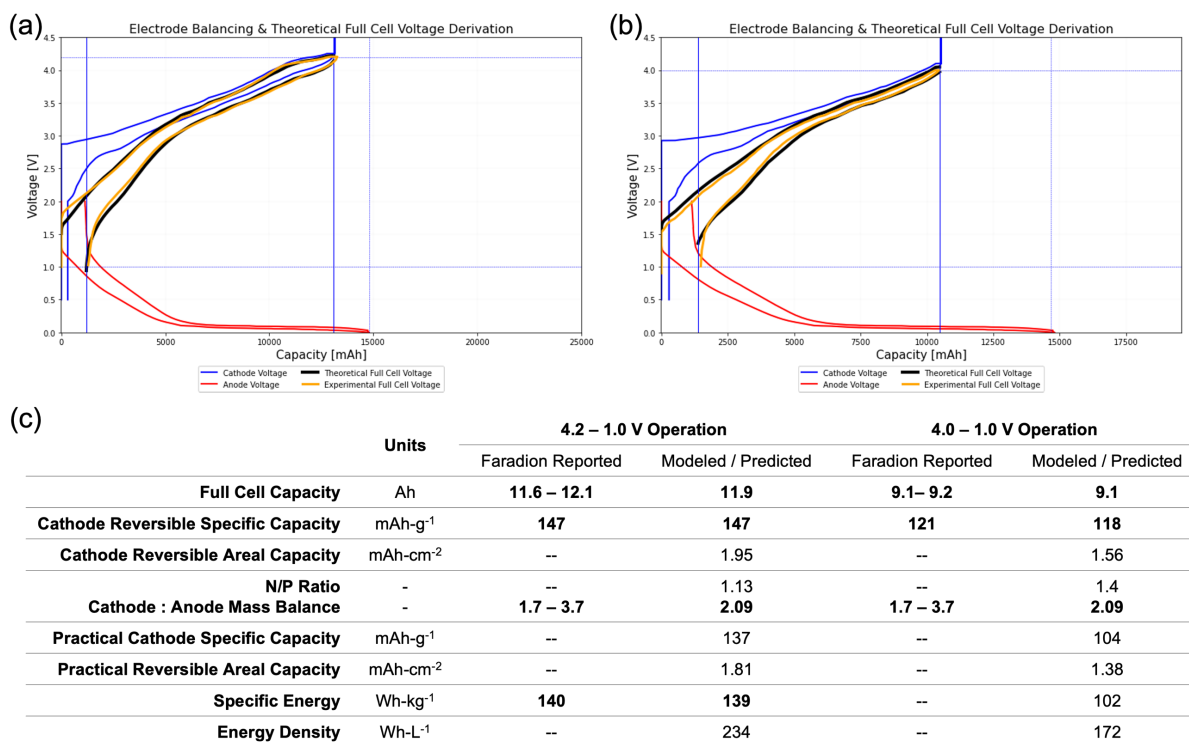

| | Units | 4.2 – 1.0 V Operation | | 4.0 – 1.0 V Operation | |
| --- | --- | --- | --- | --- | --- |
| | | Faradion Reported | Modeled / Predicted | Faradion Reported | Modeled / Predicted |
| **Full Cell Capacity** | Ah | **11.6 – 12.1** | **11.9** | **9.1 – 9.2** | **9.1** |
| **Cathode Reversible Specific Capacity** | mAh-g$^{-1}$ | **147** | **147** | **121** | **118** |
| **Cathode Reversible Areal Capacity** | mAh-cm$^{-2}$ | -- | 1.95 | -- | 1.56 |
| **N/P Ratio** | - | -- | 1.13 | -- | 1.4 |
| **Cathode : Anode Mass Balance** | - | 1.7 – 3.7 | 2.09 | 1.7 – 3.7 | 2.09 |
| **Practical Cathode Specific Capacity** | mAh-g$^{-1}$ | -- | 137 | -- | 104 |
| **Practical Reversible Areal Capacity** | mAh-cm$^{-2}$ | -- | 1.81 | -- | 1.38 |
| **Specific Energy** | Wh-kg$^{-1}$ | **140** | **139** | -- | 102 |
| **Energy Density** | Wh-L$^{-1}$ | -- | 234 | -- | 172 |

*Figure 8.3: Summary of modeling results compared with Faradion actual experimental curves and reported data figures. (a) Modeling Faradion 12Ah pouch cell from 4.2 – 1.0V. Blue and red curves produced via mathematical scaling of extracted half-cell data published by Faradion. Black curve predicted based upon cell balancing exercise. Orange curve overlay shows actual extracted voltage curves from Faradion full cells, showing good agreement with our model. (b) Same cell design but derated to 4.0 – 1.0V operation. Resulting cell voltage and capacity matches Faradion reported results. (c) Table of key cell parameters obtained via modeling and compared against Faradion reported figures. Good agreement is found for all publicly reported values from Faradion, shown in bold.*

*Table 8.1: Summary of unit cell design parameters for Faradion 12Ah cell modeling.*

| Parameter | Units | Value | Ref. |
| --- | --- | --- | --- |
| **Cathode** | | | |
| Electrode Formulation (AM : CB : Binder) | % | 89 : 5 : 6 | 21 |
| Mass Loading | mg-cm$^{-2}$ | 14.90 | |
| Calender Press Density | g-cm$^{-3}$ | 2.77 | 18 |
| Electrode Porosity | % | 30.0 | 18 |
| Electrode Thickness | μm | 53.8 | |
| Substrate | -- | Aluminum | |
| Current Collector Thickness | μm | 15 | |
| **Anode** | | | |
| Electrode Formulation (AM : CB : Binder) | % | 88 : 3 : 9 | 21 |
| Mass Loading | mg-cm$^{-2}$ | 7.14 | 21 |
| Calender Press Density | g-cm$^{-3}$ | 0.95 | 20 |
| Electrode Porosity | % | 38.0 | 18,21 |
| Electrode Thickness | μm | 75.2 | |
| Substrate | -- | Aluminum | |
| Current Collector Thickness | μm | 15 | |
| **Separator** | | | |
| Material | -- | Polypropylene / polyolefin | 20 |
| Separator Thickness | μm | 15 | |
| Separator Porosity | % | 47 | |
| **Electrolyte** | | | |
| Electrolyte Density | g-cm$^{-3}$ | 1.20 | |
| Fill amount | g-Ah$^{-1}$ | 6.0 | 20 |

## 8.2: Li-ion LFP – CATL Prismatic Cell (Tesla Model 3, E6R)

The prismatic LFP cell used in the Tesla Model 3 Standard Range model is manufactured by CATL (Contemporary Amperex Technology Limited) starting in 2020, and a detailed cell teardown is referenced in literature[22]. Based on the parameters obtained from the literature, a pouch cell representative of the same unit cell design was constructed in BatPaC, and exact cell design construction details were obtained (e.g. electrode areas, dimensions, bilayer count, etc.). The resulting cell design details are shown in Figure 8.4.

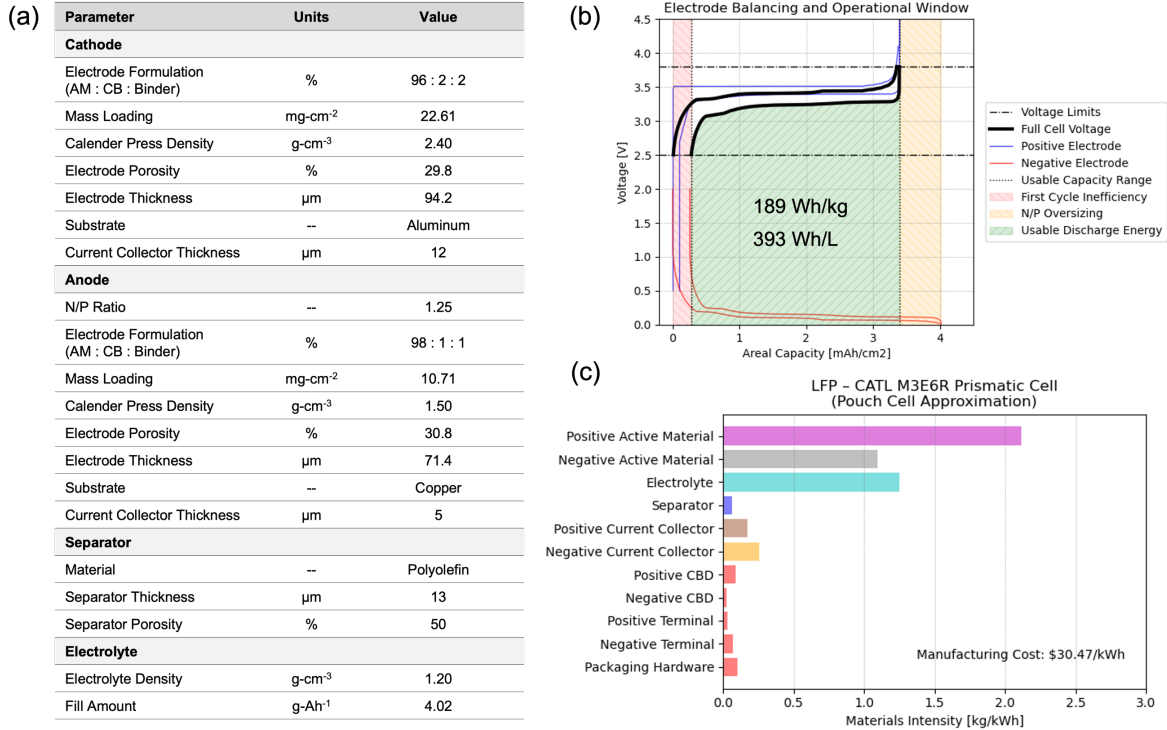

*Figure 8.4: Modeled cell design details. (a) Unit cell design parameters. (b) Electrode balancing in BatPaC-designed pouch cell representative. (c) Calculated materials intensity for BatPaC-designed pouch cell representative.*

## 8.3: Li-ion LFP – A123 18650 Cylindrical Cell

A cylindrical LFP 18650 power cell manufactured by A123 in 2009 for power tool applications is modeled here based upon a cell teardown from literature[23]. Based on the parameters obtained from the literature, a pouch cell representative of the same unit cell design was constructed in BatPaC, and exact cell design construction details were obtained (e.g. electrode areas, dimensions, bilayer count, etc.). The resulting cell design details are shown in Figure 8.5.

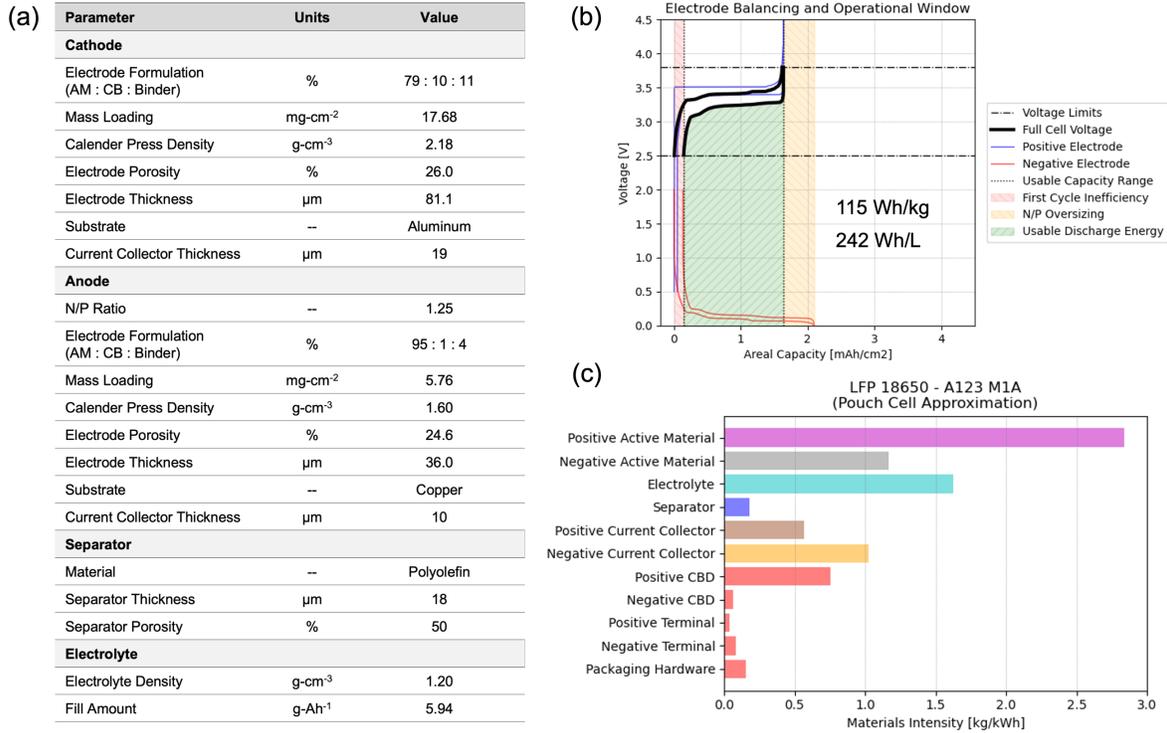

Figure 8.5: Modeled cell design details. (a) Unit cell design parameters. (b) Electrode balancing in BatPaC-designed pouch cell representative. (c) Calculated materials intensity for BatPaC-designed pouch cell representative.

## 8.4: Li-ion NMC – LGX E78 Pouch Cell (Volkswagen ID.3, Audi Q4 e-tron, others)

The pouch NMC cell used in the Volkswagen ID.3, Audi Q4 e-tron, and other vehicle models is manufactured by LG Energy Solutions, and a detailed cell teardown is referenced in literature[24]. Unlike the previous examples, this cell is already a large-format pouch cell, so we first model the actual design produced by LG Energy Solutions before converting it to a BatPaC-generated cell design. We obtain good agreement with the commercial cell but note that the total electrolyte overfill amount in addition to the actual calculated pore volume is closer to 10-20% as opposed to our assumed 80%. This makes sense as pouch cells are generally more electrolyte-starved cell designs compared to that of hard-case cells like prismatic or cylindrical formats. Then, based on the parameters that are in good agreement with the commercial cell, a pouch cell representative of the same unit cell design was constructed in BatPaC, and exact cell design construction details were obtained. The resulting cell design details are shown in Figure 8.6.

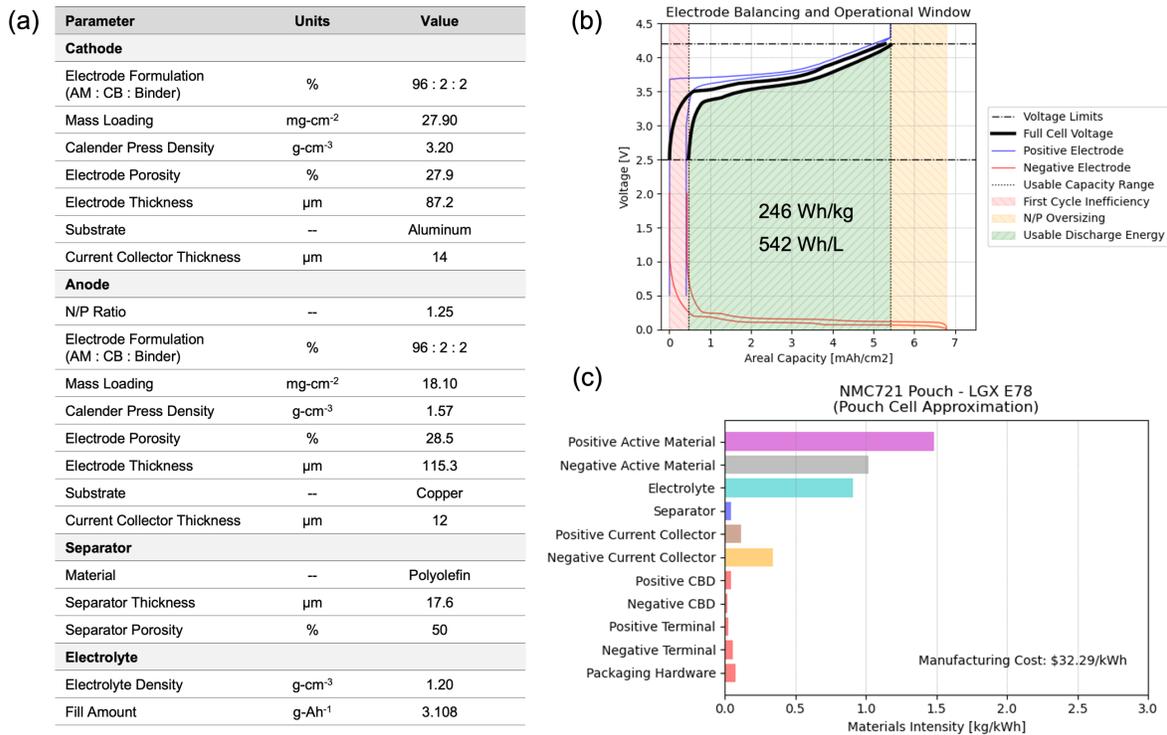

*Figure 8.6: Modeled cell design details. (a) Unit cell design parameters. (b) Electrode balancing in BatPaC-designed pouch cell representative. (c) Calculated materials intensity for BatPaC-designed pouch cell representative.*

## 8.5: Li-ion NMC – Tesla 4680 Cylindrical Cell (Tesla Model Y, other Tesla models)

The cylindrical NMC cell used in the Tesla Model Y and manufactured by Tesla and a detailed cell teardown is referenced in literature[25]. Based on the parameters obtained from the literature, a pouch cell representative of the same unit cell design was constructed in BatPaC, and exact cell design construction details were obtained (e.g. electrode areas, dimensions, bilayer count, etc.). The resulting cell design details are shown in Figure 8.7. We note here, in agreement with the authors of the cell teardown study[25], that this cell design appears to be more conservative, with a larger N/P ratio and lower calender densities than one may expect. These design decisions, however, may be more conservative considering the novelty of the 4680 cell format, and future engineering optimization can likely be rapidly implemented to increase cell-level energy densities.

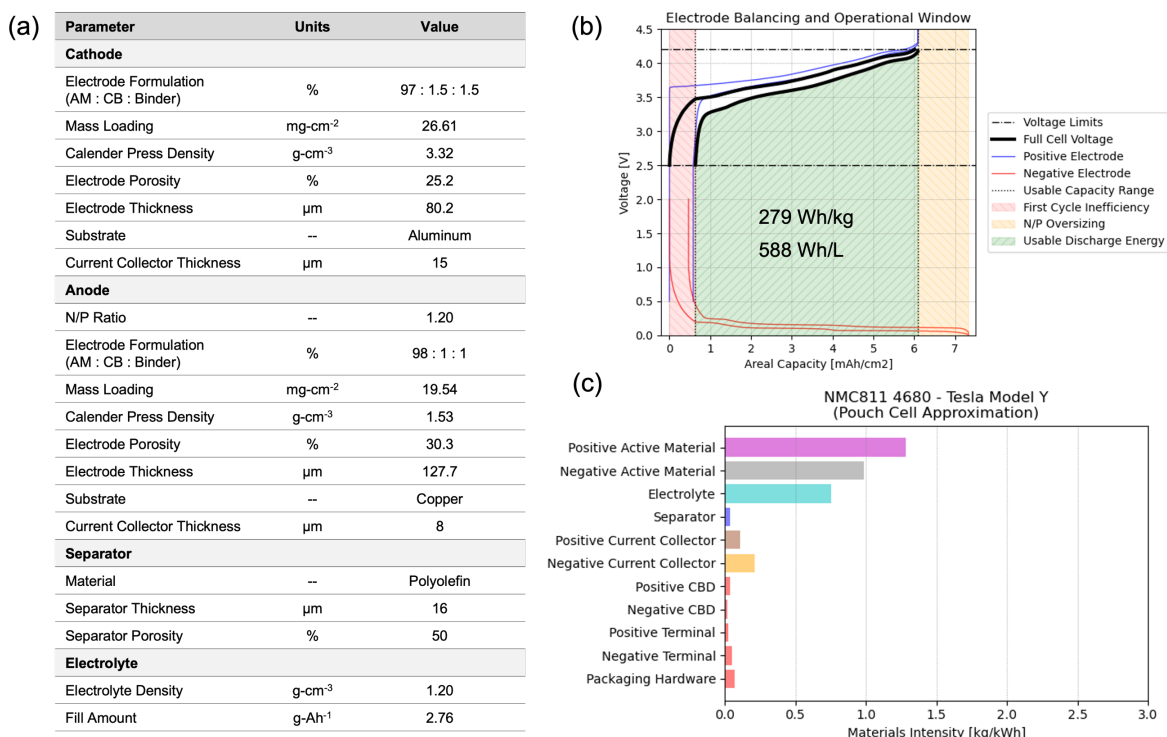

*Figure 8.7: Modeled cell design details. (a) Unit cell design parameters. (b) Electrode balancing in BatPaC-designed pouch cell representative. (c) Calculated materials intensity for BatPaC-designed pouch cell representative.*

## 8.6: Li-ion NMC – PHEV Pouch Cell

A Li-ion pouch cell representing state-of-art cells in 2014 is modeled after documents and literature gathered from around that time[26,27]. With the launch of the PHEV Chevrolet Volt circa 2011, NMC-type cells at the time tended to be power cell designs as opposed to the energy cells commonly used in BEVs today. Consistent with power cell designs, these cells had low areal loadings ~2.0 mAh-cm-2, higher electrode porosities, and thicker current collectors and separators. The resulting cell design details are shown in Figure 8.8.

(a)

| Parameter | Units | Value |
|---|---|---|
| **Cathode** | | |
| Electrode Formulation (AM : CB : Binder) | % | 96 : 2 : 2 |
| Mass Loading | mg-cm$^{-2}$ | 12.68 |
| Calender Press Density | g-cm$^{-3}$ | 3.06 |
| Electrode Porosity | % | 30.0 |
| Electrode Thickness | μm | 41.4 |
| Substrate | -- | Aluminum |
| Current Collector Thickness | μm | 20 |
| **Anode** | | |
| N/P Ratio | -- | 1.25 |
| Electrode Formulation (AM : CB : Binder) | % | 98 : 1 : 1 |
| Mass Loading | mg-cm$^{-2}$ | 7.17 |
| Calender Press Density | g-cm$^{-3}$ | 1.537 |
| Electrode Porosity | % | 30.0 |
| Electrode Thickness | μm | 46.6 |
| Substrate | -- | Copper |
| Current Collector Thickness | μm | 12 |
| **Separator** | | |
| Material | -- | Polyolefin |
| Separator Thickness | μm | 20 |
| Separator Porosity | % | 50 |
| **Electrolyte** | | |
| Electrolyte Density | g-cm$^{-3}$ | 1.20 |
| Fill Amount | g-Ah$^{-1}$ | 4.056 |

(b)
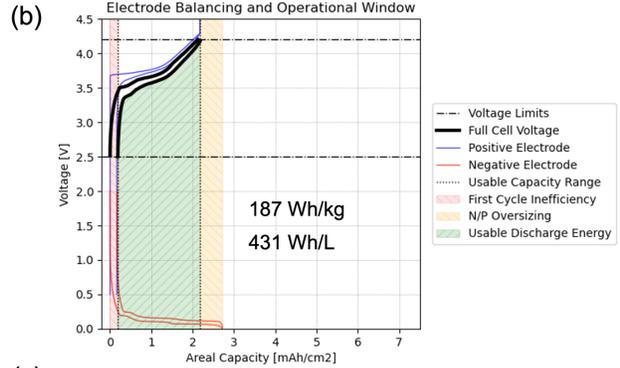

(c)
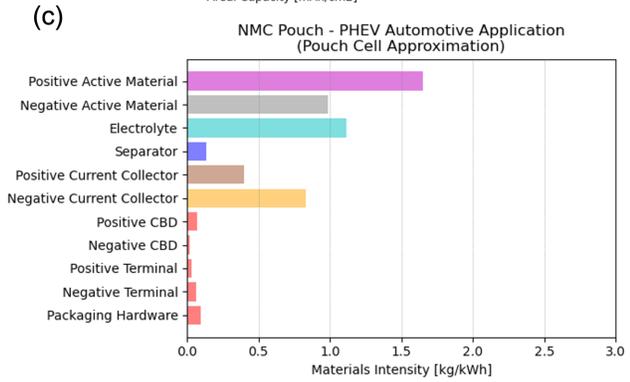

*Figure 8.8: Modeled cell design details. (a) Unit cell design parameters. (b) Electrode balancing in BatPaC-designed pouch cell representative. (c) Calculated materials intensity for BatPaC-designed pouch cell representative.*

## 8.7: Na-ion NaNM | HC – Baseline

As rapid progress is being made within Na-ion commercialization efforts, we create an anticipated Na-ion baseline cell design to represent state-of-art in 2024. This design is based upon the Faradion 12Ah cell discussed in Section 8.1: but with more aggressive design parameters. Namely, this includes increased electrode loadings and active mass fractions within electrodes. This also assumes a lower cost aqueous-based binder system using carboxymethyl cellulose (CMC) and styrene-butadiene rubber (SBR) for the hard carbon (HC) anode (as is predominantly used for graphite anodes in Li-ion batteries) instead of the polyvinylidene (PVDF) based system used in Faradion's formulations found within literature[21]. We also limit the upper voltage cutoff to 4.0V. As previously discussed in Section 8.1:, a derated upper voltage cutoff minimizes degradation reactions that have been empirically shown to increase cycle life. However, another key reason to derate voltages is to enable compatibility with state-of-art power electronics, according to feedback from several experts on our panel of industry consultants[18]. Whereas a NaNM | HC system may exhibit a cell voltage between 1.0 – 4.2V, most power conversion systems cannot handle such a wide voltage range and may be limited to $V_{min} \geq \frac{1}{2} V_{max}$[18]. Furthermore, for constant power system requirements, the current draw requirement at low states-of-charge (SOC) will substantially increase as voltage drops precipitously. We note that this is a topic that may be underappreciated and deserves much greater scrutiny. Here, we limit the voltage range of a baseline NaNM | HC system between 1.5 – 4.0V.

Compared to the Faradion 12Ah baseline cell discussed in Section 8.1:, we increase active mass fractions of both the cathode and anode electrode formulations from 89% and 88% to 96% and 95%, respectively. We further increase cathode and anode mass loadings from 14.90 mg-cm$^{-2}$ and 7.14 mg-cm$^{-2}$ to 19.77 mg-cm$^{-2}$ and 7.50 mg-cm$^{-2}$, respectively, while also reducing the electrode porosity and hence increasing the calender density of the cathode. We note that these elevated calender densities are aggressive values based on the challenging morphology of state-of-art co-precipitated NaNM-type cathode particles[18], but we prefer to err on the side of presenting a best-case scenario for Na-ion. Additionally, we drop the N/P ratio to 1.10 balanced at 4.0V to maximize material utilization. This combination of improvements increases the cell-level areal capacity by ~50% versus the Faradion 4.0V baseline. Here, we maintain the same capacity and first cycle efficiencies (FCEs) of the Faradion Gen 2 NaNM cathode and HC anode materials, as they represent state-of-art performance within their class. The result of these design changes is summarized in Figure 8.9.

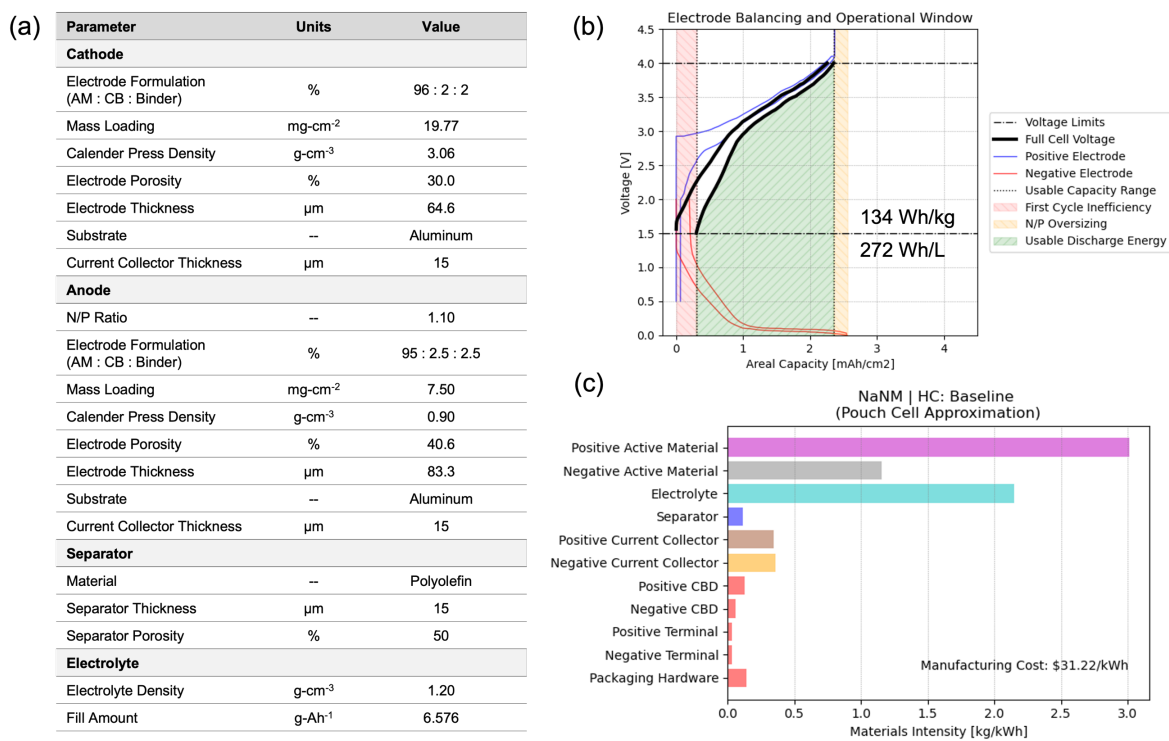

*Figure 8.9: Modeled cell design details. (a) Unit cell design parameters. (b) Electrode balancing in BatPaC-designed pouch cell representative. (c) Calculated materials intensity for BatPaC-designed pouch cell representative.*

## 8.8: Na-ion NaNM | HC – Increased electrode thicknesses

One of the most common ways to increase cell-level energy densities and reduce costs is to increase the thickness of electrodes—thereby minimizing the quantity of inactive materials. Of course, this will have a tradeoff with power density as thicker electrodes require longer distances for ion mass transport in the electrolyte phase. Therefore, there are practical limits to increasing electrode thicknesses prior to the onset of significant concentration polarization—which may result in unwanted phenomena such as sodium plating upon charging[21], not to mention challenges from a manufacturability perspective of drying thick electrodes after coating[18]. Increasing the practically attainable thickness of hard carbon anodes will result in energy density increases and cost decreases assuming rate performance is not hindered for the applications considered.

In this first baseline-improved scenario, we increase the realizable electrode mass loadings for the hard carbon anode to 10.00 mg-cm$^{-2}$, which was communicated as a realistic but challenging target by industry experts[18]. We additionally increase calender densities. Typical maximum calender densities for a hard carbon anode are below 0.9 g-cm$^{-3}$ which corresponds to ~40% porosity, but here we increase the density to 0.938 g-cm$^{-3}$ to achieve a realistically attainable 38% porosity[18,20,21]. This results in a single-sided anode thickness of 107 µm. We also assume thinner current collectors—which are yet additional common strategies associated with electrode optimization. The result of these design changes is summarized in Figure 8.10.

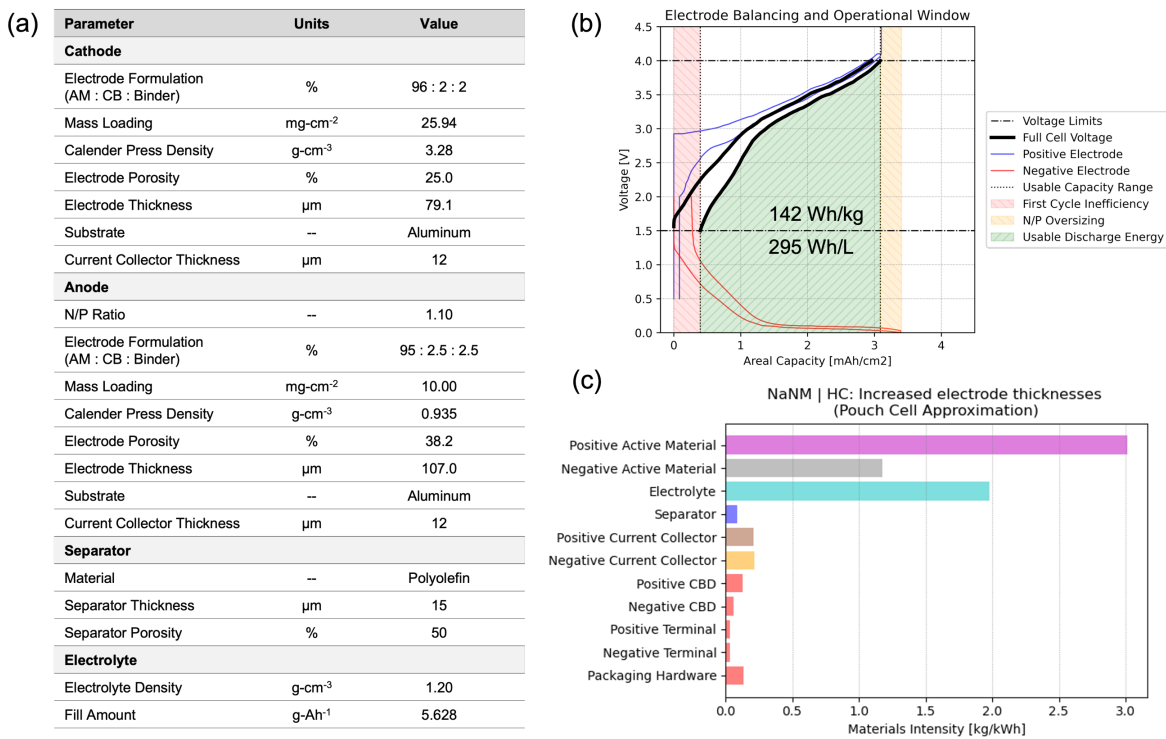

*Figure 8.10: Modeled cell design details. (a) Unit cell design parameters. (b) Electrode balancing in BatPaC-designed pouch cell representative. (c) Calculated materials intensity for BatPaC-designed pouch cell representative.*

## 8.9: Na-ion NaNM | HC – Increased anode specific capacity

Hard carbons are non-graphitizable disordered materials consisting graphene nanosheets, and a mechanistic understanding of their sodium storing capability is still not widely agreed upon[20,28–30]. Broadly, however, hard carbons are universally observed to exhibit a first slopey region in the voltage profile upon sodiation and a second flat region just above 0V vs. Na/Na$^+$, with each contributing approximately half of the overall useable capacity of the anode[20,28]. Furthermore, there appears to be general agreement attributing sodium storage via interactions with the graphene nanosheets (either intercalation between sheets or localization around defects) to the slopey region, and sodium cluster formation within micropores to the flat region[20,28]. Both phenomena appear highly dependent on the hard carbon precursor type (phenolic resin, biomass, anthracite, etc.) and synthesis conditions. Due to an inherently amorphous structure without clearly defined stoichiometry (such as $LiC_6$ for graphite), the limits to sodium storage capacities of hard carbons are not well-understood, and there is significant variability in reported values, ranging from 200–500 mAh-g$^{-1}$ [31].

In this scenario, we assume 400 mAh-g$^{-1}$ as a reach goal for useable, reversible specific capacity based upon expert opinion[18]. Given the importance of internal microporosity to the storage mechanism of hard carbons, we do not assume significant improvements to material, tap, or calender densities beyond what was discussed in the previous section to be practically realizable. We do, however, increase the calender densities on the cathode, assume thinner current collectors, and further increase the realizable electrode mass loadings for the hard carbon anode to 10.00 mg-cm$^{-2}$, which was communicated as a realistic but challenging target by industry experts[18]. This results in a single-sided anode thickness of 107 µm. Additionally, we assume the precursor for a high-capacity hard carbon material is phenolic resin-based (which has implications on assumed starting costs, see Main Text Figure 2c). This is based upon feedback regarding the variability in quality and composition for biomass precursors, and the inflexibility to micropore engineering in anthracite precursors[18]. Keeping all else the same, the result of these design changes is summarized in Figure 8.11, now with substantially higher achievable areal capacity while still being limited to 4.0V operation.

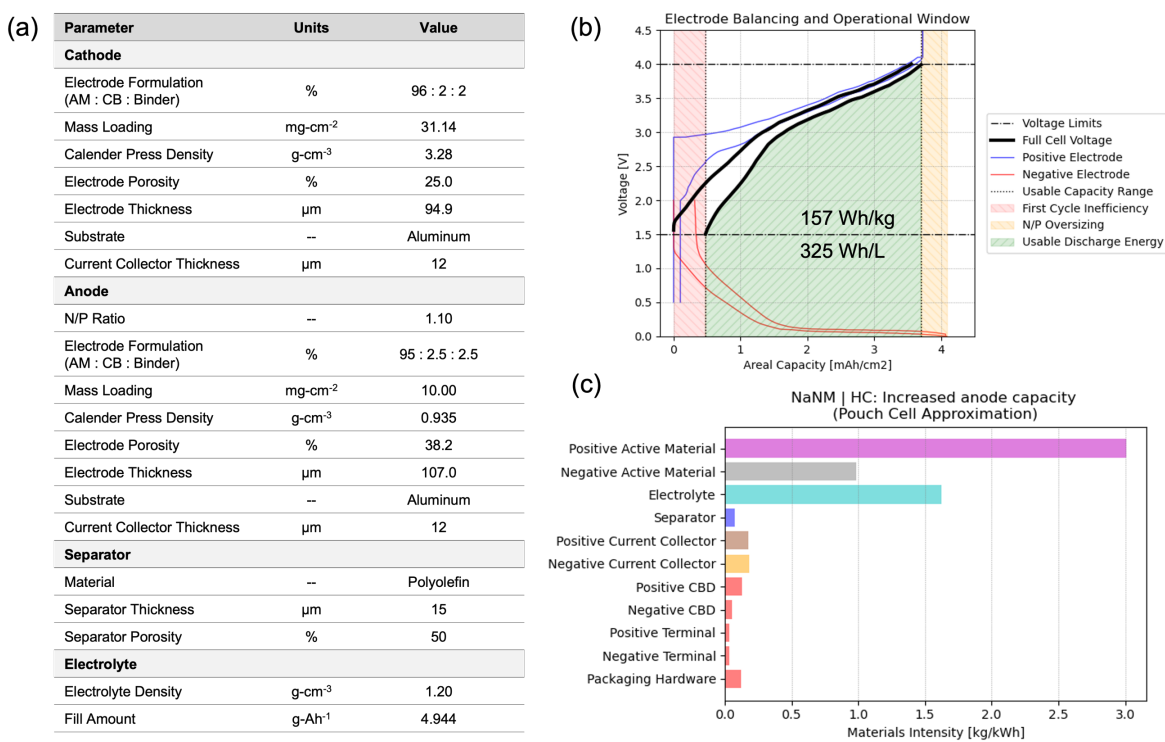

*Figure 8.11: Modeled cell design details. (a) Unit cell design parameters. (b) Electrode balancing in BatPaC-designed pouch cell representative. (c) Calculated materials intensity for BatPaC-designed pouch cell representative.*

## 8.10: Na-ion NaNM | HC – Increased cathode specific capacity

Yet another way to boost areal capacities is to increase the specific capacity of the layered oxide cathode, ideally keeping most of the capacity increase under a 4.0V upper cutoff voltage for reasons discussed above. Here, we simply evaluate a hypothetical 20% increase to realizable specific capacity in the cathode as a scenario analysis, which entails mathematically scaling the half-cell voltage curve obtained from Faradion's Gen 2 cathode. We note that actual increases to specific capacity via potential use of different transition metal redox centers, for example, will likely alter the shape of the voltage profile compared to the current NaNM representative. However, this approach will provide a reasonable estimate of the technoeconomic impact of such a development roadmap assuming the materials starting costs and learning rates will follow that of the layered oxide class of materials. Similarly to the previous scenario, we also assume increased cathode calender densities, thinner current collectors, and increased anode mass loadings. The resulting cell design is summarized in Figure 8.12.

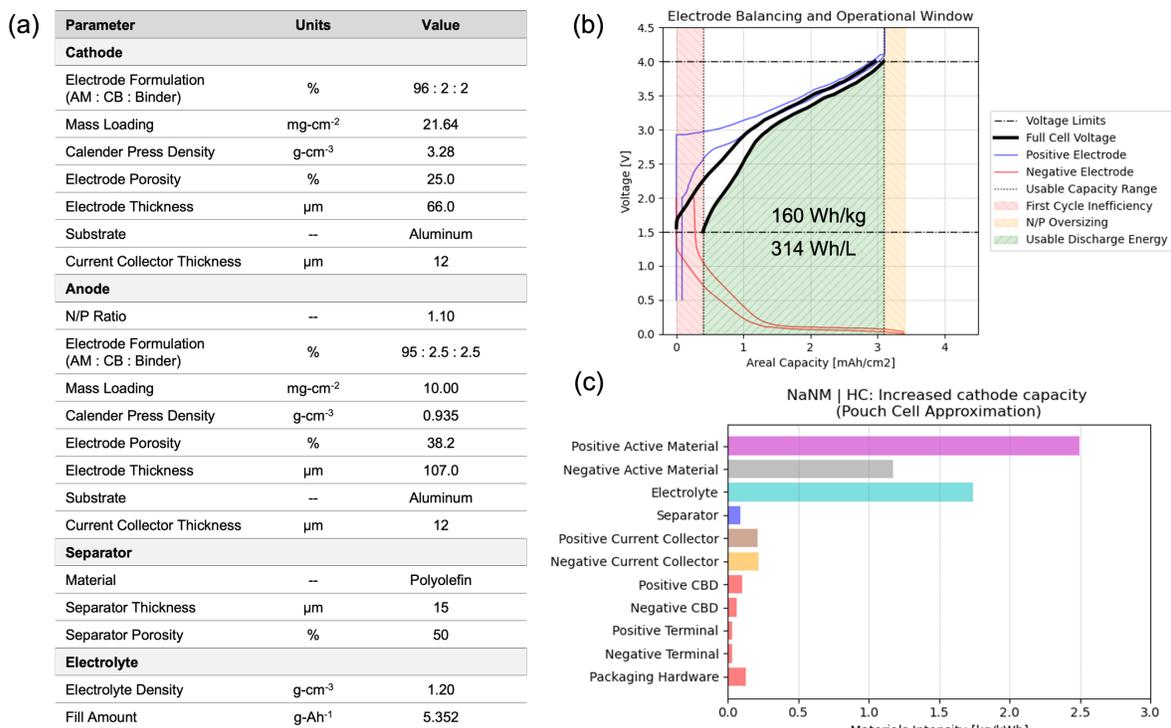

*Figure 8.12: Modeled cell design details. (a) Unit cell design parameters. (b) Electrode balancing in BatPaC-designed pouch cell representative. (c) Calculated materials intensity for BatPaC-designed pouch cell representative.*

## 8.11: Na-ion NaNM | HC – Increased cathode & anode specific capacities

Assuming improvements to the hard carbon and NaNM cathode specific capacities can occur in parallel, we combine their contributions in this scenario. The resulting cell design is summarized in Figure 8.13.

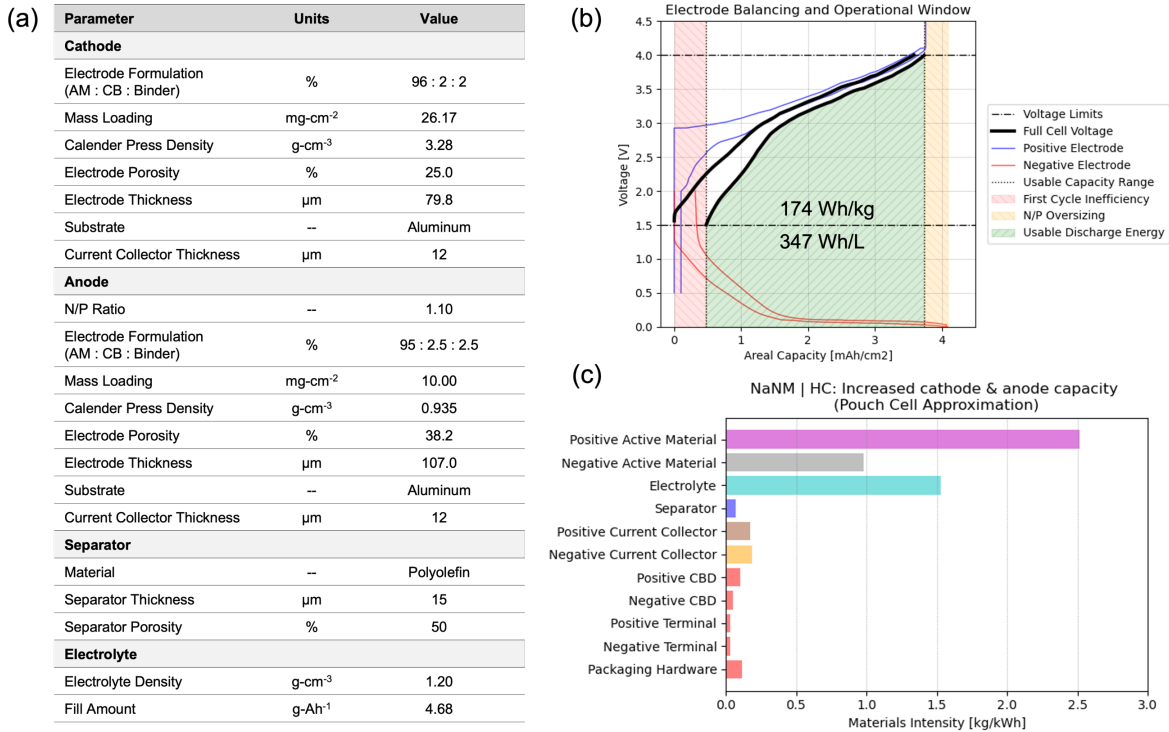

*Figure 8.13: Modeled cell design details. (a) Unit cell design parameters. (b) Electrode balancing in BatPaC-designed pouch cell representative. (c) Calculated materials intensity for BatPaC-designed pouch cell representative.*

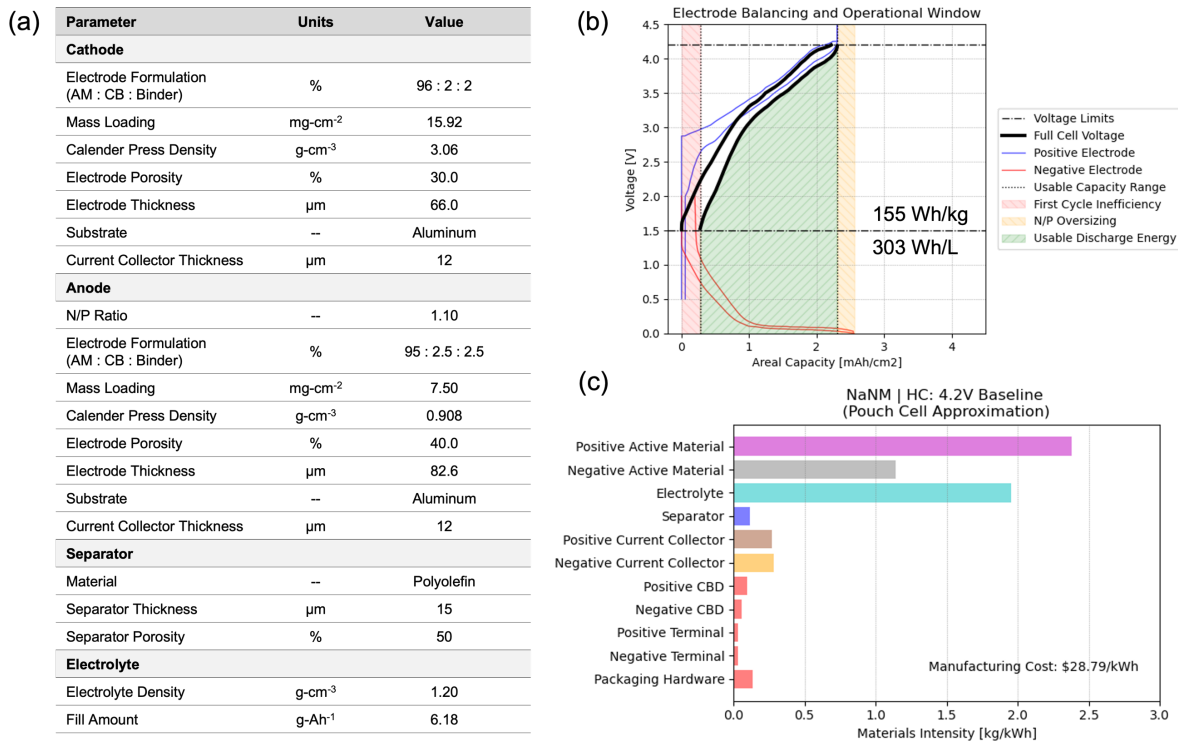

*Figure 8.14: Modeled cell design details. (a) Unit cell design parameters. (b) Electrode balancing in BatPaC-designed pouch cell representative. (c) Calculated materials intensity for BatPaC-designed pouch cell representative.*

## 8.12: Na-ion NaNM | HC – 4.2V Baseline

Assuming a wider cell voltage range is tolerable to cycle life implications and power conversion electronics, we augment the Baseline scenario from Section 8.7: by increasing the cathode voltage while still balancing the cell to a 1.10 N/P ratio. The resulting cell design is summarized in Figure 8.14.

## 8.13: Na-ion NaNM | HC – 4.2V with increased cathode & anode specific capacities

Combining the increased hard carbon specific capacity of 400 mAh-g$^{-1}$ and the 20% increase to NaNM cathode specific capacity while operating at 4.2V, we evaluate the highest energy density cell design scenario using layered oxide cathodes and hard carbons, summarized in Figure 8.15.

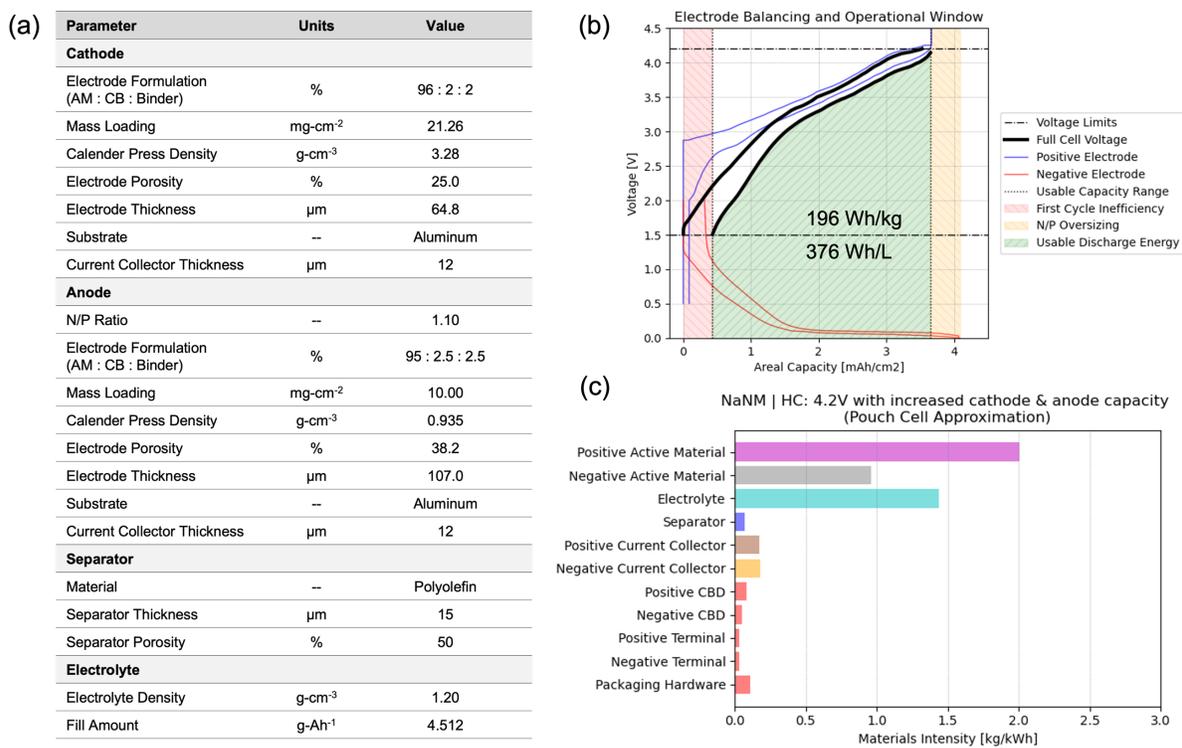

Figure 8.15: Modeled cell design details. (a) Unit cell design parameters. (b) Electrode balancing in BatPaC-designed pouch cell representative. (c) Calculated materials intensity for BatPaC-designed pouch cell representative.

## 8.14: Na-ion NaNM| | HC – Baseline with low-cost hard carbon

As discussed earlier in Section 8.8:, hard carbons can be synthesized from various precursors, and can have significantly differing performance characteristics as sodium storage is highly dependent on hard carbon microstructure. It is widely understood that the best performing hard carbons in terms of specific capacity are derived from phenolic resin, but they are typically more expensive due to additional costs of chemical feedstock and processing[18]. Alternatives to phenolic resin-based hard carbons are those derived from biomass precursors such as coconut shells or fossil fuel precursors such as anthracite, the latter of which have significantly lower costs as detailed in Main Text Figure 2c. However, as these carbon precursors are already in their disordered state, minimal microstructural restructuring can be performed with further processing, and thus little performance gain can be engineered and realized[18]. Anthracite-based hard carbons are thus known to be low-cost options with low specific capacity, approximately 250 mAh-g$^{-1}$ [31].

Here, we model cells that may use such a material, forgoing specific capacity performance to use low-cost hard carbons. Note that there are additional knock-on effects to this design decision. Not only is the anode active material intensity (in kg-kWh$^{-1}$) increased due to lower capacity, but the increased total pore volume resulting from the thicker hard carbon anode will require substantially more electrolyte, which also represents a non-trivial fraction of overall cell costs. The resulting cell design is summarized in Figure 8.16.

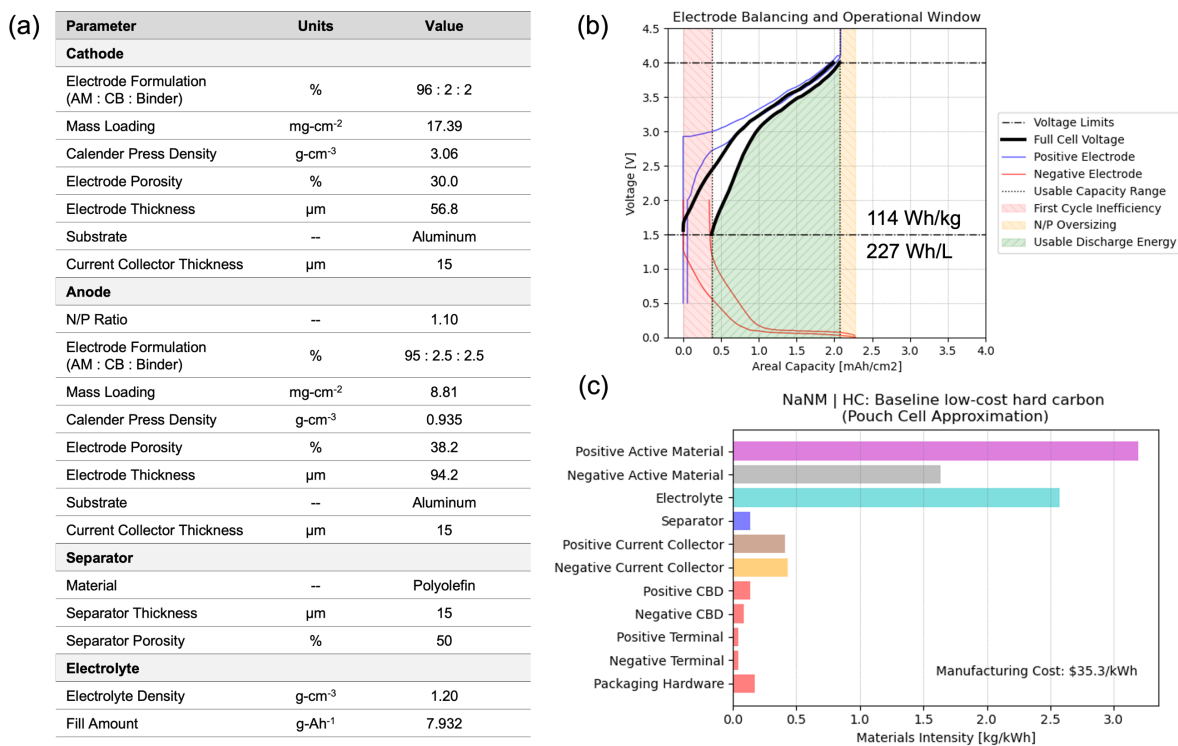

*Figure 8.16: Modeled cell design details. (a) Unit cell design parameters. (b) Electrode balancing in BatPaC-designed pouch cell representative. (c) Calculated materials intensity for BatPaC-designed pouch cell representative.*

## 8.15: Na-ion NFPP | HC – Baseline

Another class of cathodes that have gained significant attention are polyanionic materials. Specifically, within polyanionic materials, $Na_4Fe_3(PO_4)_2(P_2O_7)$ or NFPP is a vanadium-free mixed phosphate cathode that has gained recent popularity as a stable-cycling and low-cost material. However, with a relatively low specific capacity and tap density compared to layered oxide materials, the NFPP cathode becomes the limiting electrode as opposed to the hard carbon anode[18]. Similar to the NaNM baseline examples discussed above in Section 8.7: and 8.14:, we model a NFPP | HC baseline cell design here, summarized in Figure 8.17.

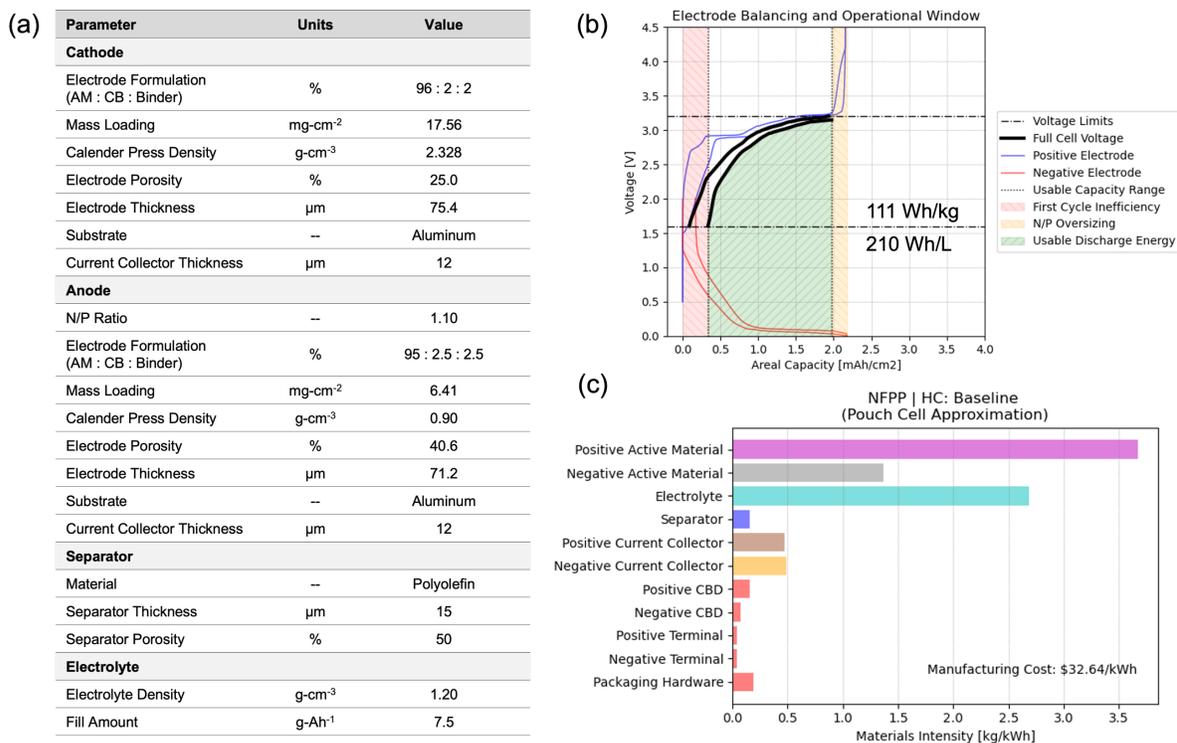

*Figure 8.17: Modeled cell design details. (a) Unit cell design parameters. (b) Electrode balancing in BatPaC-designed pouch cell representative. (c) Calculated materials intensity for BatPaC-designed pouch cell representative.*

## 8.16: Na-ion NFPP | HC – Increased electrode thicknesses

As discussed in Section 8.8:, increasing electrode thicknesses is a common strategy to increase energy densities and reduce costs. However, as NFPP cathodes are low in specific capacity and tap density, they become the limiting electrode in a NFPP | HC electrochemical couple in terms of thickness, and a post-calendered thickness of >100 µm would be required. This is substantially thicker than what state-of-art wet slurry-coated electrodes can typically achieve, and there is consensus that dry-electrode coating technology must be utilized to achieve such thicknesses for NFPP[18]. In this example, we model an NFPP thickness of 125 µm, and the design is summarized in Figure 8.18.

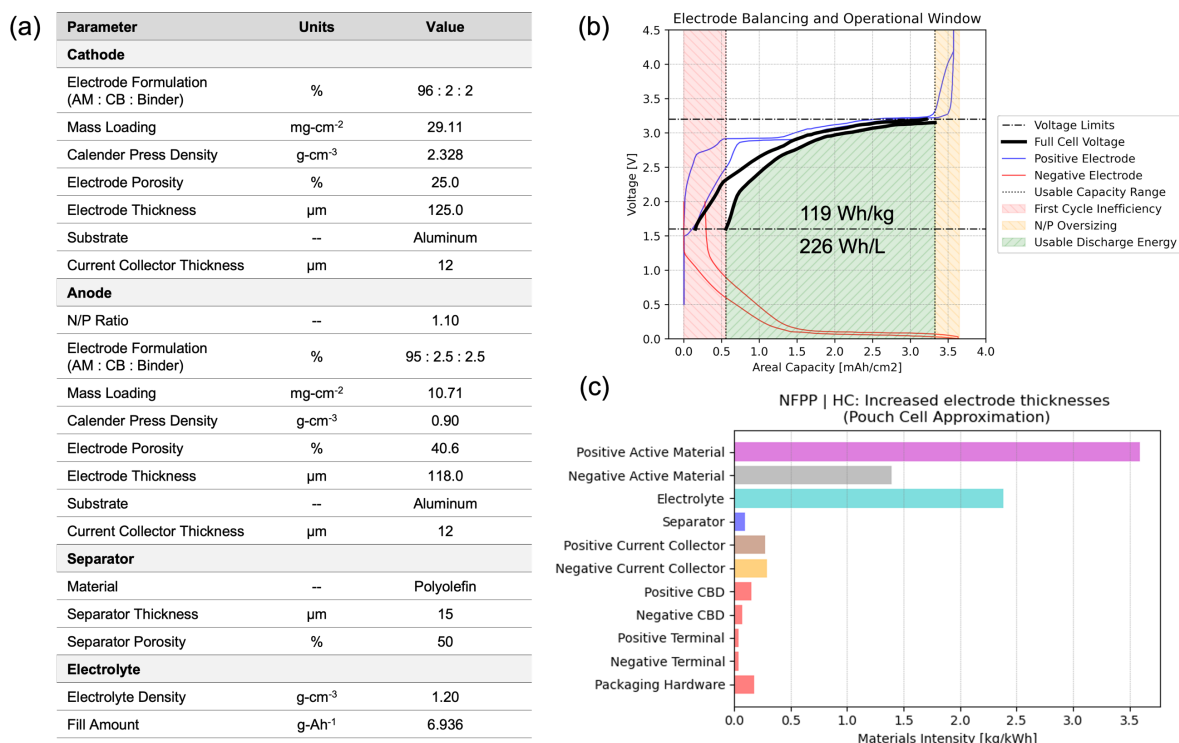

*Figure 8.18: Modeled cell design details. (a) Unit cell design parameters. (b) Electrode balancing in BatPaC-designed pouch cell representative. (c) Calculated materials intensity for BatPaC-designed pouch cell representative.*

## 8.17: Na-ion NFPP | HC – Increased anode specific capacity

As in Section 8.8:, we similarly model an increase in the hard carbon anode specific capacity. As the assumed NFPP capacity of 128 mAh-g$^{-1}$ is approximately the theoretical maximum for this material with three sodium ions transferred, improvements to the anode capacity may be the only available route for increased material-level performance. The resulting cell design is summarized in Figure 8.19.

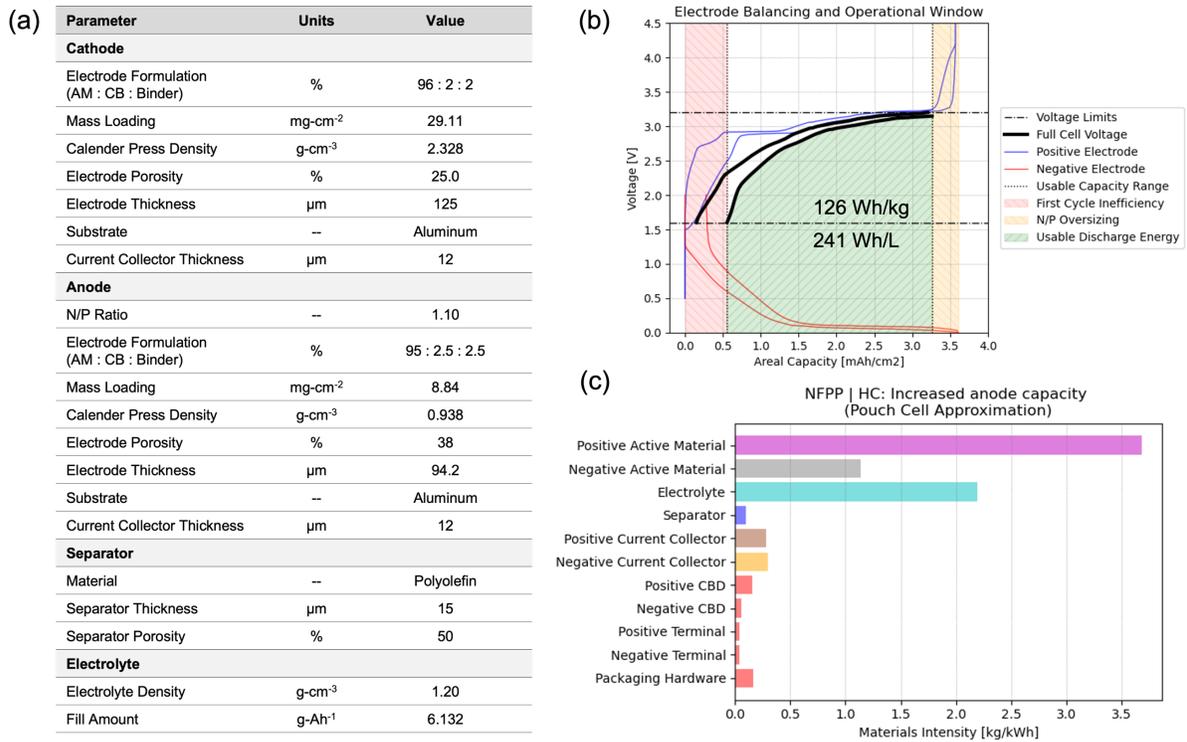

*Figure 8.19: Modeled cell design details. (a) Unit cell design parameters. (b) Electrode balancing in BatPaC-designed pouch cell representative. (c) Calculated materials intensity for BatPaC-designed pouch cell representative.*

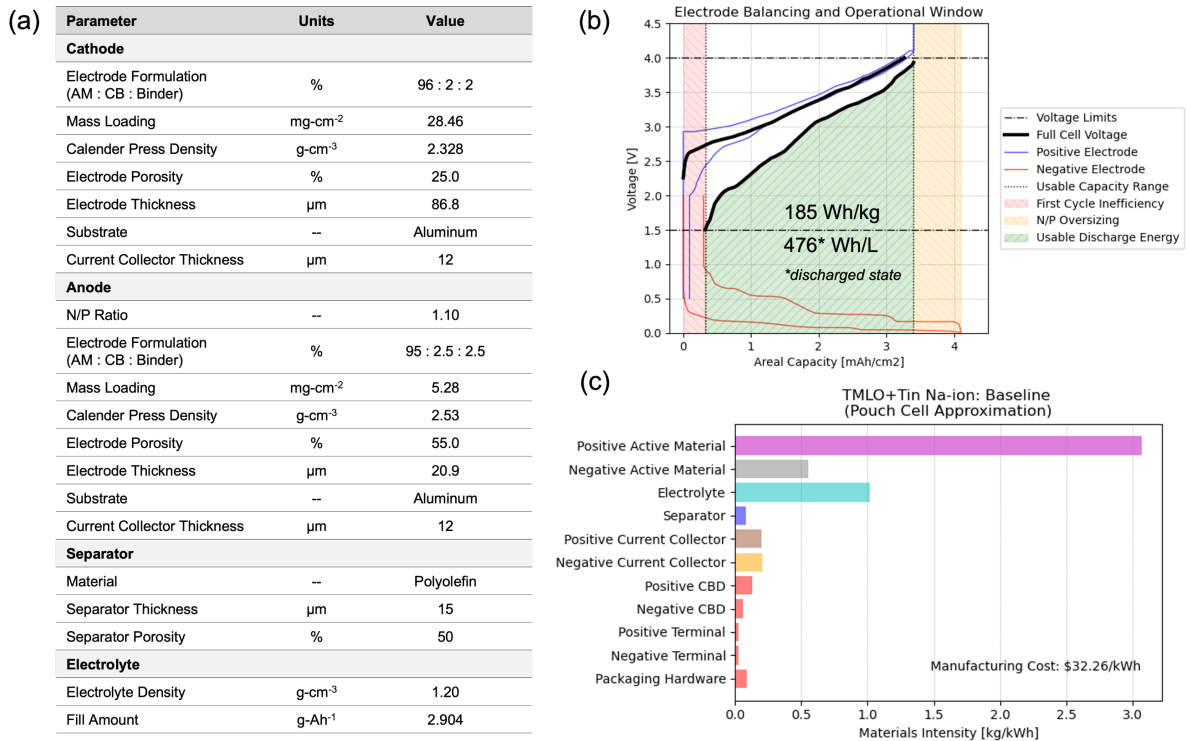

*Figure 8.20: Modeled cell design details. (a) Unit cell design parameters. (b) Electrode balancing in BatPaC-designed pouch cell representative. (c) Calculated materials intensity for BatPaC-designed pouch cell representative.*

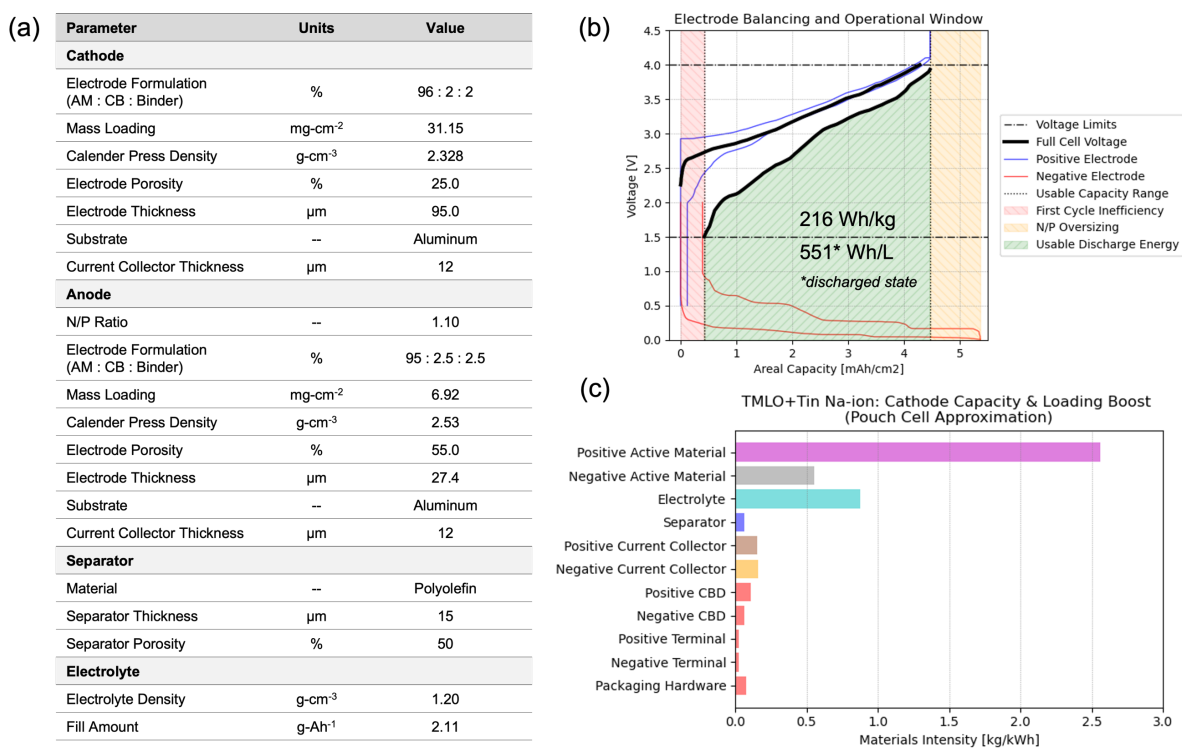

*Figure 8.21: Modeled cell design details. (a) Unit cell design parameters. (b) Electrode balancing in BatPaC-designed pouch cell representative. (c) Calculated materials intensity for BatPaC-designed pouch cell representative.*

## 8.18: Na-ion NaNM | Sn – Baseline

One potential alternative to using hard carbons as an anode material is to use micron-sized metallic tin (Sn) powder instead. Tin, like antimony (Sb) and lead (Pb) is a metal that alloys with sodium upon sodiation, but is considered the most viable candidate of the three due to its low-toxicity and relative crustal abundance[32]. As the sodium storage mechanism occurs through an alloying process versus a typical intercalation process, the specific capacity tends to be significantly higher, with a theoretical capacity of 847 mAh-g$^{-1}$ correlating with an alloy stoichiometry of $Na_{15}Sn_4$ [32]. In addition to having ultra-high density (~7.31 g-cc$^{-3}$), tin presents an interesting alternative to hard carbon capable of enabling high energy densities. However, large volumetric expansion is commonly observed with alloying reactions, and tin is known to expand by ~420% upon sodiation[32]. With repeated expansion and contraction upon cycling, the repeated formation/breakage of SEI films can often result in poor cycle life. While somewhat analogous to silicon as a candidate high-capacity anode in Li-ion batteries, tin is unlike silicon in that it is metallic and malleable, which helps mitigate some challenges associated with loss of electrical contact often experienced by silicon. Recent research[32] and commercialization efforts[18] point to the ability to form stable SEI using glyme-based electrolytes that enable substantially improved cycle lives, though their stability at higher voltages need to be improved.

Using the same principles detailed above, we model a theoretical layered oxide NaNM cathode | Sn anode baseline cell design, drawing from feedback on practical tin electrode designs from industry sources commercializing the technology[18]. Here, we assume 800 mAh-g$^{-1}$ of reversible capacity, and we increase the N/P ratio to 1.20 to minimize material pulverization. Given the high density and specific capacity of tin, the anode in such a system is no longer the limiting electrode in terms of thickness and a balanced anode can be quite thin. In fact, too thin of an electrode coating becomes challenging to manufacture with high uniformity in a slurry-based coating process, and thus a baseline NaNM | Sn design requires >3 mAh-cm$^{-2}$ capacity loading[18]. Here, we model a baseline cell design with an effective capacity loading of 3.1 mAh-

cm$^{-2}$ which correlates to single-sided anode composite thickness of 20.9 µm thickness at an assumed porosity of 55% (to accommodate material swelling[18]). The resulting cell design is summarized in Figure 8.20.

## 8.19: Na-ion NaNM | Sn – Cathode Capacity Boost

Similar to prior examples, we model an increase in NaNM specific capacity by 20% while still being limited to a 4.0V full cell system. We note that since the assumed 800 mAh-g$^{-1}$ specific capacity in the baseline example is ~95% of tin's theoretical capacity, we do not include additional increases to the anode capacity. We however do increase the electrode loading of the cathode. As the cathode is now the limiting electrode, we assume a maximum thickness of 95 µm considering practical manufacturability and liquid-phase mass transport limitations. The resulting cell design is summarized in Figure 8.21.

## 8.20: Na-ion NFPP | Sn – Baseline

Similarly, we model a Sn anode paired against a NFPP cathode as yet another strategy to increase energy densities. We first model a cell with the same NFPP cathode thickness as the NFPP | HC baseline. Note however that the resulting Sn anode is very thin due to the low areal capacity of the cathode. The resulting cell design is summarized in Figure 8.22.

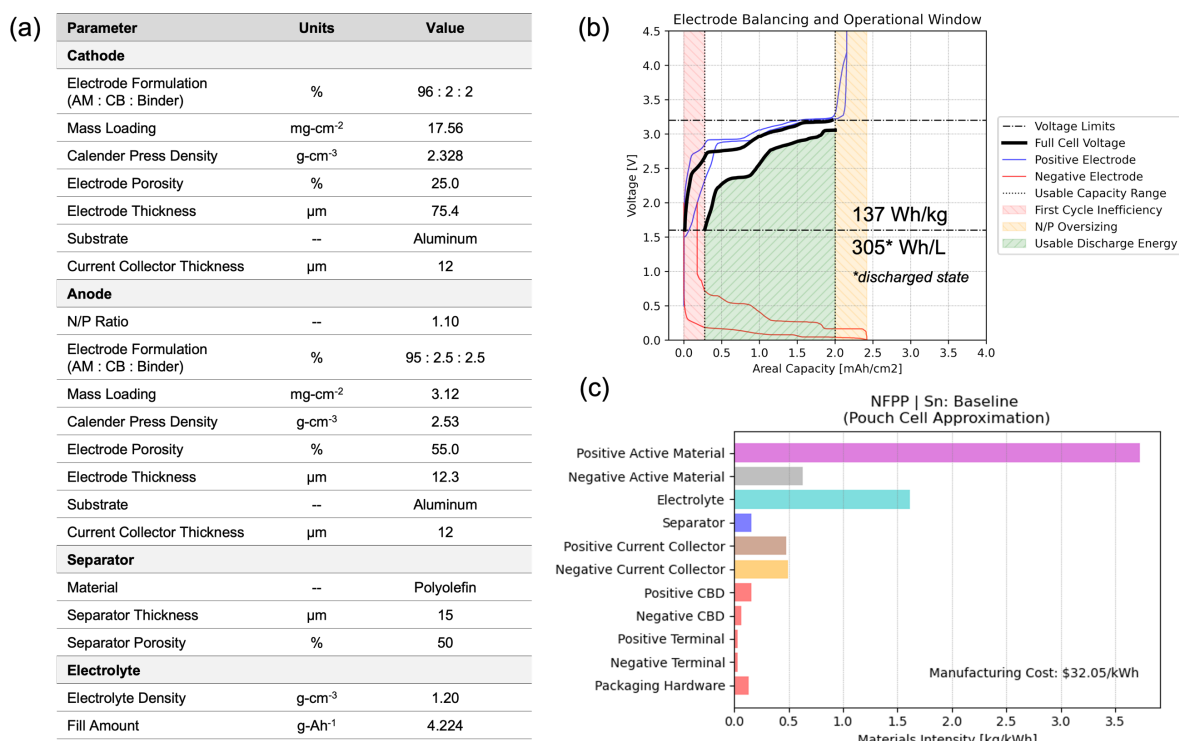

*Figure 8.22: Modeled cell design details. (a) Unit cell design parameters. (b) Electrode balancing in BatPaC-designed pouch cell representative. (c) Calculated materials intensity for BatPaC-designed pouch cell representative.*

## 8.21: Na-ion NFPP | Sn – Increased electrode thicknesses

We again assume dry-coated NFPP cathodes can enable thick cathodes of 125 µm. This increases the Sn anode back to ~20 µm. The resulting cell design is summarized in Figure 8.24.

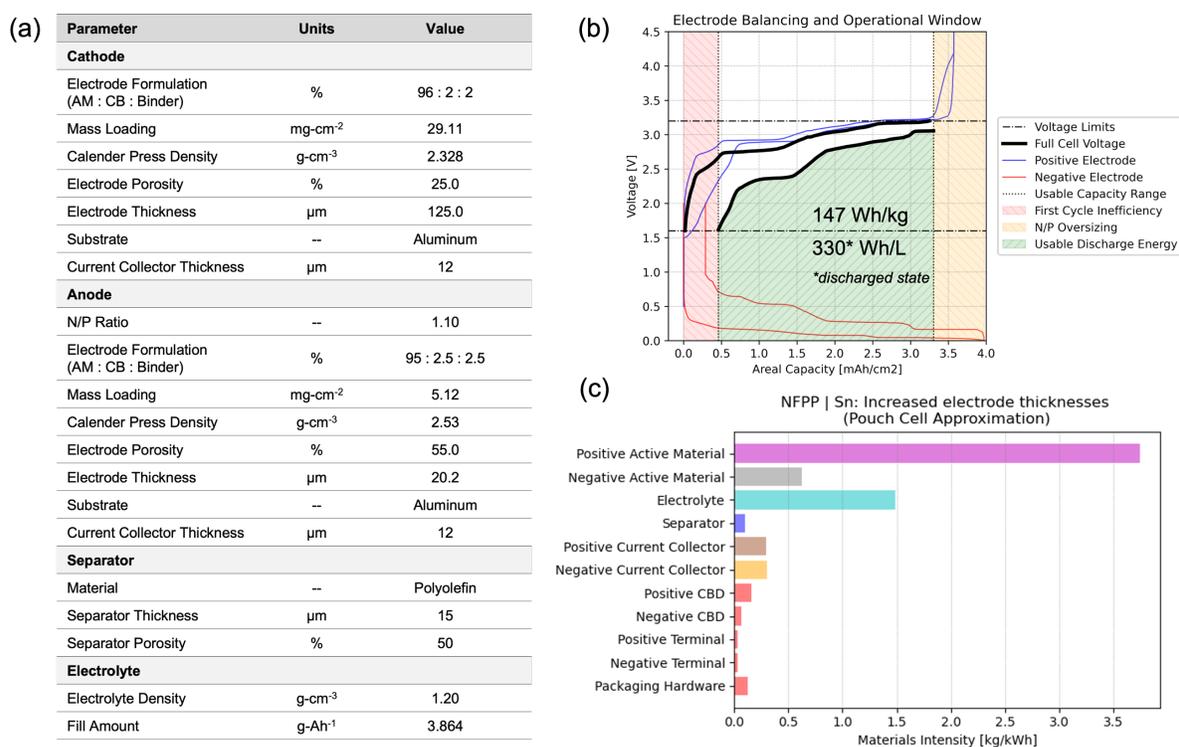

*Figure 8.23: Modeled cell design details. (a) Unit cell design parameters. (b) Electrode balancing in BatPaC-designed pouch cell representative. (c) Calculated materials intensity for BatPaC-designed pouch cell representative.*

### 8.22: Na-ion NaNM | Anode-free – Baseline

The ultimate replacement anode material is to not have an anode material at all, and instead directly electrodeposit/strip sodium metal upon a bare substrate current collector. This approach has been colloquially termed "anode-free" and has been attempted in Li-ion cell designs before. This is also notably different from a design using metallic lithium or sodium (alloy) foils, as the processing costs to handle alkali metals and the minimum attainable foil thicknesses present additional unique challenges. Therefore, in the anode-free cell configuration, the cell is assembled in the discharged state with all the working ion inventory (lithium or sodium) in the cathode active material, which are subsequently electrodeposited *in-situ* onto a bare (or treated) negative current collector upon charging and completely stripped upon discharging again. Common challenges to overcome in this configuration include mitigating excessive SEI growth on newly formed metal-electrolyte interfaces with every charge/discharge cycle, ensuring no loss of active sodium upon cycling, and suppression of dendrite growth that may eventually short-circuit through the separator. There is some preliminary evidence that electroplating sodium metal may be performed with higher Coulombic efficiency than lithium, but this is still an area of early research. What appears to be known, however, is that an anode-free Na-ion cell design will most likely require (i) a copper anode current collector due to the lower nucleation barrier overpotentials or an advanced 3D current collector to mitigate excessive swelling, (ii) an advanced electrolyte system with high cathodic stability to form stable SEI, such as a glyme-based solvent system with engineered salts and additives, and (iii) a modified separator to prevent dendrite penetration.

Assuming sodium metal can be electrodeposited as a dense film with near 0% porosity, one of the ancillary benefits of an anode-free Na-ion cell is the lack of a porous composite hard carbon anode that requires substantial electrolyte volume to saturate. Therefore, an anode-free cell design reduces the bill-of-materials significantly by eliminating the anode electrode composite and electrolyte volume but must balance those savings with the cost of advanced electrolytes, separators, and current collectors. Finally, we note that there may be additional cost savings in the manufacturing of anode-free cells if formation processes can be simplified. However, lacking a mature commercial anode-free Na-ion process today, we do not make such assumptions. The resulting cell design is summarized in Figure 8.24.

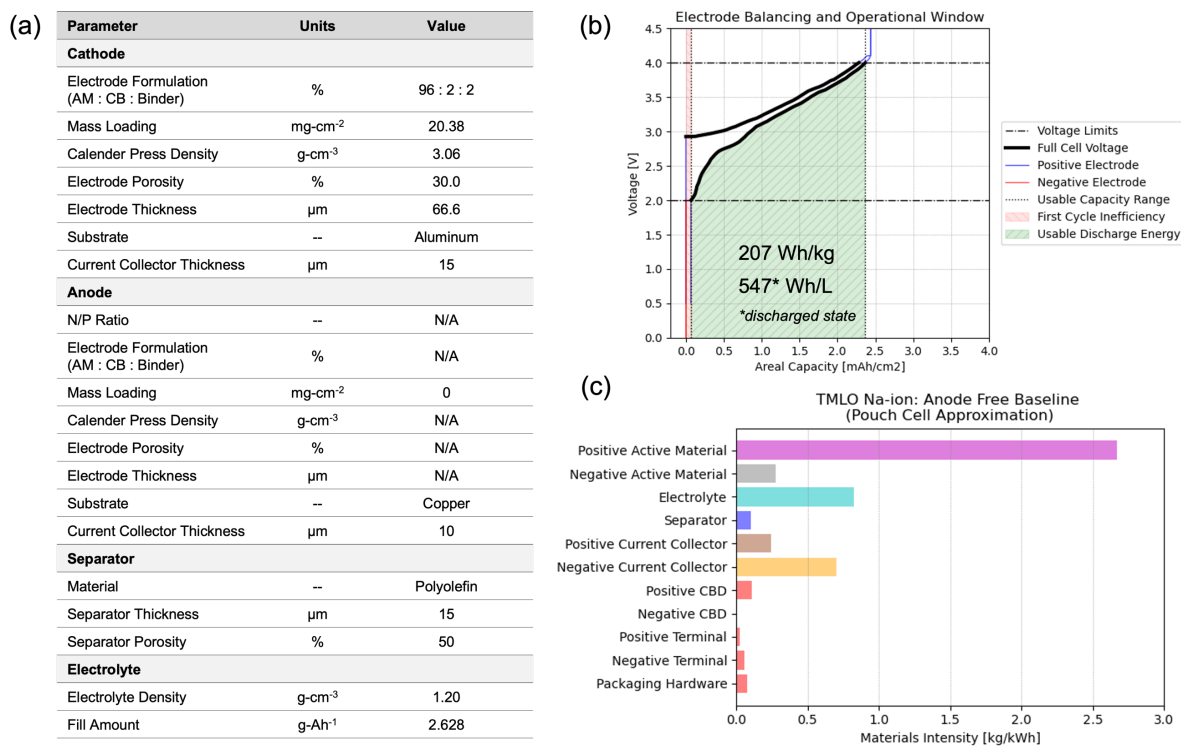

*Figure 8.24: Modeled cell design details. (a) Unit cell design parameters. (b) Electrode balancing in BatPaC-designed pouch cell representative. (c) Calculated materials intensity for BatPaC-designed pouch cell representative.*

### 8.23: Na-ion NaNM | Anode-free – Cathode Loading Boost

We further model an anode-free Na-ion cell where the cathode electrode calender density is increased along with coated mass loadings. Here, we assume the same maximum cathode electrode thickness as in Section 8.19:. The resulting cell design is summarized in Figure 8.25.

### 8.24: Na-ion NaNM | Anode-free – Cathode Capacity Boost

Finally, we model an anode-free Na-ion cell design where the cathode active material experiences a 20% increase in specific capacity while still maintaining an operating voltage range between 2.0V and 4.0V. We note that this is the highest specific energy Na-ion cell design evaluated in this entire study and would outperform the performance boundaries of LFP | graphite Li-ion cell designs. The resulting cell design is summarized in Figure 8.26.

(a)

| Parameter | Units | Value |
|---|---|---|
| **Cathode** | | |
| Electrode Formulation (AM : CB : Binder) | % | 96 : 2 : 2 |
| Mass Loading | mg-cm$^{-2}$ | 31.15 |
| Calender Press Density | g-cm$^{-3}$ | 3.28 |
| Electrode Porosity | % | 25.0 |
| Electrode Thickness | μm | 95.0 |
| Substrate | -- | Aluminum |
| Current Collector Thickness | μm | 12 |
| **Anode** | | |
| N/P Ratio | -- | N/A |
| Electrode Formulation (AM : CB : Binder) | % | N/A |
| Mass Loading | mg-cm$^{-2}$ | 0 |
| Calender Press Density | g-cm$^{-3}$ | N/A |
| Electrode Porosity | % | N/A |
| Electrode Thickness | μm | N/A |
| Substrate | -- | Copper |
| Current Collector Thickness | μm | 10 |
| **Separator** | | |
| Material | -- | Polyolefin |
| Separator Thickness | μm | 15 |
| Separator Porosity | % | 50 |
| **Electrolyte** | | |
| Electrolyte Density | g-cm$^{-3}$ | 1.20 |
| Fill Amount | g-Ah$^{-1}$ | 1.956 |

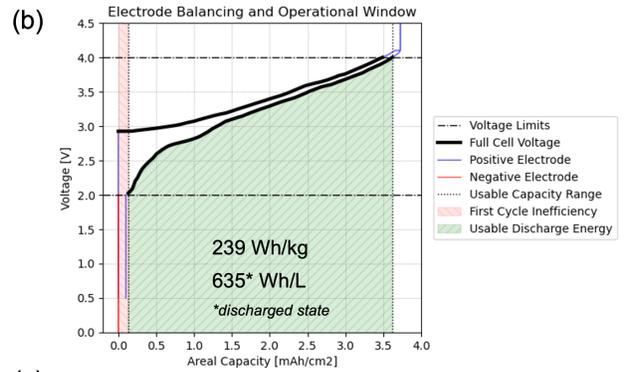
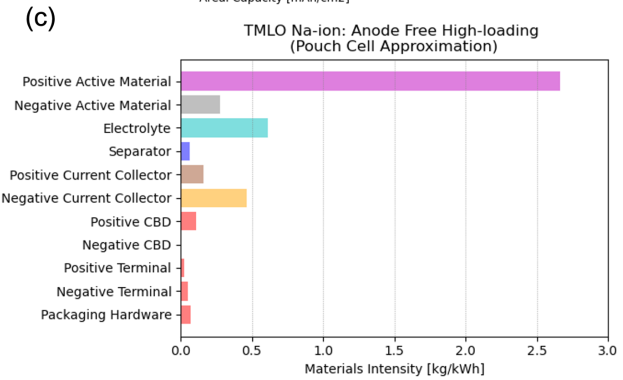

*Figure 8.25: Modeled cell design details. (a) Unit cell design parameters. (b) Electrode balancing in BatPaC-designed pouch cell representative. (c) Calculated materials intensity for BatPaC-designed pouch cell representative.*

(a)

| Parameter | Units | Value |
|---|---|---|
| **Cathode** | | |
| Electrode Formulation (AM : CB : Binder) | % | 96 : 2 : 2 |
| Mass Loading | mg-cm$^{-2}$ | 31.15 |
| Calender Press Density | g-cm$^{-3}$ | 3.28 |
| Electrode Porosity | % | 25.0 |
| Electrode Thickness | μm | 95.0 |
| Substrate | -- | Aluminum |
| Current Collector Thickness | μm | 12 |
| **Anode** | | |
| N/P Ratio | -- | N/A |
| Electrode Formulation (AM : CB : Binder) | % | N/A |
| Mass Loading | mg-cm$^{-2}$ | 0 |
| Calender Press Density | g-cm$^{-3}$ | N/A |
| Electrode Porosity | % | N/A |
| Electrode Thickness | μm | N/A |
| Substrate | -- | Copper |
| Current Collector Thickness | μm | 10 |
| **Separator** | | |
| Material | -- | Polyolefin |
| Separator Thickness | μm | 15 |
| Separator Porosity | % | 50 |
| **Electrolyte** | | |
| Electrolyte Density | g-cm$^{-3}$ | 1.20 |
| Fill Amount | g-Ah$^{-1}$ | 1.644 |

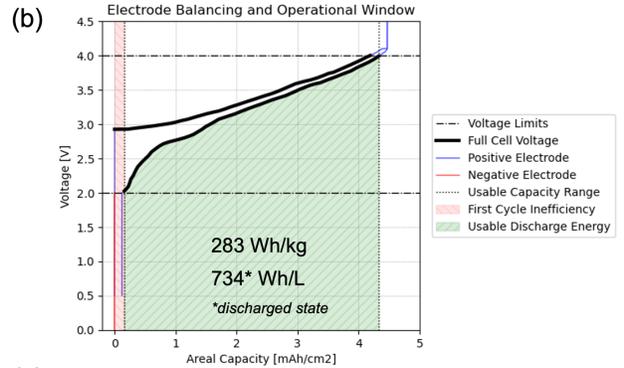
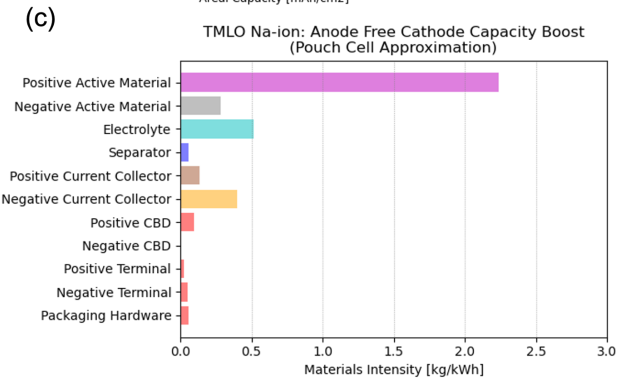

*Figure 8.26: Modeled cell design details. (a) Unit cell design parameters. (b) Electrode balancing in BatPaC-designed pouch cell representative. (c) Calculated materials intensity for BatPaC-designed pouch cell representative.*

## 8.25: Na-ion NaFM | X – Nickel-free layered oxide cathodes in various cell designs

Due to the high costs of nickel, there are efforts to reduce or fully eliminate the use of the critical metal in layered oxide cathodes by relying solely on earth-abundant metals like iron (Fe) and manganese (Mn)—as mentioned in the Main Text. Capturing these developmental roadmaps is therefore important to forecast the technoeconomic viability of future Na-ion cell designs. In Section 8.10:, we discussed how we utilized Faradion's NaNM voltage-capacity profile as representative of all layered oxides and mathematically scaled the curve to evaluate scenarios for improved specific capacity. We again note that this is an imperfect representation of actual roadmaps, as the potential use of different transition metal redox centers will likely alter the electrochemistry; once again, we caveat that this is a reasonable estimate for the sake of technoeconomic scenarios analysis. Here, we again assume that the complete substitution of nickel for an element like Fe (NaFM) can maintain our representative NaNM voltage-capacity profile, and we duplicate analysis of select roadmaps with the only difference of phasing out nickel from layered oxide cathode materials. Thus, in our NaFM | X modeled roadmap scenarios, we assume identical cell designs as their NaNM counterparts as detailed in the sections above. The only difference in the modeled roadmaps therefore lies in the minerals price floor of the cathode as nickel is gradually phased out. This nickel phase-out and minerals price floor evolution is detailed in Section 11:.

## Section 9: Modeling cell development roadmaps

The various development roadmap scenarios for Li-ion and Na-ion chemistries are detailed in the following subsections. The engineering limits for the various chemistries are tabulated in Table 9.1, which are used to calculate each of the $M_{min}$ parameters representing minimum materials intensity and $SE_{max}$ values representing maximum theoretical (engineering) limits on specific energy.

*Table 9.1: Engineering limits for development roadmaps*

|  | Units | LFP \| Graphite | NMC \| Graphite | NaN(F)M \| HC | NFPP \| HC | NaN(F)M \| Sn | NFPP \| Sn | NaN(F)M \| Anode-free |
|---|---|---|---|---|---|---|---|---|
| Nominal System Voltage | V | 3.3 | 3.7 | 3.1 | 2.8 | 3.1 | 2.8 | 3.1 |
| Cathode Max Specific Capacity | mAh-g$^{-1}$ | 160 | 274 | 200 | 135 | 200 | 135 | 200 |
| Cathode Material / Crystal Density | g-cm$^{-3}$ | 3.55 | 4.65 | 4.65 | 3.20 | 4.65 | 3.20 | 4.65 |
| Cathode Min Porosity | % | 20 | 20 | 20 | 20 | 20 | 20 | 20 |
| Cathode Max Thickness | μm | 120 | 120 | 120 | 150 | 120 | 150 | 150 |
| Anode Max Specific Capacity | mAh-g$^{-1}$ | 372 | 372 | 500 | 500 | 847 | 847 | 1166 |
| Anode Material / Crystal Density | g-cm$^{-3}$ | 2.24 | 2.24 | 1.5 | 1.5 | 7.31 | 7.31 | 0.97 |
| Anode Min Porosity | % | 25 | 25 | 35 | 35 | 45 | 45 | 0 |
| Anode Max Thickness | μm | 120 | 120 | 120 | 150 | 50 | 50 | 20 |
| Min N/P Ratio | -- | 1.02 | 1.02 | 1.02 | 1.02 | 1.02 | 1.02 | 1.00 |
| Electrolyte Density | g-cm$^{-3}$ | 1.2 | 1.2 | 1.2 | 1.2 | 1.2 | 1.2 | 1.2 |
| Positive CC Density | g-cm$^{-3}$ | 2.7 | 2.7 | 2.7 | 2.7 | 2.7 | 2.7 | 2.7 |
| Positive CC Min Thickness | μm | 8.0 | 8.0 | 8.0 | 8.0 | 8.0 | 8.0 | 8.0 |
| Negative CC Density | g-cm$^{-3}$ | 8.96 | 8.96 | 2.7 | 2.7 | 2.7 | 2.7 | 2.7 |
| Negative CC Min Thickness | μm | 5.0 | 5.0 | 8.0 | 8.0 | 8.0 | 8.0 | 8.0 |
| Separator Min Thickness | μm | 10.0 | 10.0 | 10.0 | 10.0 | 10.0 | 10.0 | 10.0 |
| Separator Max Porosity | % | 50 | 50 | 50 | 50 | 50 | 50 | 50 |
| Separator Material Density | g-cm$^{-3}$ | 1.2 | 1.2 | 1.2 | 1.2 | 1.2 | 1.2 | 1.2 |

## 9.1: Li-ion LFP Development Roadmap

Using the equations defined above, we assign the CATL prismatic LFP cell and the A123 cylindrical LFP cell to the years 2023 and 2010, respectively, approximating when they were each considered to be state-of-art. The resulting LFP development roadmap is summarized in Figure 9.1Figure 9.1, where the materials intensities of each component decreases through time and eventually plateaus at the specified engineering

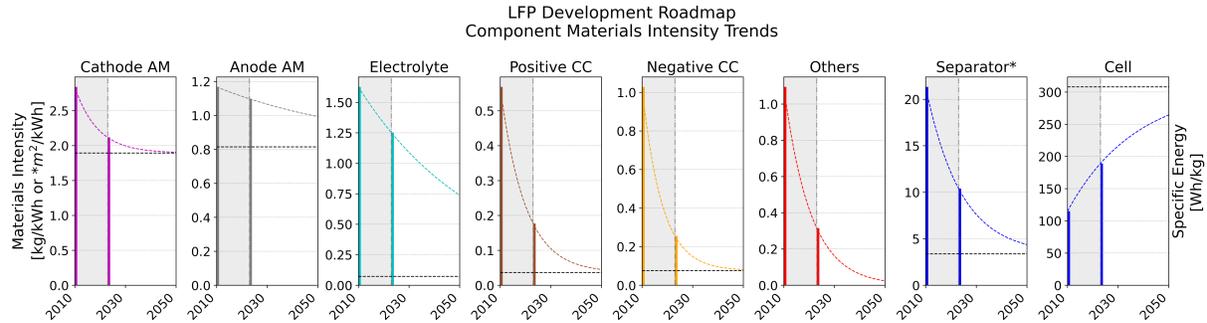

*Figure 9.1: Fitted materials intensity roadmaps showing asymptotic reduction for each component within this chemistry.*

limits.

## 9.2: Li-ion NMC Development Roadmap

For NMC, we assign the NMC PHEV pouch cell, the LGX E78 NMC721 pouch cell, and the Tesla 4680 NMC811 cylindrical cell to the years 2014, 2021, and 2025. The resulting NMC development roadmap is summarized in Figure 9.2.

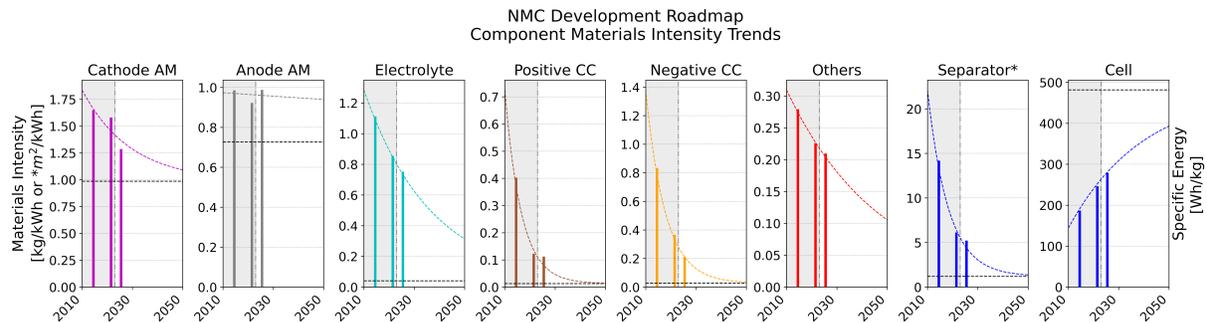

*Figure 9.2: Fitted materials intensity roadmaps showing asymptotic reduction for each component within this chemistry.*

## 9.3: Na-ion Development Roadmaps

The complete summary of technical roadmap scenarios can be found in Table 1 of the Main Text. The following figures show the materials intensity trends for each of the modeled roadmaps. All modeled roadmaps show improved cell designs are achieved by 2030.

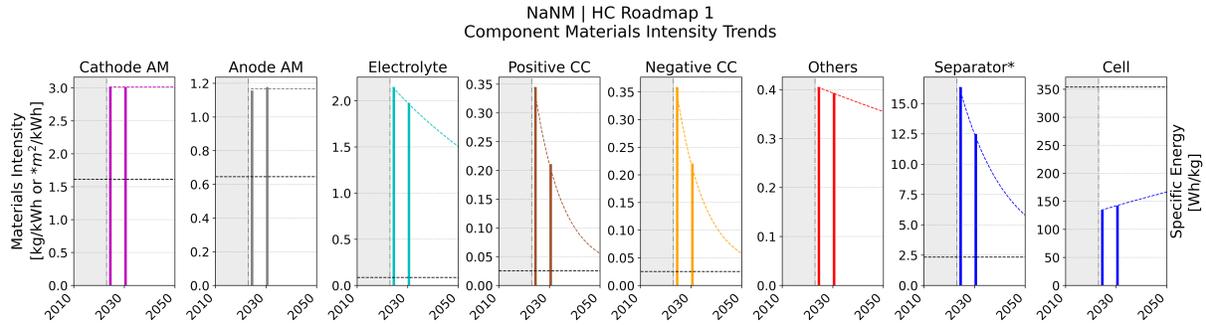

*Figure 9.3: Fitted materials intensity roadmaps showing asymptotic reduction for each component within this chemistry.*

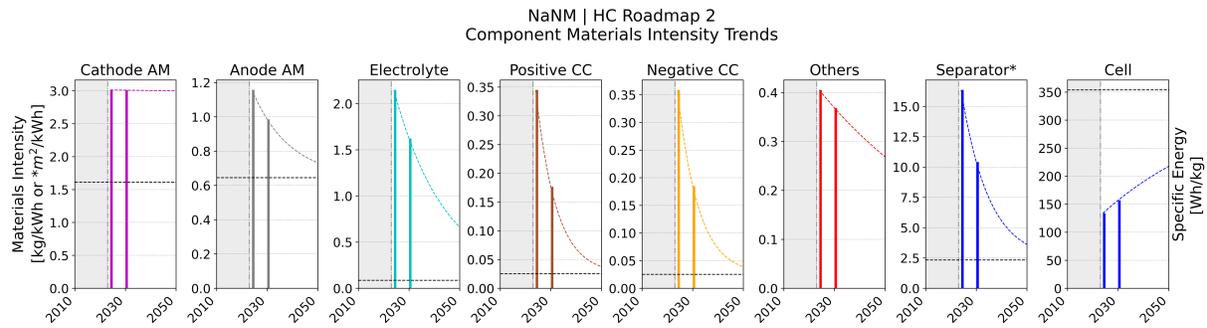
Figure 9.4: Fitted materials intensity roadmaps showing asymptotic reduction for each component within this chemistry.

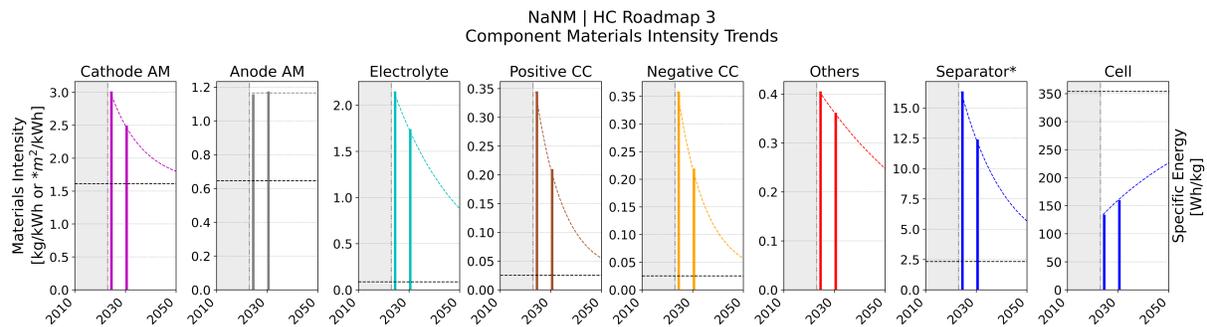
Figure 9.5: Fitted materials intensity roadmaps showing asymptotic reduction for each component within this chemistry.

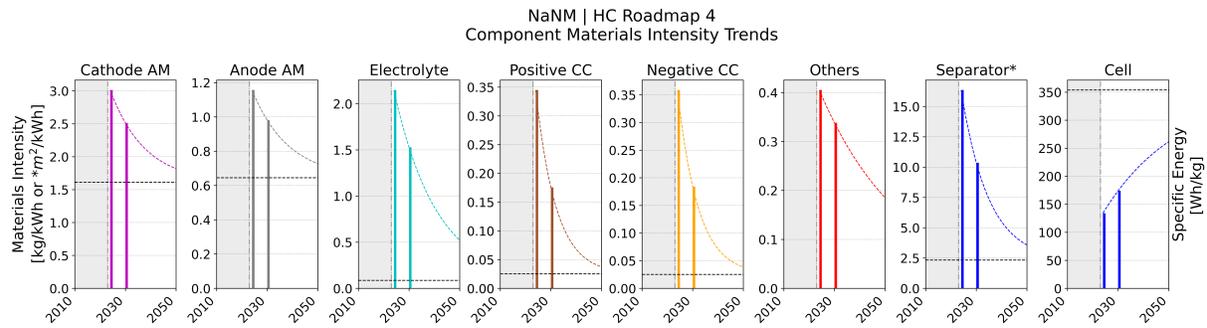
Figure 9.6: Fitted materials intensity roadmaps showing asymptotic reduction for each component within this chemistry.

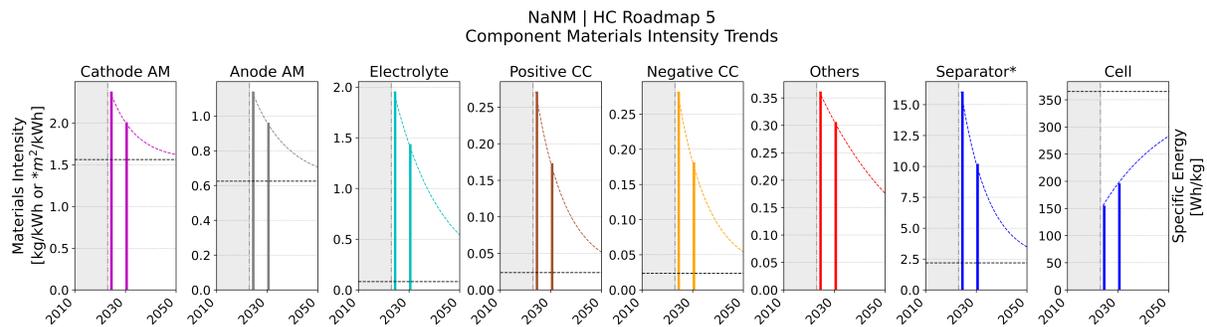
Figure 9.7: Fitted materials intensity roadmaps showing asymptotic reduction for each component within this chemistry.

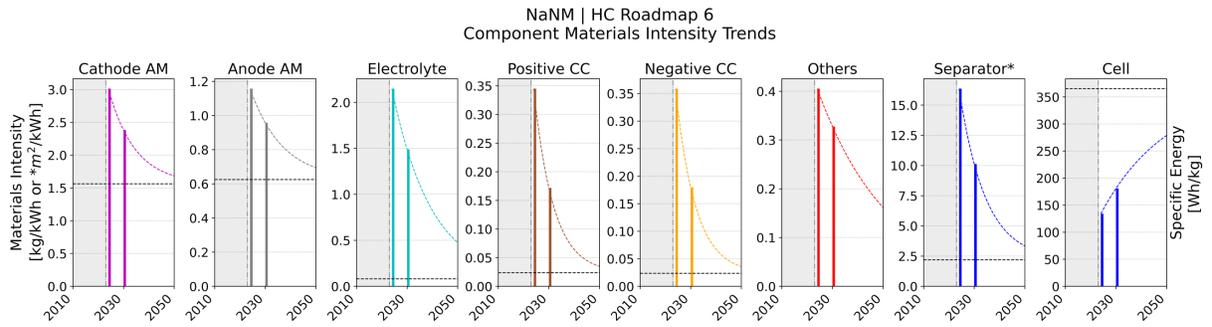

*Figure 9.8: Fitted materials intensity roadmaps showing asymptotic reduction for each component within this chemistry.*

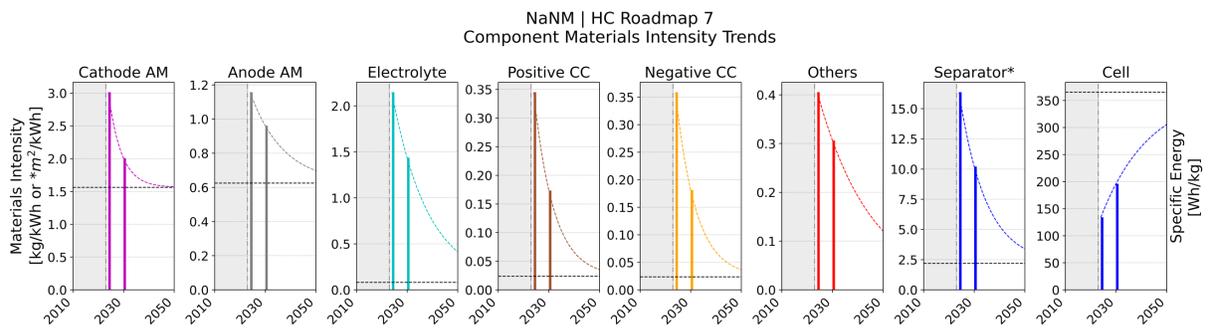

*Figure 9.9: Fitted materials intensity roadmaps showing asymptotic reduction for each component within this chemistry.*

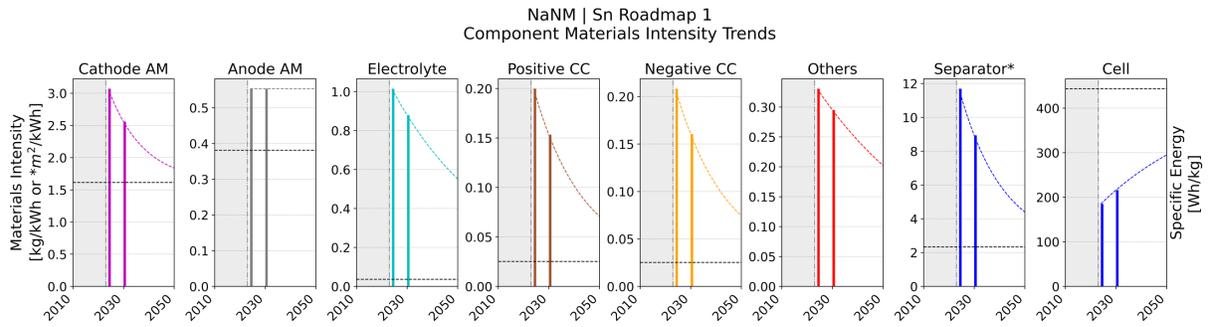

*Figure 9.10: Fitted materials intensity roadmaps showing asymptotic reduction for each component within this chemistry.*

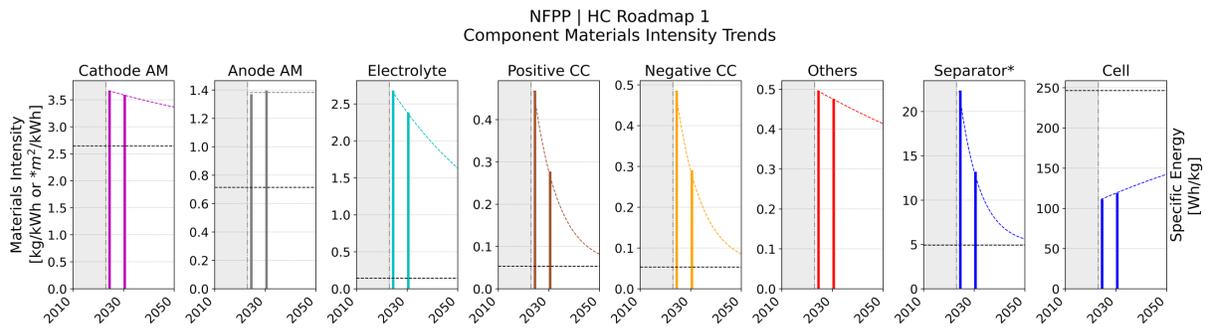

*Figure 9.11: Fitted materials intensity roadmaps showing asymptotic reduction for each component within this chemistry.*

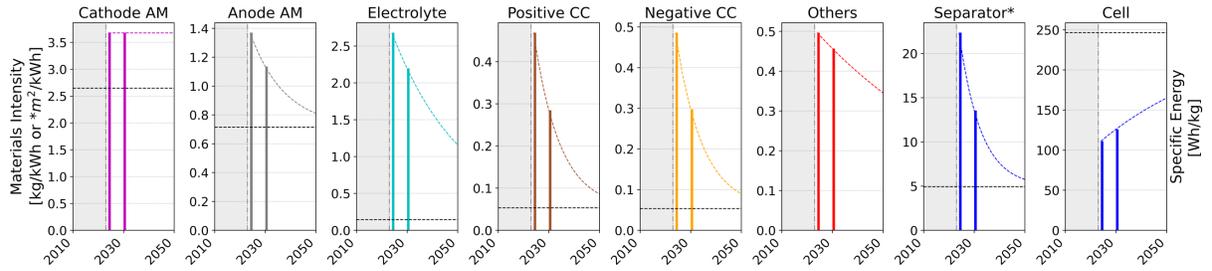

*Figure 9.12: Fitted materials intensity roadmaps showing asymptotic reduction for each component within this chemistry.*

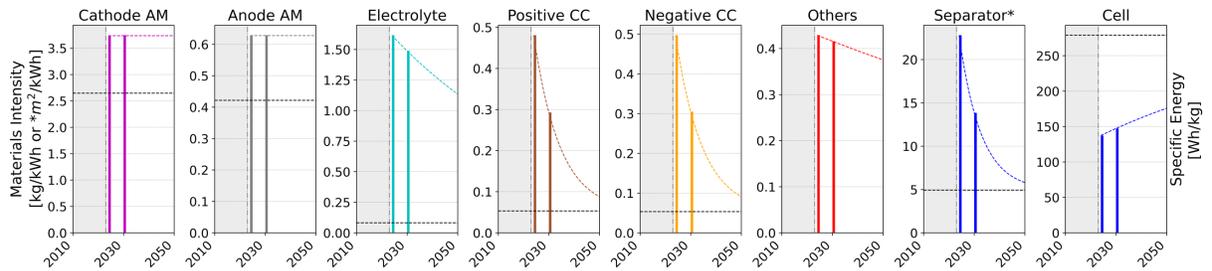

*Figure 9.13: Fitted materials intensity roadmaps showing asymptotic reduction for each component within this chemistry.*

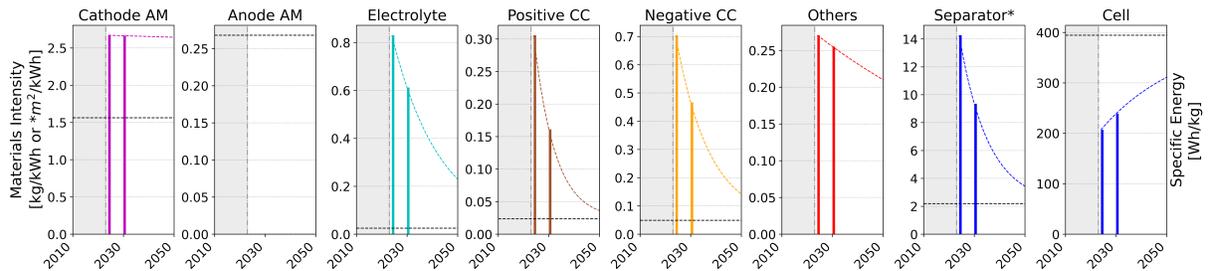

*Figure 9.14: Fitted materials intensity roadmaps showing asymptotic reduction for each component within this chemistry.*

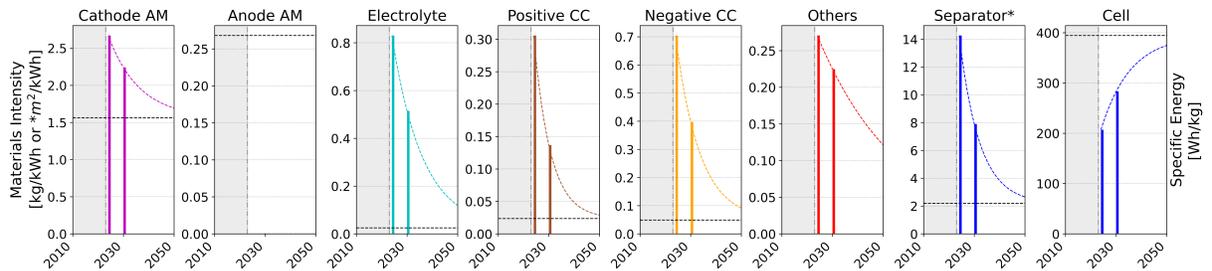

*Figure 9.15: Fitted materials intensity roadmaps showing asymptotic reduction for each component within this chemistry.*

For the roadmaps that reduce nickel content in layered oxide cathodes (NaFM), the same materials intensity roadmaps are used as their nickel-containing counterparts (NaNM). As discussed in Section 8.25:, the only difference in the modeled scenarios between nickel-containing and low-nickel roadmaps is the underlying minerals price floors.

## Section 10: NMC Baseline Price Curve Forecasts

The component curve stack-up for the NMC Li-ion baseline is shown in Figure 10.1 below showing reasonable agreement with historical component price breakdowns (bar charts[1,13]). Additionally, Figure 10.2 shows reasonable agreement with recent NMC price assessments obtained from industry data contributors.

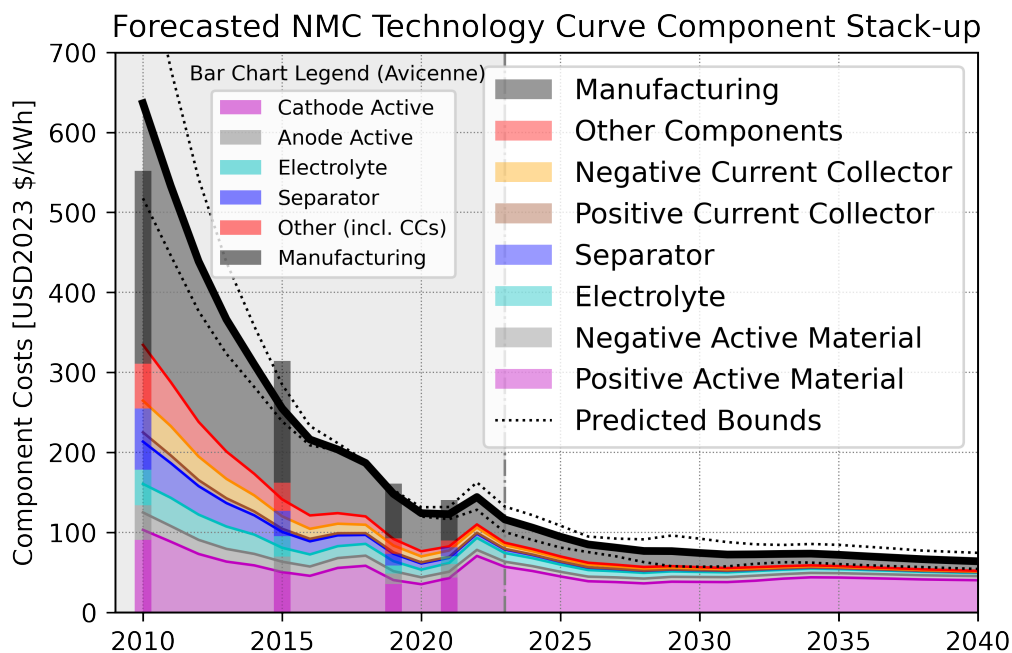

*Figure 10.1: Total price curve for NMC-type cells constructed by summing up all constituent material components based upon cell design models that inform materials intensity per energy content stored. Bar charts show cell price breakdown by component in previous years provided by Avicenne—showing good agreement in cost contributors.*

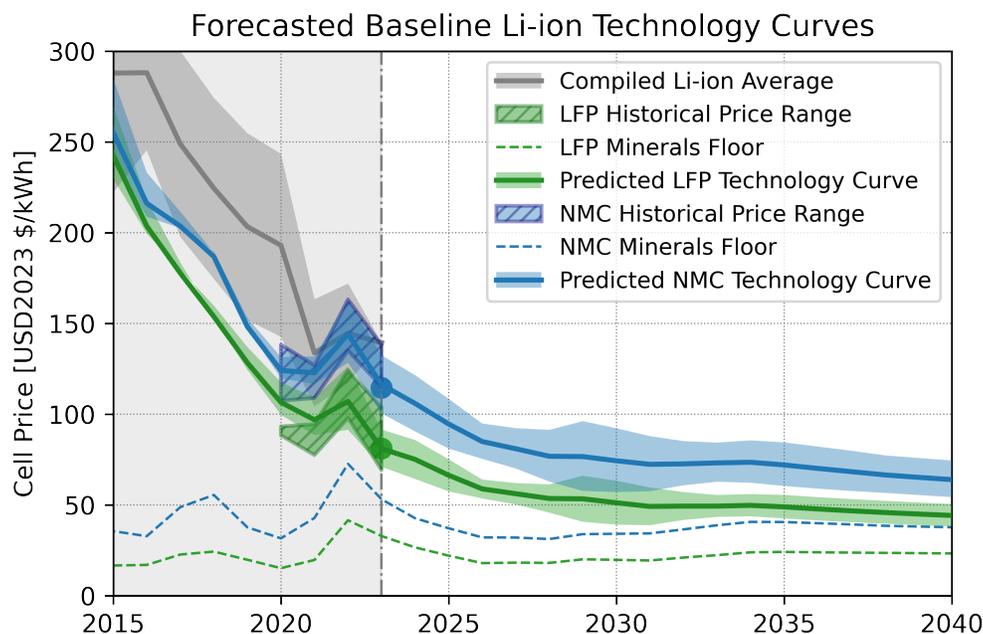

*Figure 10.2: Forecasted LFP and NMC Li-ion technology curves for the baseline scenario, showing good agreement with historical averages. This assumes the baseline minerals forecasts shown in SI Section 2.*

**Section 11: Development of nickel-free layered oxide compositions**

The first NMC Li-ion cathode stoichiometry that was commercialized was equal parts nickel, manganese, and cobalt with the formula $LiNi_{0.33}Mn_{0.33}Co_{0.33}O_2$ and abbreviated as NMC333 but more commonly NMC111, as discussed in Section 4:. Similarly, the sodium layered oxide cathode that is experiencing the most widespread initial adoption is NFM111, which refers to a stoichiometry that is equal parts nickel, iron, and manganese with the formula $NaNi_{0.33}Fe_{0.33}Mn_{0.33}O_2$ [18]. Just as NMC saw the gradual stoichiometric evolution to minimize cobalt content to reduce costs and reliance on an ethically challenged mineral (while increasing nickel content to maximize energy densities), it is anticipated that future generations of Na-ion layered oxide cathodes can gradually reduce their nickel content in a similar fashion to enable a truly low-cost chemistry free of critical minerals. To model this, we introduce a nickel-free Na-ion cathode end member, $Na_{0.67}Fe_{0.5}Mn_{0.5}O_2$ (NaFM) and gradually increase its market penetration from zero in 2024 to near complete cannibalization of its nickel-containing counterpart (NaNM) by 2040, as shown in Figure 11.1. Then, based on a market volume-weighted average of the layered oxide cathodes, we can chart the gradual evolution of nickel stoichiometry in Na-ion layered cathodes, shown in Figure 11.2. This evolution in nickel stoichiometry will manifest in a gradually reducing minerals price floor as illustrated in Figure 3 in the Main Text.

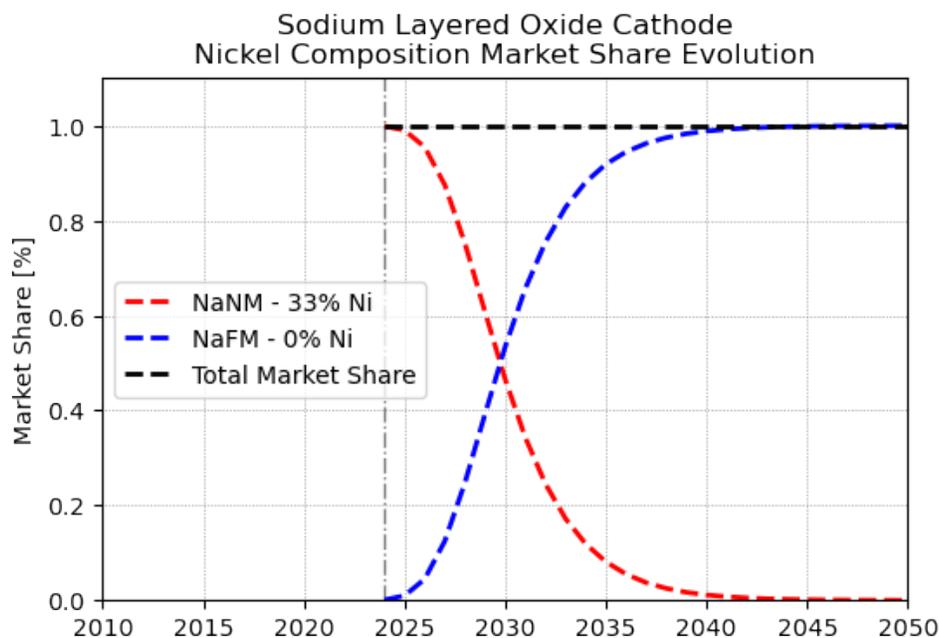

*Figure 11.1: Modeled market penetration curves for nickel-containing (33%) and nickel-free sodium layered oxide cathodes*

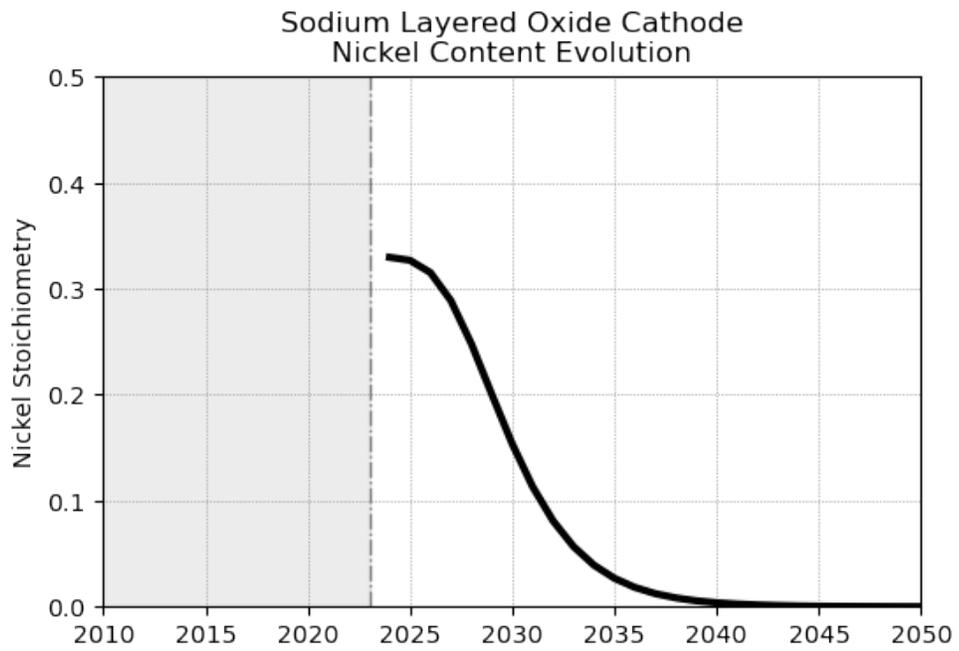

*Figure 11.2: Modeled volume-weighted nickel content evolution in sodium layered oxide cathodes..*

**Section 12: Defining "cumulative experience"**

In forecasting componential learning curves for Na-ion, there is a critical nuance in the assumptions of experience that inform future price trends. For example, Na-ion's drop-in compatibility with existing Li-ion manufacturing infrastructure is perceived as an advantage. While it may be impractical to substitute one for the other on the same production lines from a logistics, equipment uptime, yield, and contamination standpoint[18], it is generally accepted that the same manufacturing processes and learnings from Li-ion apply to Na-ion as it is not a fundamentally new process. Therefore, the cumulative experience applied to the manufacturing cost component of Na-ion batteries should scale with the cumulative production experience of *both* Li-ion and Na-ion batteries. In contrast, for new Na-ion-specific material components such as hard carbons, the cumulative experience growth rate is expected to be much greater. This is because the total deployed quantity of hard carbons to-date is substantially lower and is therefore expected to see market doublings at a quicker pace, thereby seeing a greater rate of price reduction even if learning rate values are similar. Considering transition metal layered oxide cathodes that may predominantly be manufactured using the same co-precipitation process pioneered by Li-ion layered oxides, the expected pace of cumulative experience growth for cathode synthesis may therefore be more modest. In our subsequent scenarios modeling, we vary this assumption and toggle whether this applies to the anode material only or to the cathode material as well to test the sensitivity to outcomes. Assuming Na-ion batteries see large-scale commercialization this year (2024), the normalized cumulative experience growth curves are shown in Figure 12.1 for the various Na-ion market penetration scenarios.

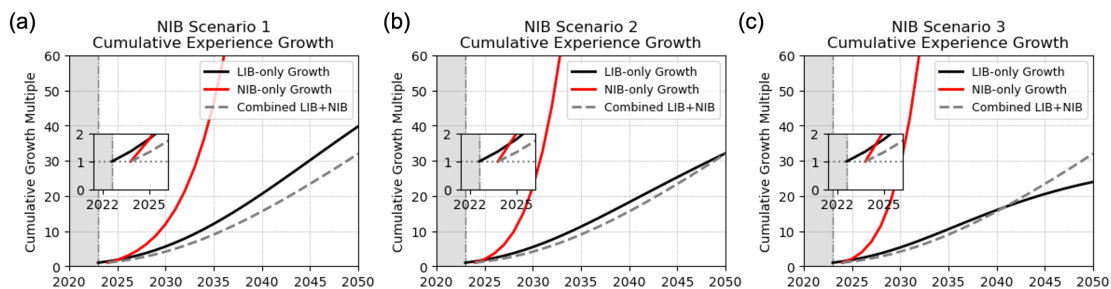

*Figure 12.1: Cumulative experience growth curves for each of the Na-ion market penetration scenarios discussed in Section 6:. Note that the cumulative experience growth of Li-ion-specific components is more modest due to the large cumulative deployed capacity to-date in comparison to an emerging technology like Na-ion.*

## Section 13: Scenarios-based Modeling Outcomes

The complete set of modeled scenarios are tabulated below in Table 13.1.

*Table 13.1: Summary of modeled scenarios*

| Parameter | Scenarios | Count |
|---|---|---|
| **Technical Scenario (Roadmap)** | • Technical Roadmaps detailed in Table 1 of Main Text | 28 |
| **Na-ion Market Penetration** | • NIB Market Penetration 1<br>• NIB Market Penetration 2<br>• NIB Market Penetration 3 | 3 |
| **Lithium Minerals Pricing** | • Lithium High<br>• Lithium Baseline (Aggregated forecasts)<br>• Lithium Low | 3 |
| **Nickel Minerals Pricing** | • Nickel High<br>• Nickel Mid<br>• Nickel Baseline (Aggregated forecasts) | 3 |
| **Graphite Supply Shock** | • No Supply Shock<br>• Supply shock in 2027 | 2 |
| **Cumulative Experience Scenarios** | • Novel growth for Anode only<br>• Novel growth for Anode & Cathode | 2 |
| **Na-ion Learning Rates** | • Normal LRs – based on Li-ion analogues<br>• Aggressive LRs for Na-ion cathode, anode, and electrolyte | 2 |
| | **Total Count of Scenario Combinations** | **6,048** |

### 13.1: Na-ion Price Parity Summary

Despite many scenarios modeled showing long timelines to Price Advantage, or even no Price Advantage condition at all (Main Text, Figure 4), it does not mean Na-ion is not competitive against LFP. Figure 13.1 shows a complementary plot to Main Text, Figure 4 by illustrating the timelines to Price Parity. As depicted via the noticeably higher density of darker colored markers which correspond to a nearer term time frame, almost all Na-ion development roadmaps lead to a price competitive scenario by the mid 2030s.

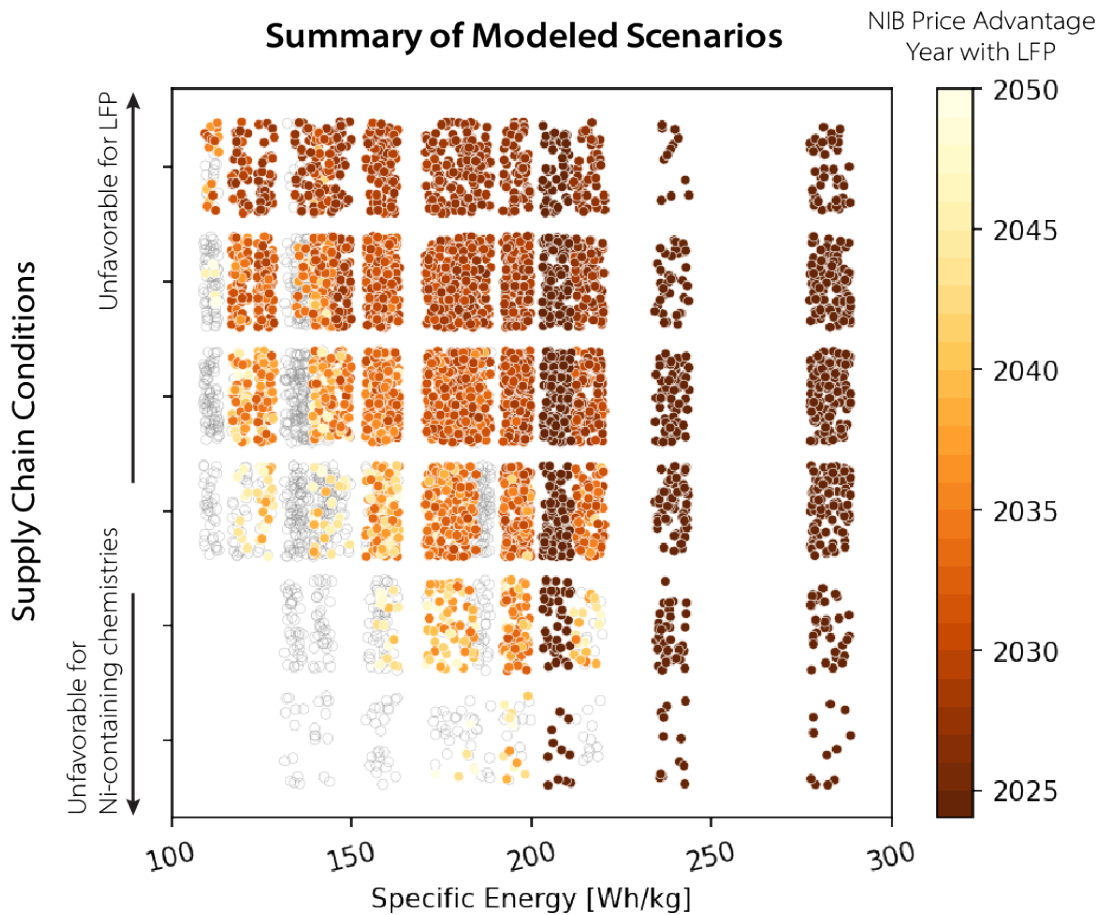

*Figure 13.1: Heatmap showing timelines to price parity. Supply chain conditions ranked categorically here via summation of integer increments/decrements of lithium, graphite, and nickel supply chain conditions according to Na-ion favorability—with increased nickel commodity pricing being unfavorable to only nickel-containing Na-ion layered oxide cathodes. For example, a high lithium price scenario (+1) coinciding with a graphite supply shock (+1) will produce a supply chain score of +2. The x-axis is ranked by the specific energy density targets achieved by 2030 in the various roadmaps.*

**Section 14: Simulating graphite supply shock**

While there is substantial supply chain concentration within China for the mining, refining, processing, and manufacturing of many Li-ion components, few are as concentrated as the graphite supply chain. 92.6% of battery-grade graphite for Li-ion anodes came from China in 2023[33], including both synthetic and natural graphite. As the dominant anode material within Li-ion by far, graphite's concentrated supply chain poses a significant risk amongst rising geopolitical tensions. On December 1, 2023, China's Ministry of Commerce and the General Administration of Customs imposed export controls on graphite in order to protect the country's "security and interests"[33]. While this move only requires additional permits and does not outright ban graphite exports, it is largely seen as a retaliatory response to the U.S. policy on restricting semiconductor technology exports to China[34], and may foreshadow actionable threats to cut off Western access to this critical mineral.

Given the recent policies and the vulnerability to this risk, we choose to model scenarios in which the graphite supply chain does in fact experience a supply shock. However, the means to model such a shock are notably different from the high fixed lithium price scenario discussed in Section 2.1: where the underlying mineral experiences sustained high prices. Here, instead of affecting the minerals pricing, a graphite supply shock would more likely affect the conversion costs associated with graphite production. We specifically refer to synthetic graphite as it is the dominant form (accounting for 70% of anodes in 2023 versus just 25% for natural graphite[33]), where the minerals derived from petroleum coke are relatively geographically widespread. As the graphitization process of carbon-rich precursors requires use of Acheson-type furnaces that can attain extremely high temperatures (>2,800ºC) for extended periods of time, the capital and operational expenditures associated with graphitization dominates the costs of conversion[18,35]. In addition to the need to address sulfur emissions ($SO_x$) from graphitization, volatile organic compounds (VOCs) from calcination steps, and considering other factors such as labor safety, it is expected for the conversion costs of graphite production ex-China to be substantially higher[18]. Like any other process, the conversion costs are expected to fall due to learning rates, but such a graphite shock would manifest as a step-change in the price of graphite followed by the same characteristic price reduction trend. This would resemble a "sawtooth" pattern in the price curve.

Of course, we cannot predict if or when such a supply shock could occur, but we perform a scenario analysis on a hypothetical supply shock in 2027 where the conversion costs of graphite ex-China are three times higher. We note that in addition to the above operational cost disadvantages, the U.S. and Europe currently lack large-scale graphite production and would need to incur substantial capital expense to stand up production. The resulting graphite price curve subject to these scenarios are illustrated in Figure 14.1

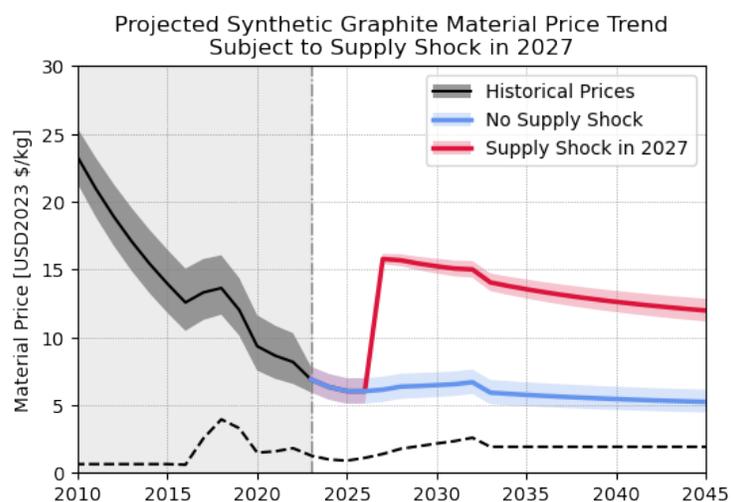

*Figure 14.1: Synthetic graphite prices with and without hypothetical supply shock scenario.*

**Section 15: BatPaC modifications**

The Battery Performance and Cost model, BatPaC v5.1 (last update August 2023)[17], developed by Argonne National Lab was used to generate cell designs and manufacturing cost overheads. To reconcile BatPaC with the unit cell electrode-level balancing discussed in 0 some modifications were made to BatPaC to ensure the cells featured the same unit cell design.

For the new chemistry couples modeled here that do not come pre-loaded in the default version of BatPaC, such as NaNM|HC, NFPP|HC, NaNM|Sn, NFPP|Sn, new columns were added in the "Chem" tab per the instructions of the BatPaC Manual[36]. Additionally, the "Open circuit voltage" section ("Chem" tab, rows 95-106) is updated to match outputs from our models. Once balanced unit cell designs were obtained from voltage curve matching, the "Positive electrode active material capacity", "Positive electrode porosity", "Positive electrode active material weight %", and "Maximum thickness limit for positive electrode" fields were overridden (per instructions of the BatPaC Manual). The specific capacity, formulation, and porosity fields of the negative electrode were similarly adjusted. Then, the "Negative-to-positive capacity ratio after formation" field was tweaked to ensure the final "Selected electrode Thicknesses" ("Battery Design" tab, rows 281-283) matched our model. In some cases where the assumed electrode thickness is higher than normal (such as for high-loading NFPP cathodes, see 0), the "Design electrodes to meet charging requirements?" toggle was turned off in the "Dashboard" tab, and in some cases, the "Optional positive electrode thickness override" on the same tab was utilized. The resulting dimensions of the cathode, anode, and separator sheets were obtained from the "Cell Dimensions" section in the "Battery Design" tab, rows 301-313, and the calculated areas were used to scale the modeled areal capacity loadings. Additional component masses, such as tabs/terminals were also obtained from the same "Battery Design" tab. We also adjust the "Electrolyte per cell" value in the same tab (row 66) to account for the realistic overfill amounts beyond the calculated volume of pores that would result in an underestimation of required electrolyte volume. Some values of the "Cost Input" tab were similarly adjusted, namely the "Direct labor rate" (row 93). The default value of $25-hr$^{-1}$ is an overestimation compared to actual labor rates in present-day manufacturing, especially considering the regions where such manufacturing is located. To adjust for regional differences, we rely on studies showing the relative regional share of global production for NMC and LFP cells[37], and obtain labor rates for those respective regions[38–41] to calculate a globally weighted average. The globally weighted average labor rate for NMC and LFP type cells is $13.41-hr$^{-1}$ and $7.35-hr$^{-1}$ in 2023, respectively. The higher average labor rate for NMC results from the higher share of manufacturing based in Europe, Korea, Japan, and the U.S., whereas LFP is almost exclusively manufactured in China. For all our Na-ion modeling, we assume a labor rate equivalent to that of LFP.

Apart from labor, all manufacturing costs calculated as a function of capital expenditures (depreciation, overhead, SG&A, R&D, and financing) were further discounted based upon the significantly lower cost of capital equipment in China. As the majority of Li-ion and Na-ion cells are currently produced in China, modeling Chinese prices would be most representative of present-day actuals. Based on reporting in early 2023, the average capital cost of a North American / Western gigafactory was ~$105M/GWh capacity, whereas plants in China ranged from $55M/GWh capacity on the low end to $72M/GWh capacity on the high end[42]. Based on industry feedback, the low end capex figures correspond to LFP plants whereas the high end figures correspond to NMC plants[18]. Thus, for polyanionic cathode chemistries that we model (LFP and NFPP), we apply a discounting factor of 0.52 to the BatPaC calculated capex figures (which all correspond to >$100M/GWh, as expected), and for layered oxide chemistries (NMC, NaNM, NaFM), we apply a discounting factor of 0.685.

Finally, the cell capacity is obtained from the "Summary of Results" tab (row 100), and matched with our model. In general, there is very good agreement between our voltage-based electrode balancing approach and the outputs of BatPaC. The only exception is the calculated "Cell total energy, kWh" ("Summary of Results" tab, row 101).

We found that while the total energy calculations from BatPaC accurately model the energy within Li-ion chemistries, the characteristically slopey voltage profiles of some Na-ion chemistries cause this value to deviate. Therefore, we use our integrated area underneath the discharge voltage curve from our analysis as the basis for energy, which can deviate by up to 3% from BatPaC values.

Finally, although there are inputs for materials costs in the "Chem" tab, we do not utilize this feature as we rely on our dynamic materials pricing informed by our componential learning curves. We only extract the manufacturing costs (discounted, as discussed above) associated with cells manufactured in the present-day to inform our manufacturing overhead cost component, which subsequently experiences its own unconstrained learning rate (see Methods section in Main Text).

## Section 16: Industry feedback on pre-print via open portal

A pre-print of this manuscript was [published on arXiv](#) on March 20, 2024. Thereafter, the authors solicited feedback from industry via an open portal on a publicly accessible web page ([steer.stanford.edu](#)). The written comments/feedback in the following sub-sections are reprinted with the permission of each of the formal respondents, either with or without direct attribution by name or organization. Additional feedback received without express permission to reprint is not published.

### 16.1: Colin Wessells, CEO & Founder, Natron Energy

1. There is a big difference between rated/nameplate $/kWh and useful $/kWh for different cell chemistries. NNM vs. HC cells operate over a wide voltage range because the potential profiles of each of the NNM and HC electrodes are steep, as shown in SI section 7.7 [original pre-print] onwards. The result is that to access more than about 80-85% of total cell energy, the cell voltage must fall below ½ of the maximum charge voltage. This is the lower limit of what most off-the-shelf power conversion systems (inverters, UPS, etc.) can handle. It is generally noneconomic to extend to Vmin < ½Vmax because doing so requires much more copper to carry the higher current associated with maintaining a constant power at lower voltage. Therefore, the fraction of useful energy for a NNM vs. HC cell will be lower than the fraction of useful energy for either any Li-ion chemistry containing a graphite anode or a PB vs. HC sodium cell. I suggest that the authors consider a metric for "useful "cell energy and $/kWh for all economic comparisons since the nameplate energies are not apples-to-apples for real world applications. This would also be relevant for the learning curves, e.g. the extent to which additional NNM capacity is achieved by cycling over a wider potential range may negatively impact the effect of the higher specific capacity on useful cell energy.

2. Na-ion learning rates: these are likely to be different from what was observed historically for Li-ion for different reasons than discussed in the manuscript. For its first big markets (consumer and EV), Li-ion did not have any strong incumbent competitors. Learning proceeded as fast as tech innovations could be converted into new production capacity across the supply chain. In contrast, Na-ion will have to compete with extremely strong Li-ion incumbent products. Therefore project development for Na-ion production capacity would only be financed if the resulting products are fairly mature and competitive with Li-ion (unless policy decisions significantly influence project financing). Therefore Na-ion will "hit the market" at a higher level of maturity than Li-ion originally did both in terms of core tech, cell design, and manufacturing processes. Learning rates will be modest compared to what we have seen for Li-ion. The implication of this is that if the core tech and cell design are limiting Na-ion competitiveness, then we would not see projects financed until after more learning occurs in those areas, and then we would start to see Na-ion giga-factories and materials plants. So the market would see a lag in product performance and cost and the subsequent learning curve, in comparison to a tech/cell design learning curve at lab/pilot scale that occurs sooner.

3. The modeling key assumptions deserve further clarification. For example, does drop-in compatibility mean that existing Li-ion plants would be converted over to Na-ion production if the resulting product were competitive in consideration of that point on the learning curve for each tech, plus the materials costs/supply chain status at that time? Or, does it mean that new production capacity would be built greenfield using the same tools as a contemporary new build of a Li-ion plant? This matters because if an existing plant is repurposed then the Na-ion products will be made with previous-generation manufacturing tools that will have a higher cost structure. To counteract that effect, an even more aggressive Na-ion tech/cell design learning curve would be needed.

4. The entire analysis makes an implicit assumption that Na-ion and Li-ion cells are drop in replacements for one another at the product/system level. Is this true? For example, will there really be the same balance of systems for protective circuitry and BMS? Or is one tech likely advantaged over the other? It is worth stating this assumption.

**16.2: Boston Consulting Group (BCG)**
1. We appreciate that there is some tempering of excitement on sodium ion. This paper delivers that, and potentially steers the community to more useful directions in sodium R&D.
2. Love the clarity of this takeaway: "Here, we highlight the importance of using such a techno-economic toolkit to *drive* research decision making as opposed to applying it *post facto*."
3. Pg2 [original pre-print], Figure 1: LCO = Lithium cobalt oxide, <u>not</u> carbonate
4. Pg8 [original pre-print]: Consider replacing "more inexpensive" with simpler language of "lower cost"?
5. Figure 4 [original pre-print] axes are challenging to interpret. Consider adding Roadmap number to the x-axis, at minimum.
6. Figure 4 [original pre-print]: How does this work ensure that the supply scenarios across elements are treated equivalently? E.g. is a moderately elevated nickel price scenario precisely equivalent to a graphite supply shock? This also will drive a more nuanced discussion of the sensitivity of each of these factors, such as whether an xx% elevation in Ni cost or xx% elevation in Li costs is more significant to NIB price advantage timeline? This will be particularly insightful in informing which aspects of the supply chain are particularly influential.
7. Following Figure 4 and 5 [original pre-print], consider summarizing the performance minimums that need to be achieved to reach price advantage. E.g. agnostic of cell design, without meeting certain cell level KPIs of energy density XXX, there is no path to price advantage.
8. SI section 6.1 [original pre-print] has Nickel + Iron-based Li-ion market share = 100% through 2050. Has there been appropriate consideration of emerging LIB chemistries (LMO/other Mn-rich compounds, LiS)?
9. Cell designs: Electrolyte loading (figures 7.xx and 8.2in SI [original pre-print]) for baseline designs seem very high for both lithium ion and sodium ion. Main text sweeps electrolyte usage down to small numbers (0.2 kg/kWh) – Fig 2a, so it probably covers the range of developments, but starting point is a bit too high here.
10. Consider a note about how the paper aggregates benchmark information across China and non-China markets, knowing that there are markedly different cost bases today.
11. Consider some mention about how recycling of LIB materials can put further downward pressure on price and mitigate supply risk.
12. Pack-level advantages, if any, for sodium ion (better low temp performance, fast charge) have potential to bridge overall system gap with LFP (just like how LFP gap was bridged with NMC). We recognize pack-level is outside the scope of this study. But we ask you to consider making such a statement in final thoughts.

**16.3: Andrew Wang**

Discussion points:

- Recommendation for Na-ion is that Govt. intervention is best applied at R&D level, is this because the inverse (subsidize demand/production) applies for Li-ion?

- General discussion on whether the IRA/BIL Govt. intervention is enough or is it missing a zero - where does it compare with what China has sunk into the industry to get to this point?

Preprint comments:

- "True nominal" voltage in the cell model appears to be calculated from integrating the half-cell balanced voltage curves, does it deviate much from 3.1V nominal, there appears to be industry signals that 3.3-3.6V nominal NNM+HC systems are being commercialized - unsure how much +0.2V is likely to alter the study outcome though.

- SI Page 33 Section 7.15 [original pre-print], multiple sections refer to Figure 7.18 [original pre-print], in the text instead of separate individual figures (7.15-7.19 etc), consider adding cell scenario name more clear in the figure caption/figure title.

**16.4: Spencer Gore, Founder & CEO, Bedrock Materials**

1. **We believe the study needs to add a reduced nickel or nickel free layered transition metal oxide to the technology roadmap.**
   In particular, iron, copper, and manganese dominant layered oxides, when engineered with appropriate countermeasures, are able to offer energy density similar to NNM without the drawback of nickel costs. Some formulations that ought to be modeled are: $NaNi_xCu_yFe_zMn_aO_2$ (HiNa, Ni<0.3) and maybe $Na_{0.67}Fe_{0.5}Mn_{0.5}$ (P2 type) as a limiting case.

   $Na_{0.9}[Cu_{0.22}Fe_{0.30}Mn_{0.48}]O_2$ is an interesting example formulation.

   Other common formulations include [external reference]

   When we look at bookended formulations $Na_{0.67}(Fe_{0.5}Mn_{0.5})O_2$, we model cathode bill of materials costs as low as $0.25/kg and processed costs as low as $2.36/kg or ~$7/kWh.

   We think this is probably the most impactful change to the forward looking sodium-ion cost roadmap.

2. **We encourage the author to revisit electrolyte cost assumptions.**
   Our market intelligence shows spot prices for LIB electrolytes at ~$9.60/kg. The study estimates SIB electrolyte costs of $11/kg. This doesn't make much sense, given that SIB electrolyte omits $LiPF_6$, the most expensive component. Golden Dragon Capital's report estimates the cost of SIB electrolyte as 25% that of LIB electrolyte, so our modeling accounts for $2.40/kg.

   While it's possible spot prices today are high, that may just be because $NaPF_6$ production has not reached scale. We would be surprised if prices did not fall quickly as production volumes increased

3. **We believe the study should take biomass hard carbon prices from cheapest/best-in-class sources (like Zichen or Chengdu) rather than incorporate previous generation material costs into price averages**

   In general, our industry intelligence suggests hard carbon prices of $7-8/kg for cornstarch and potato starch derived hard carbon. The study currently accounts for $10/kg

**16.5: Tyler Evans, Nicholas Singstock, Heather A.S. Platt, Mana Battery**
1. General comments: This is a sober look into the competitiveness of sodium battery cell technology. With the correct development strategies, Na-ion will be a strong player in the future energy storage landscape. The technoeconomic analysis highlights some of the key limitations of current Na-ion technology, including those that are less discussed – the voltage curves of today's hard carbon materials and their limiting effects on energy density and $/kWh, the outsized impact of electrolyte price when electrode material costs are low (electrolyte comprises 16-20% of Na-ion cell $/kWh), and more.
2. It is noted that this analysis does not consider pack level implications of sodium-ion cells. A number of the benefits of sodium technology may be considered at the pack level, though they remain to be proven in scaled demonstrations for most sodium battery designs. Potential benefits include safety (particularly with NFPP and other polyanion chemistries), 0 V discharge/storage/shipping, and temperature range/ stability. If proven, these value propositions of sodium battery technology would have significant cost implications. The scope of this study should be specified, and research to enable improved pack designs should be encouraged. For example, non-flammable or self-extinguishing cell designs/ electrolytes, or cell designs/electrolytes which do not result in dangerous levels of hazardous gas, can allow for more cost competitive pack designs.
3. It feels unclear if the same assumptions/rationalization in the augmented learning curve model are applied across each chemistry, and if this methodology is fair in the generated materials intensity curves. While the trajectory for NMC cells is laid out in Supplementary Section 4 [original pre-print] and can be traced back to the 2000s (with research on increasing Ni content and addition of silicon into anodes), what prospective advancements substantiate the validity of the model moving forward and the future materials intensity curves in Supplementary Section 8 [original pre-print] (LMFP, Si dominant anodes, etc.)? For sodium batteries, the future learning curve may include the advancements described in Table 1 [original pre-print], low-cost electrolytes, implementation of higher capacity cathodes, higher capacity hard carbons, and more. What would the same table look like for LFP? Graphite/ silicon anodes? NMC? It would be interesting to have some more discussion on what future technologies substantiate the forecasted advancements of the learning curve for each material class. Some set of technologies in R&D today will be the commercial technologies in 10+ years, such that we should have some insight into the validity of these learning curves.
4. The conservative case for Na-ion market growth of "100% penetration into ESS by 2035" feels aggressive, when cost parity (Figure 5 [original pre-print] ) may not occur until after 2035. Should the conservative case include EV cells, given the capacity announcements in China for Na-ion EV cells?

**16.6: Katherine He, Investment Associate, TDK Ventures**

Thanks for sharing the latest research and modeling of Na-ion commercialization. I love the approach you guys are taking, and totally agree that the fundamental techno-economic analysis is the key to understand the timeline of cost/performance advantage and large scale adoption of Na-ion.

Happy that you are using some of the more latest development like V-free NFPP as the modeling base, and I love the visualized chart and graphs to show the modeling results. However, I do want to share that in our trip to China visiting several key Na-ion battery manufacturers, the data we collected are showing even more exciting results compared with your modeling results, that some Na-ion cells are achieving price parity than LFP very soon this year. The reasons are various, mostly based on the consideration of application (cycle life) and manufacturing/supply chain integration.

Eg. Some of the NFPP we saw are already>110Wh/kg and >200Wh/L, and if you consider cycle life, the very stable NFPP type cells could achieve >10,000 cycles and much lower LCOS compare with LFP. Note that this is before considering all the continuous improvement, for example, we see company going to implement dry electrode coating to make much thicker and denser electrode and achieving 40% less manufacturing cost. For some companies that also integrated their own hard carbon manufacturing that achieved $3-4/kg of HC, the cost model they showed us will be better than LFP in the second half of 2024.

For the discussion of Sn anode, I agree it could boost the energy density, but it is like Si, I used to work on Si anode a lot, the specific capacity is great, but the cycle life is challenging, and it really limits the application.

So, overall, I would suggest the modeling to also consider the end application and performance not only on the energy density, also the adoption like dry electrode that intrinsically solved many challenges of Na-ion and boost ED and lowering the manufacturing cost.

## 16.7: Kara Rodby, Technical Principal, Volta Energy Technologies

I found the paper to be very interesting, nice work and thank you for sharing. I honestly have very few comments.

My only small point:
- I was a bit confused reading Table 1 [original pre-print]. In particular, confused why lines 4 and 5 have the same texted under "Improved Cell Design." I would've thought line 4 should have been solely "+@4.2V" to show how that affects things alone and then build in the other additive benefits in line 5 as written. Perhaps I'm misunderstanding.

My big points are just around potential added context I think would be interesting to at least reference, as it adds to the picture:
- More context on how LIBs evolved (i.e., via high-value, performance-seeking markets of consumer electronics), and how this compares to the prospect for NIB development if the markets of applicability (stationary, EV) are extremely slow-changing and margin-squeezed markets.
- Near the end you say investment in NIB development is needed. You also acknowledge some opportunity for government support with R&D. But I think there's an opportunity to 1) acknowledge that a lot of this innovation, particularly in the West, is being led by venture and 2) therefore, how do venture fund stipulations (e.g., exit-ability - which often relies on revenue, which relies on cost parity at least - needs to be viable in a fund lifetime, i.e. ~7 years) interact with these relatively far-out timelines for NIB cost-parity and then what is needed to help technological advancement in the meantime.

**16.8: Charlie Parker, Ratel Consulting**

Thank you for the opportunity to review this analysis and write-up, I learned a great deal and particularly liked the broken-out learning rates by major components. My comments reflect my background (more market, less technical), recent interactions with battery technology clients, and OEM procurement considerations.

**Application Considerations For Modeling:** Mobility and more specifically passenger vehicle applications will be the dominant market in energy storage moving forward. Because learning rate is based on deployment (i.e. sales) it is warranted to understand OEM procurement considerations. While the cell level $/kWh is the key metric throughout the analysis and one an OEM considers, OEMs generally look at unit level pack costs that meet minimum specifications. A scenario can occur where a cell is far less expensive on a $/kWh basis but incur a heavy pack cost penalty to operate safely and ultimately be more expensive on a $/pack basis.

Adding this consideration can become unwieldy and bloat the scope of the analysis quickly if not done in an efficient manner. This may be a good module to standardize at a later date.

**Diminishing Returns On Density Ceilings For Active Materials:** The frequently cited [Determinants of lithium-ion battery technology cost decline](#) shows how great of an impact active material costs have played in past cost reductions. As certain cell chemistries approach their maximum theoretical energy density it is likely to become more difficult and costly to develop active material innovations that will contribute to further cost declines.

It may make sense to consider where each chemistry is at in relation to its maximum theoretical energy density as a broad measure of potential innovation handicaps.

**Consideration For Developing Chemistries Not Analyzed:** There are other promising low-cost chemistries (LMFP, LXMO, etc.) that may impact Na-ion market uptake.

I don't have a specific recommendation other than to possibly note chemistries that have reached a certain TRL/CRL which may factor into future low-cost analysis.

### 16.9: Mrigi Munjal, MIT

**Pros/Standouts:**
1. The paper was a **terrific read** and I loved the breadth of analyses! All the figures do an excellent job of conveying the information to the audience!!
2. It's very exciting to see learning curves comprehensively taking into account mineral price floors and theoretical capacity limits and engineering limits.
3. It's useful to see conclusions like higher learning rates being correlated with processing complexity.
4. Quite interesting to see the reasoning behind componential learning rates being lower than overall LIB learning rates. It underscores that batteries are a game of innovation at the lab scale and not small optimizations during manufacturing.
5. Putting a number of years value (~7.4 years) for Price Parity is very helpful for businesses working downstream from cells in battery packs for stationary storage or otherwise that need to plan for which Li-ion-like chemistries could their systems leverage in the short and long term.
6. The discussion on what each chemistry specifically lags behind on is super useful to see! It seems somewhat distinct from the rest of the analyses but it's very insightful to read.

**Suggestions/Comments:**
1. Cycle life for sodium-ion batteries is a very important characteristic. Especially because for applications like grid-scale storage, cost effective solutions with high cycle life are greatly preferred even if they lag somewhat behind in energy density. Making use of a metric like $/kWh does not take into account this important attribute in the discussion. It might be beyond the scope of this paper but a new metric like $/kWh(lifespan) might be more useful and reflect improvement in cycle life.

$$\$ / \int_0^{Cycle\ at\ 80\%\ init.\ cap.} kWh\ (cycle) * dcycle$$

   When financially modelling for battery systems for stationary storage with software like NREL's SAM, design life /cycle life is a very important parameter that NPV and IRR are super sensitive to.
2. It would be interesting to see other relevant chemistries in Sodium-ion like Prussian Blue Analogues and Solid-state sodium-ion batteries. One of the Sodium-ion battery startups is trying to commercialize Prussian Blue Analogues because of higher C-rates possible which are not captured in the $/kWh. This opens up completely new market segments where quick response rates are required like at data centers or for grid frequency regulation. Segmenting the market based on performance characteristics could be nice.
3. It could be interesting to consider a modelling scenario for P2/O3-type layered metal oxides that don't use nickel like $Na_1Mn_{0.5}Fe_{0.5}O_2$ or $Na_{0.67}Mn_{0.5}Fe_{0.5}O_2$. These chemistries are being explored for their use of abundant elements only (Na, Mn, Fe).
4. For drop-in compatibility, the authors rightly mention that there are very limited manufacturing learning requirements but it's important to qualify that there may be significant learnings involved for the pre-cell manufacturing steps. Cathode active material synthesis for Sodium-ion battery cathodes can be quite different than their layered lithium-ion batteries cathode active materials (NMC) counterparts. While both systems could use co-precipitation methods at scale, this synthesis method is studied far less at scale for sodium-ion batteries. Moreover, many layered metal oxide chemistries for NIB cathodes (particularly P2) can be air and moisture-sensitive and can require dry room handling of the cathode active material even at the cell manufacturing steps.
5. NIBs can prove to be super useful for the planned gigafactories in the Global South (like in India) since it might be easier to secure supply chains for NIB precursors. It would be nice to hear the authors' thoughts on that.

6. Figure 4 [original pre-print] axes don't have numbers so it's unclear how specific yearly projections are aligned with the axes.
7. The error bars for most predictions in this paper are not mentioned. Would be helpful to have error bars and CIs. A more explicit sensitivity analysis could be cool too.
8. As per my knowledge, BatPaC does not include cost modeling unit operations for active material synthesis and only takes that in as input. I was wondering if there are details available on how the granularity in cathode active material manufacturing cost was dealt with to reflect an accurate $/kg figure.
9. The abstract lists very generic insights gathered that could be said about any emerging technology. The paper is more specific than that and that could be reflected better in the abstract.

# Section 17: STEER Na-ion Commercialization Forum Event Notes[43]

A Na-ion Commercialization Forum event was held on April 15th, 2024 hosted by the authors at Stanford University. The Forum invited industry, government, and academia participants and was held under Chatham House Rule. Therefore, the takeaways summarized below do not attribute any information to a particular individual or source.

Pre-Panel 1
- Battery minerals markets are highly volatile and sensitive to shocks because production is highly concentrated, and the time to establish new mines is greater than 10 years. Futures markets are needed to stabilize against this volatility.
- Assuming net zero carbon emissions and 86% EVs by 2050, be prepared for Co, Cu, Ni, and Li prices to stay high for the next decade. In the "business as usual" case, prices stay low. The impact of technology development and policy makes this very hard to predict.
- Minerals are also unlike most economic commodities because the supply elasticity is usually low. A 10% increase in demand for Co causes a 3.3% increase in supply. Li is the exception, with a 10% increase in demand causing a 22% increase in supply.
- Unlike the oil market where there are global producers and the maximum market share by a single country can be up to 20%, minerals are strongly impacted by trade restrictions.
    - Example: if China + Russia and EU + USA form respective blocs, graphite costs in the EU + USA bloc will rise by >150%, and Li costs in China + Russia bloc will rise
- To insure against uncertainty, the following are needed:
    - Deeper futures markets, which provide short-term liquidity
    - Setup of multilateral green corridors for the transport of critical minerals
    - Global data sharing for supply and demand for critical minerals

Panel 1
- The business case for EVs is strongly dependent on how committed we are to the energy transition. They can't stand on their own against ICEs due to the impact of mineral availability and price on the supply chain. The domestic supply chain cannot stand on its own.
- ESG is not the biggest problem for the current Ni mines being set up in the US. Permitting in the US, Australia, and Canada takes much longer than it should. There is a strong appetite for investment from OEMs and battery makers - for IRA benefits they need to invest in mining projects in the US/Australia. Panasonic has invested in NMG in Canada, GM is also investing in mines. There should be no tolerance for OEMs in the US for incorporating Chinese Li - some OEMs are gaming the rules and taking taxpayer benefits to do this.
- China became the dominant force through heavy subsidies. Foreign companies were shut out of participating in China. Similarly, we need to protect the IRA, or else fossil fuels will stay cheap. Minerals discovery requires high-risk capital, so the industry needs security.
- Country risks are massive (changing governments, etc.). Infrastructure also matters - railroads in NV would help with a lot of mining efforts. Non-diluting grants help.
- Phosphate rock and Mn mining capacity (if mid-Ni chemistries kick-off) are not talked about enough.
- 30% of global soda ash is mined in Wyoming.
- The consumer side is highly sensitive to supply pressures. When Co became really expensive, Ni started becoming more prevalent. Similarly, LFP came along when Ni started becoming expensive.
- Fluorine is critical for nearly every component of batteries ($LiPF_6$, PVDF, natural graphite purification using HF, etc.). The US and Mexico have good reserves of fluorspar.

Pre-Panel 2
- Small improvements in Na-ion performance can lead to large changes in the $Li_2CO_3$ breakeven point, as cost and energy density are coupled at scale.
- Just like how Li-ion did not replace Pb-acid and instead opened up a new market, Na-ion should not try to replace Li-ion and instead try to open up new applications by playing to its competitive advantages, such as low-temperature utility, safety, high rate capability, etc. In the short term, one potential use for Na-ion could be as starter batteries in cold climates, instead of Pb-acid.
- To capture the diurnal market, only C/10 charge and discharge required, which may be possible with mm-thick electrodes.
- There are no standards for "battery grade $Na_2CO_3$" yet. Most are using food grade.
- NaOH cannot be used for sodiation like LiOH due to its hygroscopic nature.
- Currently, Na-ion technology manufacturing borrows a lot from the high Ni NMC space. Dry rooms for moisture sensitivity and material corrosion to furnace refractory ceramics are all similar. This causes dissonance because high-end equipment used to manufacture premium products like NMC 811 need to be used to make supposedly cheap Na-ion cathode products.
- Rather than drop-in compatibility, better to think of it as trickle-down compatibility.

Panel 2
- Qualifying materials at the same time as scaling up is very hard. Li-ion got to solve these problems separately.
    - CAM grade determines the quality of the cells.
    - Batch-to-batch variability exists, and accurate root-cause analysis is tough.
    - A good suite of physical and chemical characterization doesn't exist yet for Na-ion materials, but it does for Li-ion
    - High-volume coating is very hard to master, material characteristics change the coating properties drastically
    - Process parameters like pH and calcination temperature are critical to control
- Cathode synthesis and processing do not scale easily.
    - Cannot just make vessels (co-precipitation reactors and kilns) bigger - you need the same equipment but multiplied, which changes the cost depreciation. The kilns do not last long for Na-ion. In Chinese manufacturing plants, CapEx is cheap to enable three RHKs – one for pre-, and post- heat treatment as well as the actual calcination step. The West cannot afford that.
- If coating electrodes, expect a <50% yield rate for at least the first year of cell manufacturing. Experimenting at the production scale is very expensive.
- Recipe/chemistry swaps reset yield and productivity. Will take weeks for recipe changes and months for a chemistry change at the gigafactory scale.
- Li-ion has consolidated the energy storage market much faster than expected 10 years ago. Expect only incremental changes going forward.
- Problems for academia
    - 4-hour formation instead of 4-days
    - Make better binders
    - Cheaper raw materials than TM sulfates
- Current capacity probably exceeds demand. Expecting oversupply of Na-ion and LFP for the next 5 years.
- Dry room costs are very high on small scales. The dew point depends on the scale.
- Equipment suppliers are all Korean, Japanese, and Chinese. They have 30 years of experience. Massive lead times (2 years for a good coater).

Pre-Panel 3
- Announced Na-ion capacity to-date: 76% layered oxide, 20% polyanion, 4% PBAs. 100% of anodes are hard carbons.
- Hard carbon production limits Na-ion cell properties.
- Most announced capacity is from Tier 2 and 3 manufacturers in China

Panel 3
- 50% of the cost of ESS is cells, and 75% cost of EV batteries is cells.
- Important to consider the levelized cost of storage.
- LFP generates very large pack-level imbalances which are absent for NMCs.
- Roughly inverse relationship between safety and energy density. If chemistries are safer (thermal propagation, flammable gasses, etc.), larger cells can be made.
- The unique combination of cycle life, safety, and price advantage keeps Na-ion competitive.
- Stationary storage has fewer requirements (maximum C/2 discharge rate) than EVs - Na-ion can start out here and learn.
- Important to stack all tolerances together and explore edge cases. E.g. EVs not working in Chicago winters are an example of a situation where different corner cases converge.
- Advantage for Na-ion:
    - Transmission losses are a bottleneck and necessitate storing batteries closer to consumers. Safety makes Na-ion more attractive, allowing thicker electrodes which also leverage better ionic conductivity.
- Due to the slopiness of voltage profiles, Na-ion uses only a small section of the entire voltage range. Non-Ni cathodes might be favorable despite worse performance, such as $NaCrO_2$. Conductivity without Ni is a problem, but carbon coating can be carried out for non-Ni cathodes.
- Need to explore the tradeoff between eating energy density losses from using only a small portion of the voltage profile and the cost of additional electronics required to be able to invert the full voltage range.
- For ESS, weight is not a limit, but volumetric energy density is critical. NFPP and polyanions come out very unfavorably here due to their poor crystal and tap densities.
- Scaling in China is not based on what is the best chemistry, but instead on what is easiest to scale. This might explain why Ni, Mn, and Fe-based cathodes are being scaled, as there is a lot of know-how already from using these elements in Li-ion cathodes.
- China doesn't work on Cr because they are a net importer. It is mined in India, South Africa and the eastern EU, where there is no established battery industry.